# Clues for Solar System Formation from Meteorites and their Parent Bodies

**Bernard Marty**

Université de Lorraine, CNRS, CRPG, F-54000 Nancy, France

bernard.marty@univ-lorraine.fr

**Katherine R. Bermingham**

Department of Earth and Planetary Sciences, Rutgers University, Piscataway, NJ 08854, USA

**Larry R. Nittler**

Carnegie Institution of Washington Earth and Planets Laboratory

5241 Broad Branch Road NW, Washington DC 20015, USA

*Now at:*

School of Earth and Space Exploration, Arizona State University

Tempe, AZ 85287, USA

**Sean N. Raymond**

Laboratoire d'Astrophysique de Bordeaux

Université de Bordeaux, CNRS

B18N, allée Geoffroy St Hilaire, 33615 Pessac, France

# ABSTRACT

Understanding the origin of comets requires knowledge of how the Solar System formed from a cloud of dust and gas 4.567 Gyr ago. Here, a review is presented of how the remnants of this formation process, meteorites and to a lesser extent comets, shed light on Solar System evolution. The planets formed by a process of collisional agglomeration during the first hundred million years of Solar System history. The vast majority of the original population of planetary building blocks (~100 km-scale planetesimals) was either incorporated into the planets or removed from the system, via dynamical ejection or through a collision with the Sun. Only a small fraction of the original rocky planetesimals survive to this day in the form of asteroids (which represent a total of ~0.05% of Earth's mass) and comets. Meteorites are fragments of asteroids that have fallen to Earth, thereby providing scientists with samples of Solar System-scale processes for laboratory-based analysis. Meteorite datasets complement cometary datasets, which are predominantly obtained via remote observation as there are few cometary samples currently available for laboratory-based measurements. This chapter discusses how analysis of the mineralogical, elemental, and isotopic characteristics of meteorites provides insight into (i) the origin of matter that formed planets, (ii) the pressure, temperature, and chemical conditions that prevailed during planet formation, and (iii) a precise chronological framework of planetary accretion. Also examined is the use of stable isotope variations and nucleosynthetic isotope anomalies as constraints on the dynamics of the disk and planet formation, and how these data are integrated into new models of Solar System formation. It concludes with a discussion of Earth's accretion and its source of volatile elements, including water and organic species.

# 1: ORIGIN AND DIRECT OBSERVATIONS OF SMALL BODIES

Cosmochemistry – the science of meteorites – has opened a window into the origin of the Solar System and the timeframe of its evolution. Outstanding advancements via laboratory-based analytical means now enable the exploration of the composition of primitive material down to the atomic scale. Coupled with astronomical observations and numerical modeling, cosmochemistry has provided a comprehensive framework for not only the inner Solar System but also for the comet formation region for which data are scarce and mostly obtained remotely. Besides remote observation, the laboratory analysis of cometary material was restricted to that of a few grains recovered by the Stardust mission and to some dusty grains inferred to be of cometary origin (i.e., some of the so-called interplanetary dust particles). In contrast, meteorites are abundant and diverse in composition. Aside from waiting for a future comet sample return mission, meteorites are the best material available to explore the diversity of Solar System material and its stellar precursors, the timing of disk formation and evolution, and the exchange of material between the outer and the inner regions of the disk.

This chapter provides an overview of cosmochemistry with emphasis on the composition and evolution of outer Solar System bodies, the precursors of today's comets. We also discuss how cosmochemical data are integrated into models of Solar System formation and evolution and provide constraints on the origin of volatile elements on inner planets and their potential relationship with comets. Our chapter complements others in Comets III: Chapter 1 on the Sun's birth environment (Bergin et al in this volume), Chapter 2 on the physical and chemical properties of protoplanetary disks (Aikawa et al in this volume), Chapter 3 on planetesimal (and cometesimal) formation (Simon et al in this volume), Chapter 4 on how cometary reservoirs were sculpted by the dynamics of the early Solar System (Kaib & Volk in this volume), Chapter 23 on the Asteroid-Comet continuum (Jewitt and Hsieh in this volume), and Chapter 18 on dust (Engrand et al in this volume).

## 1.1 Origin of small bodies

The Solar System formed from a portion of a molecular cloud (Adams, 2010) that was composed of ~99 % gas and ~1% dust. The fragment spun into a protoplanetary disk surrounding a nascent star, the proto-Sun (Turner *et al.*, 2014). Within this disk, grains aggregated into increasingly large planetary bodies (Hayashi, 1981) (**Figure 1**). The demographics of planet-forming disks in stellar clusters of different ages suggest that gaseous disks typically last for a few million years before evaporating (Haisch *et al.*, 2001; Mamajek *et al.*, 2009). While the detailed physical structure and evolutionary pathway of the disk remain debated, models agree on the central steps in the process. The disk around the nascent Sun underwent an episode of viscous spreading to larger orbital radii, reaching a radial extent of at least ~50 AU, judging from the outermost planetesimals thought to have formed locally in the cold classical Kuiper belt. From that point onwards, the gaseous disk evolved in a relatively quiescent manner. Current thinking is that the disk slowly dissipated by draining onto the central star, while a small portion of the outer disk expanded to conserve angular momentum. The mechanisms of angular momentum transfer remain a subject of active research (e.g., (Armitage, 2011; Turner *et al.*, 2014)). The final phases of evolution were likely dictated by photo-evaporation driven by energetic photons (X-rays and UV) from the Sun, with a potential contribution from nearby massive stars (Ercolano and Pascucci, 2017). The inner parts of the disk were evaporated first, and the inner edge of the disk swept outward in a final swan song. Here, the inner disk refers to the region inboard of where the gas giants formed, and the outer disk refers to the region outboard of the gas giants. Due to the establishment of a thermal gradient towards the Sun, water ice could not be stable close to the central star, defining a so-called snow line beyond which ice survived. Likewise, organic matter would tend to disintegrate close to the Sun, defining a limit labelled the tar line. Each of the major volatile species (e.g., $CO_2$) admits a line/distance from the Sun marking a phase change from solid to gaseous.

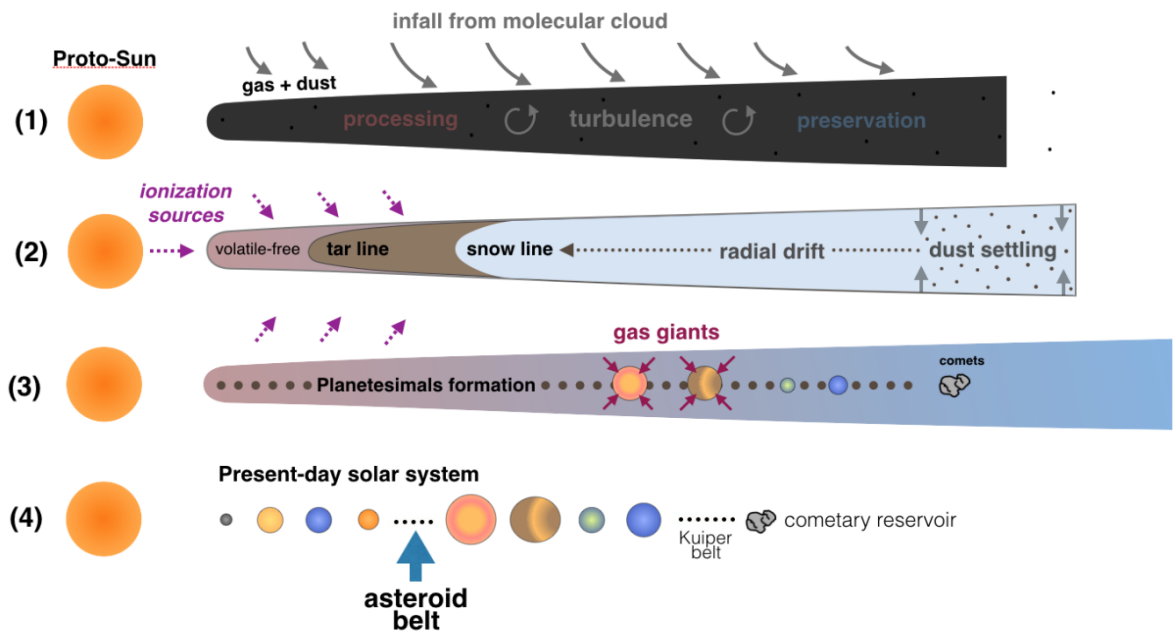

*Figure 1*: *Schematic evolution of the Solar System. (1): A protoplanetary disk formed around a central protostar from the collapse of a molecular cloud core made of gas and dust ~4.56 Ga. (2) The protoplanetary disk was chemically and thermodynamically zoned, with an inner region close to the Sun where volatiles could not condense, a tar line where refractory organic material could be stable, and a snow line beyond which water could condense. This temperature gradient determined the composition of volatiles in small planetary bodies. Those bodies that accreted between the tar line and the snow line did not accrete substantive water ice, whereas those that formed beyond the snow line accreted water ice. (3) Small planetary bodies evolved to yield planetesimals and planetary embryos. Gas giants formed from the accretion of planetesimals and from capture of the protosolar nebula gas. In the inner part of the Solar*

*System (the region inboard of where the gas giants formed), planets grew essentially dry as thermodynamic conditions and/or timing (i.e., dissipation of the gas) did not permit gravitational trapping of nebular gas. (4) Meteorites originate from the main asteroid belt located between Mars and Jupiter (adapted from Broadley et al., 2022).*

Material that did not accrete to the Sun was either ejected into interstellar space or formed planetesimals. Some of these planetesimals grew into planets whereas others remained as small bodies (asteroids, Trojans, comets), and these thus preserve a record of the epoch of Solar System formation. The surviving planetesimals in the asteroid belt are the primary source of Earth's meteorites.

From the analysis of meteorites, it was realized that small bodies can be divided into two groups: "primitive" or "undifferentiated" bodies and "differentiated" bodies (Weisberg *et al.*, 2006) (**Figure 2**). Both groups variably comprised the building blocks of the planets. Primitive bodies, samples of which are also known as chondrites, are those that did not undergo wholescale melting (i.e., differentiation) and thus preserve a mixture of early formed Solar System solids. These meteorites provide a unique opportunity to document the first stages of Solar System evolution through the analysis of their components. Comets, for which few samples exist, are examples of primitive bodies that formed in the outer bounds of the Solar System. Differentiated bodies, samples of which are known as achondrites, are small bodies that underwent wholescale melting at high temperature, which resulted in segregation of the body into a silicate-rich crust and mantle, and a metal-rich core. Although these planetary bodies have lost their component's original petrologic characteristics, they provide insight into the earliest stages of planet differentiation.

The highest taxonomic division in meteorite classification defines material as either "carbonaceous chondrite, CC" or "non-carbonaceous, NC" (Warren, 2011) (**Figure 2**). Advancing earlier work by Trinquier et al. (2007; 2009), Warren (2011) compared the isotopic composition of bulk meteorites and found that all meteorites consistently fell into one of two distinct groups, NC or CC (see subsection 2.6.3.2). This bimodal grouping has become known as the "NC-CC isotopic dichotomy". The NC-CC notation describes the isotopic or "genetic" character of a parent body regardless of how the meteorite is classified by its mineralogy and petrogenesis. With the addition of several new chondrites and achondrites, it is now realized that meteorites, almost without exception, can be classified as either CC or NC.

From chronological studies of chondrites and achondrites, small bodies were accreted within a few (≤5) Ma of the birth of the Solar System which is marked at time zero or $t_0$. This attests to the rapidity at which a stellar system with its cortege of planets can form (Morbidelli *et al.*, 2012), where such information can only be gained from laboratory-based meteorite analysis. Other primitive materials include comets, outer Solar System objects. The lack of $^{26}$Al (a useful short-lived chronometer which is described in section 2.6.1) in cometary grains returned by the Stardust mission, and the possible genetic link between CR chondrites and outer Solar System objects suggest a protracted formation of related bodies, but such evidence is extremely weak. Dating cometary matter, however, is of outmost importance and will be a key priority of future cometary missions.

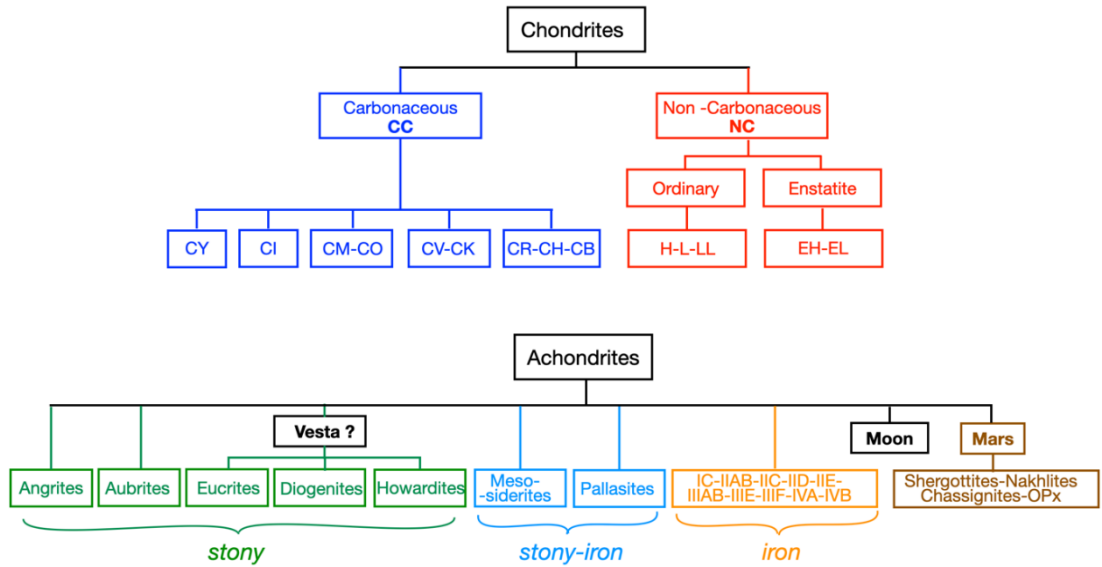

*Figure 2. Classification of meteorites, simplified after Weissberg et al. (2006), Warren (2011) and Bischoff et al. (2011). Some of the meteorite groups (e.g., R-chondrites) are not represented. The classification between carbonaceous chondrites (CC) and non-carbonaceous chondrites (NC) is given for primitive meteorites only (see section 2.3.). Almost all the achondrites can be genetically linked to NC or CC (see Bermingham et al., 2018a, Spitzer et al., 2021, and refs. therein)*

### 1.2. Knowledge of small bodies from observations

#### 1.2.1. Telescopic and Spacecraft Observations

Telescopic and spacecraft observations of the composition and structure of small bodies provide the most informative approach to determining the diversity of bodies that formed from the protoplanetary disk. The 1801 discovery of 1Ceres by Giuseppe Piazzi ushered in a string of telescopic discoveries of other small planetary bodies orbiting beyond Mars, which were named "asteroids" by William Herschel in 1802. More than a million asteroids have now been identified, many through large surveys designed to track potential impact hazards. Most orbit the Sun between Mars and Jupiter in the main asteroid belt, but a small fraction have been gravitationally perturbed into orbits that come much closer to the Sun. Those whose perihelion distance is less than 1.3 AU are termed "near-Earth" objects. In addition, some asteroids, called "Trojans," share Jupiter's orbit, either leading or trailing the largest planet by $60°$.

Telescopic and spacecraft observations of asteroids have revealed a wide diversity of colors, albedos, and spectroscopic features reflecting the variety of parent body compositions in the Solar System. The classification scheme derived is based on distinct spectroscopic types of parent body. Notably, the different asteroid types are not uniformly distributed in the asteroid belt (**Figure 3**). For example, E- and S-types dominate the inner asteroid belt, while C- and D-types dominate further out. S-types show spectral features of silicate minerals olivine and pyroxene and have been long suggested to be related to ordinary chondrites, the most common type of meteorite found on Earth. (**Figure 2**), This link has been proven by the Hayabusa mission that sampled a S-type asteroid and returned to Earth ordinary chondrite – type material. In contrast, C- and D-type asteroids are darker, spectrally flat, and thought to be related to carbonaceous chondrite meteorites. Such a compositional gradient likely reflects initial gradients in the protoplanetary disk, but there is clear evidence for substantial mixing of different compositional types throughout the asteroid belt (DeMeo and Carry, 2014).

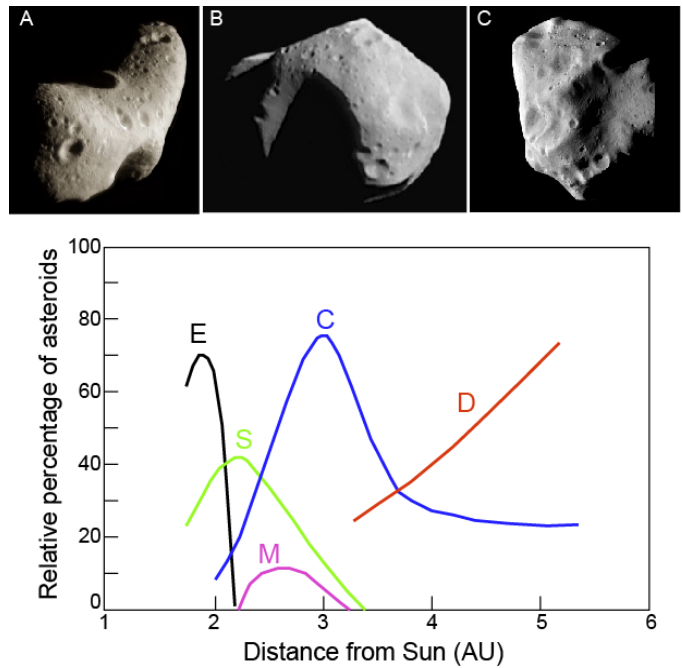

*Figure 3:* *Asteroid classification and distribution. Top: A) S-type Eros (image from NEAR-Shoemaker, NASA), B) C-type Mathilde (NEAR-Shoemaker, NASA), C) M-type Lutetia (Rosetta, ESA). Bottom: A schematic representation of distribution of major asteroid types (after Gradie and Tedesco, 1982). The symbols refer to different asteroid spectral types; C-types dominate the mass budget of the belt (DeMeo and Carry, 2013).*

### 1.2.2. Sample return missions

Several space missions have returned rocks from planetary bodies, yielding exceptional insight into their nature and their evolution (see the chapter by Snodgrass et al. in this volume). The return of samples from the Moon during the Luna (Russia) and Apollo (USA) missions, although not the main objective of these missions – which was geopolitical – proved the value of sample-based laboratory analysis. Three decades later, NASA's Genesis mission sampled Solar wind ions implanted into pure target materials over a period of 27 months and returned its precious cargo in 2004 (Burnett and Team, 2011). Analysis of these samples resolved two of the most pressing problems in cosmochemistry: the triple isotope oxygen ($^{16}O$, $^{17}O$, $^{18}O$) composition of the Sun (McKeegan *et al.*, 2011) and the nitrogen isotope variability in the Solar System (Marty *et al.*, 2011).

The era of small planetary body sampling came with NASA's Stardust mission (**Figure 4**), which sampled grains from the coma of comet 81P / Wild and returned them to Earth in January 2006. The analysis of the cometary samples revealed the presence of high temperature phases (metal, chondrules, CAI, olivine, pyroxene) such as those found in primitive meteorites. For the first time, this demonstrated the exchange of matter at the protoplanetary disk scale between the inner and outer regions of the Solar System (Brownlee *et al.*, 2006).

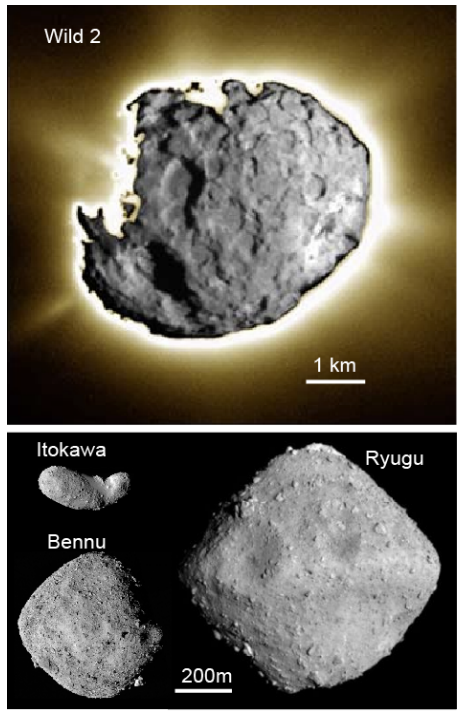

*Figure 4. Small body targets of sample return missions. Top: Comet Wild 2; from which coma grains were sampled by the NASA Stardust spacecraft and returned to Earth in 2006. Bottom: asteroid targets of sample return missions. Thousands of micro-sized grains were sampled from S-class Itokawa by the Japanese Space Agency (JAXA) Hayabusa spacecraft and returned in 2010. About 5 g of dust and pebbles was sampled on Ryugu, a C-type asteroid, by the JAXA Hayabusa2 spacecraft. The sample return capsule landed in December 2020. The NASA OSIRIS-REx mission sampled about 0.5-1 kg of material from asteroid Bennu, another C-type asteroid, and the sample canister is expected to land in 2023. Photo credits: NASA and JAXA.*

The first asteroidal sample return came with the Hayabusa mission from The Japan Aerospace exploration Agency (JAXA). This technologically complex mission was only half-successful due to engine and sampling problems, however, a few micrograms of dust from S-class asteroid 25143 Itokawa was returned to Earth. A total of 1543 particles, ranging in size from 3 to 40 micrometers have been analyzed so far. The sample analysis confirmed that the chemical composition and structure of the particles were similar to those of the most common meteorites on Earth, ordinary chondrites (**see Section 2.3**), which originate from S-type asteroids such as Itokawa (**Figure 4**) (Kawaguchi *et al.*, 2008).

The overall success of the Hayabusa mission paved the way for Hayabusa2 which used a similar spacecraft architecture, propulsion, and navigation systems, but with an improved sampling design (Watanabe *et al.*, 2017). Hayabusa2 was a huge technological success and a sophisticated sampling mission that permitted two different sampling sites, of which one was excavated before sampling with an explosive charge. Hayabusa2's target was 162173 Ryugu, a C-type (carbonaceous) asteroid. Unlike S-class Itokawa visited by the first Hayabusa probe, C-type asteroids are thought to be related to carbonaceous chondrites and likely to contain organic materials and volatiles and are, therefore, a target of choice for investigating the relationship between terrestrial atmosphere and oceans and primitive material. Ryugu is diamond-shaped with a diameter of about 875 meters and a rotation period is 7.63 hours (**Figure 4**). Its albedo of 0.047 is very low, possibly indicating a high abundance of organic matter (Watanabe *et al.*, 2019). The mission was launched on December 3$^{rd}$, 2014, and the sample return capsule landed in an Australian desert on December 6$^{th}$, 2020. It returned 5.4 g of asteroidal sample, fifty times more than the expected mass of 100 mg. Hundreds of scientists across the world are currently proceeding to elemental/isotopic analysis with the latest instrumentation and methods.

OSIRIS-REx (Origins-Spectral Interpretation-Resource Identification-Security-Regolith Explorer) is a NASA mission aimed at investigating the B-type carbonaceous asteroid Bennu and returning a sample to Earth (Lauretta *et al.*, 2017). The mission launched Sept. 8, 2016, from Cape Canaveral Air Force Station, the spacecraft reached Bennu in 2018 (Lauretta *et al.*, 2015; 2019) and sampled up to 1 kg of pebbles, a great success with respect to the planned nominal mass of 100 g. The sample return capsule will land in September 2023.

## 2. METEORITES

### 2.1. A little bit of history

Although meteorites have been known from the pre-Roman epoch, it took considerable time to recognize their extraterrestrial nature. Iron meteorites were often used for forging arms and decorations since the antiquity, and the origin of these bolides, which observed falls were often associated with formidable events (lightnings, detonations) was for long thought to be supernatural. The first documented fall of a meteorite in Europe took place near the locality of Ensisheim in Alsace (France) on November 7, 1492. The detonation was heard at several tens of km around and witnesses reported the fall of several stones in the nearby fields. Maximilien of Habsburg, under whose authority Ensisheim was placed, interpreted this extraordinary event as a favorable omen, and went to war against Charles VII of France who had taken his wife away from him a few years before. In 1794, Ernst Florens Friedrich Chladni (also known for pioneering work on acoustical physics) proposed that (i) masses of iron and stone fall from the sky, (ii) they create balls of fire and (iii) they come from space (Chladni, 1794). On April 26, 1803, a rain of stones fell over L'Aigle, a small town in Normandy, France, with remarkable environmental effects observed by numerous witnesses. The French Academy of Science sent Jean-Baptiste Biot, a young promising scientist (also known for his works on electromagnetism that led to the Biot and Savart law), to investigate the cause of this extraordinary event. Through a careful field investigation including collection of numerous pieces and the interview of peasants, Biot could demonstrate not only the extraterrestrial nature of the stones but also could evaluate their trajectories, a method which is now widely used to infer meteoritic origins in space and initial velocities. This study definitively established that meteorites are natural objects within the order of things (Gounelle, 2006a).

## 2.2. Falls and finds

Meteorites are rocks and metal alloys originating from planetary bodies that survive their atmospheric entry and reach the surface of the Earth. They provide the most direct access to the chemical and isotopic composition of the nascent Solar System. Fragments of planetary surfaces, either large bodies like Mars or the Moon, or smaller ones like asteroids or comets, are ejected through gravitational perturbations or impact events. When the meteor enters the atmosphere, heating, mechanical disruption, and chemical interactions with the atmospheric gases cause it to heat up and radiate energy. Consequently, meteor entries are often accompanied by strong visual effects, such as fireballs, and by detonations. Fresh samples can be recovered after impact if the landing site is found by witnesses. Recovered meteorites make only a few percent of the pre-atmospheric masses, the rest being destroyed during passage through the atmosphere (Zolensky *et al.*, 2006). Meteorite chunks obtained immediately after impact are called "falls", however, most meteorites are found after some time has passed by passers-by or during dedicated meteorite hunts in hot or cold deserts (e.g., deserts located in North Africa, South America, Australia, Antarctica). Estimates for the modern flux of meteorites are variable, of the order of a few thousands to a few tens of thousand tons per year, depending on the type of objects considered (Zolensky *et al.*, 2006) (see the chapter by Quanzhi Ye et al. in this volume for more insight into small bodies delivery).

## 2.3. Meteorite classification

### *2.3.1. Chondrites and achondrites*

Chondrites are primitive or undifferentiated stony meteorites that have not been modified by differentiation of their parent body. They are formed from an assemblage of components, including refractory inclusions and chondrules, which are discussed in more depth in Section 2.5, that existed before chondrite parent body accretion. As such, chondrites preserve some of the earliest Solar System-formed materials. Carbonaceous chondrites, a well-studied chondrite group, preserve stardust or circumstellar condensates, which are micrometer-sized mineral grains that condensed in stellar outflows or ejecta before the Solar System formed.

Achondrites originate from differentiated parent bodies that have undergone wholescale melting to form a core, mantle, and crust. These materials do not preserve components such as chondrules (hence their name), CAIs, or presolar grains, although these components may have originally been present in the parent body prior to differentiation. Silicate achondrites are from the crusts and/or mantles of differentiated asteroids. For example, eucrites, howardites and diogenites are considered to have been ejected from asteroid 4Vesta (Binzel and Xu, 1993; McSween et al., 2010). Iron meteorites are made mostly of metal (Fe and Ni alloy), and many of these sample cores of differentiated asteroids. Stony irons are meteorites are made of both silicate and metal, and include mesosiderites which are breccia resulting from impacts and pallasites which are made of large olivine crystals encapsulated into metal and thought to come from core-mantle interfaces or impacts (Yang *et al.*, 2007; 2010; Scott, 1977). Primitive achondrites are called primitive because they are achondrites that have retained much of their original chondritic properties, including relic chondrules and chemical compositions close to the composition of chondrites. Their petrology indicates have clearly experienced melting, but not substantial separation of silicate and iron melts leading to differentiation. A lunar origin for a specific meteorite clan has been attributed based on their close chemical and petrographic ties with samples returned by the Apollo missions. "Martian" meteorites have been identified as originating from the red planet thanks to their relatively young ages and remnants of Martian atmospheric gases trapped in glassy veins. The different types of meteorites are described in **Figure 2** and some examples are presented in **Figure 5**.

The formation of chondrite parent bodies was protracted by a few million years, typically 1 to 5 Myr relative to CAIs (MacPherson, 2014; Villeneuve *et al.*, 2009). Surprisingly, chondrites were not the first planetary bodies to accrete. They were preceded by the formation of small differentiated planetesimals, some of which formed within the first 1 Ma of Solar System history (e.g., Kruijer *et al.*, 2014; 2017). These early formed planetary bodies were destroyed by collisions and only remnants of their cores remain as iron meteorites to testify to their occurrence.

Although meteorites are the easiest to recover and provide the most mass for laboratory study, these cosmochemical samples make up only a small fraction of the flux of extraterrestrial materials on Earth. The peak of the meteorite mass flux to Earth corresponds to grains of a few hundred micrometers in diameter. Particles of this size are usually highly heated during atmospheric entry (producing the commonly observed meteors or "shooting stars") to the point of vaporization or extreme melting. Those that survive to Earth's surface are termed micrometeorites **(Figure 5E)**. Melted micrometeorites were first found in deep sea sediments in the 1870s and termed "cosmic spherules." In recent decades, melted and unmelted micrometeorites have largely been recovered by melting and filtering snow and ice in Antarctica (Duprat *et al.*, 2010, **Figure 5E**), though some can be found even in urban environments. Many primitive micrometeorites are chemically and mineralogically similar to larger meteorites, especially carbonaceous chondrites, and are thus likely asteroidal in origin, though some are extremely C-rich and may have originated from comets (Duprat *et al.*, 2010). Even smaller grains (<100 micrometers) are collected in Earth's stratosphere by special collectors on aircraft (Brownlee, 1985) or recovered from Antarctica (Noguchi *et al.*, 2015; Taylor *et al.*, 2020), and are known as interplanetary dust particles. These also are considered to originate in both asteroids and comets and are discussed elsewhere (see the Chapter by Engrand *et al* in this volume).

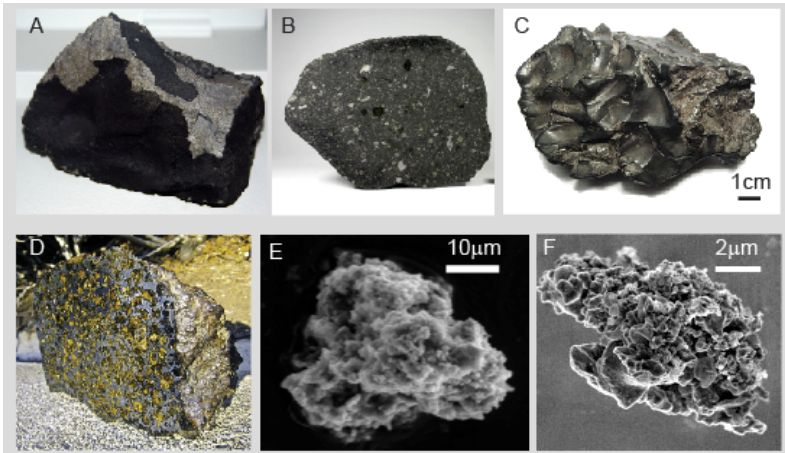

***Figure 5.*** *Types of meteoritic material from asteroids. A) Ordinary chondrite (Ochansk https://en.wikipedia.org/wiki/Ordinary_chondrite#/media/File:Ordinary_chondrite_(Ochansk_Meteorite).jpg) B) Carbonaceous chondrite (Allende, https://en.wikipedia.org/wiki/Allende_meteorite#/media/File:Allende_Matteo_Chinellato.jpg) C) Iron meteorite (Sikhote-Alin) (photo by H. Raab, licensed under CC BY-SA 3.0) D) stony-iron meteorite (3-kg piece of Brahin pallasite) (photo by S. Jurvetson, licensed under CC BY 2.0) E) Antarctic micrometeorite (courtesy of J. Duprat); F) Hydrated interplanetary dust particle (IDP) (courtesy S. Messenger).*

### *2.3.2. Constituents of chondrites*

Primarily, chondrites consist of three main components: chondrules, refractory inclusions, and matrix (**Figure 6**). Chondrules (from the ancient Greek word "chondros" for grain) are spherical objects that dominate the volume of ordinary and enstatite chondrites and are a major fraction of most carbonaceous chondrite groups (with the notable exception of CIs). Typically, they range in size from 0.1 mm to a few mm, and different chondrite groups showing distinct characteristic sizes. They are mineralogically and texturally diverse, but in general are composed of crystalline olivine and pyroxene minerals surrounded by Al-rich silicate material that is often glassy, and lower and variable amounts of Fe sulfides, FeNi metal, and/or other minerals. Their spherical shape and textures imply that they formed by rapid (hours) solidification of tiny molten droplets in space. How and where chondrules formed is still mysterious and highly debated. Some models advocate a nebular origin like incomplete fusion of solid precursors (Tenner *et al.*, 2017), or recycling of early condensates (Marrocchi *et al.*, 2018), whereas others call for planetary processes such as collisions between protoplanets (Libourel and Krot, 2007) or impact splash (Lichtenberg *et al.*, 2018; Sanders and Scott, 2012). Their ubiquity in undifferentiated asteroids suggest they played an important role in planet formation.

Refractory Inclusions (RIs) are sub-mm to cm-sized irregular objects consisting of minerals that form at higher temperatures than chondrule constituents. The most common type of refractory inclusions is made of calcium-aluminum-rich minerals (CAIs), amoeboid olivine aggregates (consisting of Mg-rich olivine mineral forsterite - $Mg_2SiO_4$, the most refractory ferromagnesian silicate - Al-diopside, anorthite and spinel), and refractory metal nuggets (RNM, alloys of highly siderophile elements). Calcium-aluminum-rich inclusions are considered to be the first solids to condense from the cooling of an initially hot nebular gas of Solar composition (Grossman and Larimer, 1974; MacPherson, 2014), and their chronometric age of CAIs of 4,567.30 ± 0.16 Ma constitutes the anchor point ($t_0$) of Solar System chronology (Connelly *et al.*, 2012). Ages of events in the forming Solar System are either expressed in absolute ages from the present or with reference to $t_0$ taken as the formation of CAIs. The proto-Sun and its parent molecular cloud core were obviously older than CAIs, but, in the absence of samples of their formation, events having occurred before the condensation of CAIs rely on analogies with stellar proxies and numerical astrophysical models.

The matrix surrounding CAIs and chondrules is a very fine-grained (sub-micrometer to a few micrometers) mixture of various phases, including tiny silicates, sulfides, metal grains (including fragments of refractory inclusions), presolar grains, and carbonaceous material. The amount of matrix varies from a few % in ordinary chondrites to nearly 100% in CI chondrites. Whether the matrices of different chondrite groups represent a common material that was added to the refractory inclusions, or each group has a distinct matrix is debated. Most carbonaceous chondrites also contain hydrated minerals which attest for liquid water once present in their parent bodies. Notably, both CC and NC stony meteorites contain organics in their matrix, which are more abundant in the CCs than in the NCs. Chondrites CB and CH are exceptions since they do not contain interchondrule matrix. This implies that NCs formed in a region between a snow line, beyond which water could condense, and a tar line, within which organics could survive (**Figure 1**).

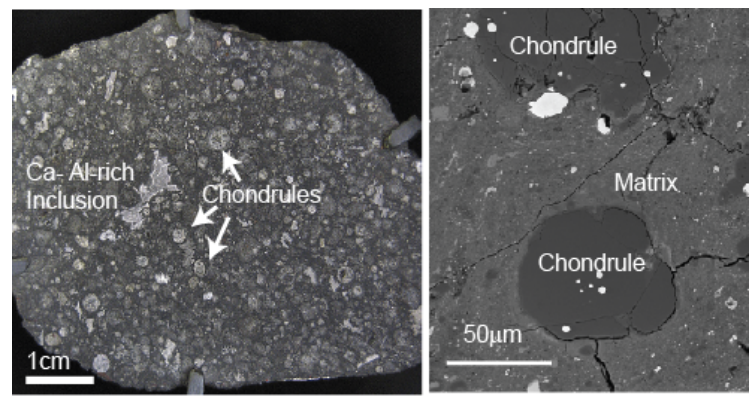

*Figure 6.* Left: Allende CV chondrite (CC by 2.0, Wikipedia). Right: Close up of Asuka 12169 CM chondrite (Nittler et al., 2021).

Presolar grains are preserved in the matrix of some carbonaceous chondrites. Presolar grains are microscopic solids that condensed in the outflows or ejecta of stars. These grains survived their passage through the interstellar medium and the protosolar parental molecular cloud, and the early stages of Solar System formation before they were accreted into parent bodies (**Figure 8).** Although only some rare meteorites preserve intact grains, presolar grains were inherited by all celestial bodies in varying proportions. Interstellar grains are also found in chondrites; however, these formed in interstellar space and have no direct link to specific stellar events. The term "presolar grains" includes both circumstellar condensates and interstellar grains, however, much of the literature uses "presolar grain", "circumstellar condensate", and "stardust" interchangeably, without reference to interstellar grains. The existence of presolar material surviving in chondrites was first indicated by isotopically unusual noble gas compositions. This led to a long process of trial-and-error chemical dissolution experiments, spearheaded by Ed Anders at the University of Chicago, that culminated in the discovery of bona fide presolar grains of nanodiamond, graphite and SiC (see Anders and Zinner, 1993, for a history of the discovery of presolar grains). Later technological advances such as the development of the NanoSIMS ion microprobe, led to the discovery of many additional presolar phases including oxides and acid-soluble silicates (which can only be identified by time-consuming brute-force SIMS analyses). The defining feature of presolar grains is their highly unusual isotopic compositions compared to everything that formed in the Solar System. As discussed below, stable isotope variations for materials that formed in the Solar System vary over relatively narrow ranges (tens of percent for O isotopes, factor of a few for N isotopes, parts per $10^6$ to $10^4$ for many heavy elements) reflecting the large degree of homogenization that took place during the early stages of planet formation. In contrast, presolar grains have isotopic ratios that range over many orders of magnitude for C, N, and O and tens of percent or more for heavier elements, reflecting almost-pure nucleosynthesis signatures. Because presolar grains are essentially pristine products of the stellar environments where nucleosynthesis occurs, they have proven invaluable for probing detailed processes of stellar evolution, nucleosynthesis, stellar dust production, interstellar dust processing, and the initial starting materials of the Solar System (Nittler and Ciesla, 2016; Zinner, 2003; see chapter by Bergin *et al*. in this volume).

### 2.4. Chemical analysis of meteorites

The elemental and isotopic compositions of planetary materials reflect both the mixture of their composite building blocks and the products of parent-body processes (e.g., thermal and aqueous alteration processes). Chemical characterization of meteorites is thus of utmost importance. Telescopic, or spacecraft-based reflectance spectroscopy is used to obtain some mineralogical and element abundance information about the exposed surfaces of asteroids. Not all minerals, however, exhibit spectral features, and many processes complicate the interpretation of remotely detected spectra. For example, the interaction of airless planetary surfaces with the space environment (space weathering) can induce very strong spectral modifications. Based on the results of the NEAR and Hayabusa missions, S-type asteroids are related to ordinary chondrite meteorites, however, spectra of the former are much more steeply sloped towards longer wavelengths ("reddened") than laboratory chondrite spectra due to space weathering. Remote X-ray and gamma-ray spectroscopy can be also used to directly determine the elemental composition of planetary surfaces from orbiting spacecraft, albeit with limitations in both sensitivity and spatial resolution. Only one small body has been successfully characterized in this way: 433 Eros by the NEAR-Shoemaker mission, from which X-ray and gamma-ray data provided additional evidence that this S-type asteroid is a space-weathered ordinary chondrite (Nittler *et al.*, 2001; Peplowski *et al.*, 2015). Consequently, by far most of the detailed geochemical information we have on small bodies in the Solar System comes from laboratory analysis of meteoritic and returned samples.

The goal of laboratory-based sample analysis is to obtain chemical, isotopic, and textural information about a sample to determine its origin. Planetary scientists attempt to derive the composition of celestial bodies and the origin of the Solar System from the small fractions of the cosmos available to us in returned samples, whether they have been delivered to Earth naturally or via spacecraft. Although this approach is limited by small samples (grams of material or less), this is somewhat mitigated by the collection of exacting compositional data obtained and interpretation of these data in the context of geochemical and cosmochemical principles that describe element behavior during planet evolution.

The past 30 years has seen a revolution in instrumentation and associated analytical accuracy and precision that is available to planetary scientists. Given the variety of cosmochemical samples available for study, flexibility in instrumentation is required. Samples come in the form of either bulk specimens of a whole rock or its individual components (e.g., chondrules, CAIs). Mineral fractions or chemical leachates of either type of sample can be studied. Analyses of individual chondrules or refractory inclusions provide information about the physical and chemical conditions in the protoplanetary disk before they were accreted to their parent body. Analyses of whole rock samples provide information about the bulk mixture of phases comprising a sample. This informs us about the bulk composition of the parent body and thus the region of the disk that it sampled. The composition of differentiated asteroids cores can be constrained by iron meteorites, whereas silicate achondrites tell us about the silicate portion. Note, the composition of undifferentiated asteroids can typically be constrained by a sample from any part of the parent body.

### 2.4.1. Non-destructive analytical techniques

Compositional and textural information can be obtained from meteorites at the bulk sample scale to the grain scale using non-destructive techniques that permit analysis with little or no damage to the samples. At the bulk rock scale, simple visual examination of hand samples helps determine the texture and componentry, thus informing one where to sample further. At the grain scale, the primary instruments for non-destructive characterization are electron microscopes. This includes the scanning electron microscope or SEM, often referred to as an electron microprobe analyzer when optimized for precise elemental abundance measurements. In an SEM, a finely focused beam of electrons (from a few keV to a few tens of keV energy) is scanned over a sample. Various signals are detected, including secondary and backscattered electrons which reflect topography and average atomic number, X-rays emitted by the atoms present, and cathodoluminescent light emitted by some minerals. X-ray spectrometers allow the quantitative determination and mapping of elemental abundances on scales from 0.5 to hundreds of micrometers. Together with optical microscopy, SEM data are crucial for overall characterization of the mineralogy, textures, and major-element chemistry of planetary samples.

Other widely used non-destructive techniques include laser Raman spectroscopy, which uses incoherent scattering of light off molecular vibrations to characterize certain minerals and phases, synchrotron-based X-ray micro-spectroscopies and transmission electron microscopy, which uses high-energy electrons to probe the atomic structure and composition of materials at nanometer to micrometer scales. Note that even if a technique is non-destructive, often some sample destruction is necessary to prepare it for analysis, e.g., through polishing or preparation of thin sections for bulk rocks or slicing of dust particles with a diamond knife attached to an ultramicrotome.

### 2.4.2. Destructive analytical techniques

*2.4.2.1. Sample preparation*

Destructive analytical techniques involve the partial or complete destruction of a sample to obtain elemental and/or isotopic data. These techniques require careful preparation of the sample prior to introduction into the instrument. To limit contamination of the sample during sample preparation, this work is often performed in a clean laboratory which is designed to control airborne contaminants by reducing their concentration in the air. A typical city environment contains 35,000,000 particles per cubic meter (0.5 micrometer and larger in diameter), whereas clean lab operations may require environments of 1,000 particles per cubic meter. Some laboratories that specialize in siderophile (iron-loving) element work demand further design restrictions by requiring the laboratory also to be metal-free.

Preparation of a bulk rock sample requires isolation of a rock chip from the main meteorite mass which is typically housed at a museum or university. If the sample is friable (e.g., carbonaceous chondrites), a section can be broken off by hand. Most other samples, however, require cutting into the main mass using a saw blade, and subsequent crushing into a powder. Alternatively, meteorite components can be isolated from a sample by gently crushing it and physically picking out phases of interest using tweezers and a binocular microscope for magnification of the worksite. Micro-drilling or micro-coring isolates meteorite components by drilling around the desired phase until it can be physically removed. Meteorite components can also be separated using freeze-thaw disaggregation where the sample is cycled multiple times through immersion in ultrapure water, freezing, and thawing until it naturally fragments and components can be physically removed.

If individual minerals of a bulk rock or meteorite component are of interest, they can be separated via gravity using heavy liquids to isolate minerals of different densities. Alternatively, if minerals have different shapes, they can be separated based on their shape. At the microscopic level, a focused ion beam (FIB) system can be used to extract specific site-selective materials for analysis. This is used frequently for transmission electron microscopy (TEM) and/or synchrotron methods where very thin samples (<<1 mm) are required. In FIB, a finely focused beam of ions (usually Ga or Xe) is used to sputter material with nm-scale precision.

In some cases, extracting microscopic phases that are distributed throughout a sample requires a brute force approach. Some types of presolar grains are isolated by sequentially dissolving or "digesting" the bulk meteorite sample in concentrated mineral acids (e.g., hydrofluoric acid, hydrochloric acid, nitric acid, and perchloric acid) to remove the main mass until a residue remains (methods pioneered by Ed Anders; e.g, Lewis *et al.*, 1987). This residue contains chemically resistant presolar grains (e.g., nanodiamond, nanospinel, corundum, SiC, graphite) which can be subsequently analyzed as a slurry or individual phases **(Figure 7)**. Similarly, refractory organic matter that dominates the carbon budget in primitive meteorites can be chemically purified through dissolution of inorganic phases (Alexander *et al.*, 2017).

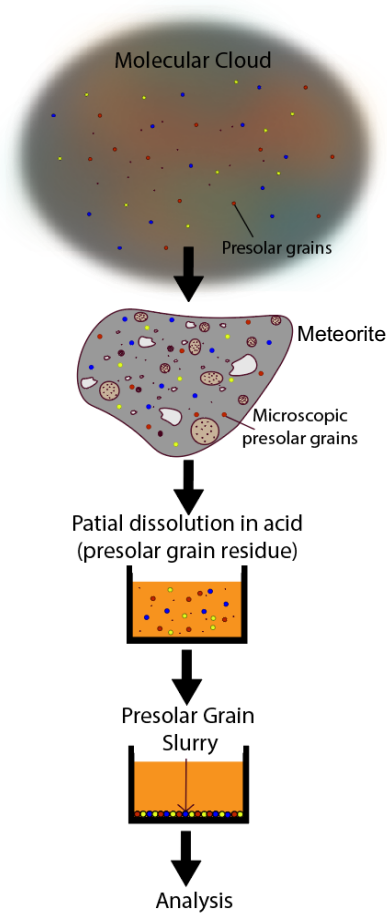

*Figure 7:* *Schematic depicting the history of presolar grains. Presolar grains produced in stellar events that proceeded the formation of the Solar System accumulate in the parent molecular cloud. The protoplanetary disk accretes from a fragment of the molecular cloud. Meteorites which sample the protoplanetary disk thus also sample the mixture of presolar grains that were in the molecular cloud. Presolar grains are isolated from chondrites by dissolving all other phases in the meteorite to leave a slurry of presolar grains. These gains are then analyzed as a mixture of grains or individual grains using high resolution mass spectrometry. Figure based on Nittler (1996)*

Some instruments require the sample to be liquified for introduction into the instrument (e.g., thermal ionization mass spectrometry TIMS, and inductively coupled plasma mass spectrometry ICP-MS). Liquification is achieved by digesting the rock sample in concentrated mineral acids, the type of which is dependent on the type of sample being dissolved **(Figure 8)**. Dissolving metal-rich samples can be done using concentrated hydrochloric acid, or mixture of concentrated hydrochloric acid and nitric acid ("aqua regia"). Silicate rock samples are typically dissolved using a combination of concentration hydrofluoric acid and nitric acid. Hydrofluoric acid or cesium fluoride (Brooks, 1960) breaks silicon-oxygen bonds resulting in dissolution of the silicate phases.

Some instruments require elements to be preconcentrated or "purified" from the sample prior to introduction into the instrument. This may be because the element is low in concentration such that its signal would fall below the detection limit of the instrument or there are interferences on the element of interest caused by accompanying elements in the bulk sample solution (Potts, 1992). A commonly used method for isolating elements from a dissolved sample is ion exchange chromatography **(Figure 8)**. The principle of ion exchange chromatography is to separate ions via reversable exchange with a solid resin based on the ions' differing affinities to the resin (Potts, 1992). Briefly, the sample is dissolved in acid prior to introduction to the resin bed which is packed into a Teflon or glass column. An ion's affinity to the resin changes by modifying the acidic environment of the column. Once the sample is loaded onto the resin, elements are eluted from the resin sequentially by washing the column with acids that were selected based on calibration tests run prior to the column chemistry. Multiple elements of interest can be isolated from a single elution scheme, each of which are sufficiently pure to analyze using mass spectrometry. The efficiency of ion exchange chromatography makes it one of the most commonly used methods of element purification of meteorite samples.

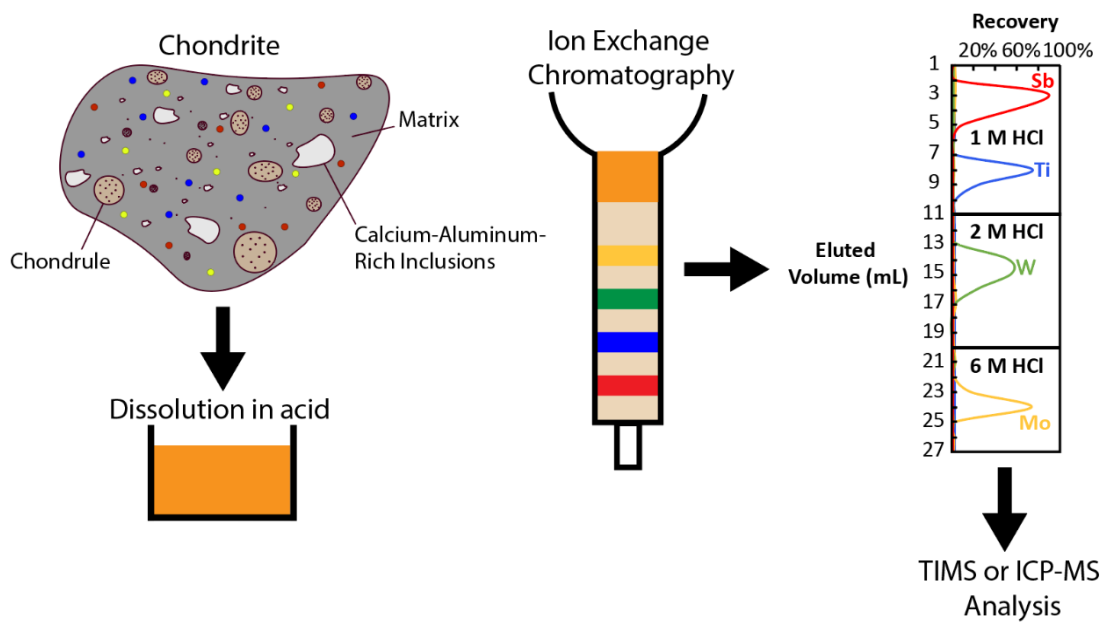

***Figure 8:*** *Schematic depicting complete dissolution (including presolar grains) of a meteorite in concentrated acid. To isolate elements of interest, the resultant solution is passed through a resin bed housed in a column. A sequence of acids is then washed through the column to elute elements of interest that wash off the column in bands. Solutions can then be used for mass spectrometric analysis.*

### *2.4.2.2. Bulk rock elemental chemistry X-ray fluorescence spectrometry (XRF)*

Major element (Na, Ma, Al, Si, P, K, Ca, Ti, Mn, Fe) and trace element (Ba, Ce, Co, Cr, Cu, Ga, La, Nb, Ni, Rb, Sc, Sr, Rh, U, V, Y, Zr, Zn) compositions of bulk samples can be determined using X-ray fluorescence spectrometry (XRF). This analytical technique was commonly used in silicate rock analysis in the late 1960s, following development in the commercial industry (Norrish and Hutton, 1969; Leake *et al.*, 1969; for review see Potts, 1992). Briefly, X-ray fluorescence spectrometry requires a sample to be powdered and fused into glass disks or compressed into pellets. The sample must be thoroughly homogenized to obtain representative bulk rock elemental measurements. Without this, the nugget effect can occur, where there is an uneven sampling of phases that are concentrated in some elements leading to artificially high abundances of those elements in the bulk rock data. After sample preparation, the pellet is exposed to X-ray radiation which causes ionization of discrete orbital electrons. After exposure, these electrons de-excite to the ground state and emit characteristic fluorescence X-rays particular to the element present. Measurements of the unknown samples are compared to those of previously well characterized rock standards with similar matrices to determine the absolute abundances of elements in the sample. Accuracy of this method, therefore, is reliant on a close fit between the standard, the sample matrix, and the composition.

### *2.4.2.3. Mass spectrometry*

Mass spectrometry is a commonly used analytical method that generates isotopic data via the separation of a sample's atoms or molecules based on their mass/charge ratios. Modern mass spectrometry is based on designs by Alfred Otto Carl Nier, who is widely seen as the "father of modern mass spectrometry" (De Laeter and Kurz, 2006). Many of Nier's determinations of the isotopic composition of minerals are still in use today. Mass spectrometers measure the isotopic composition of solids, liquids, and gases. This section focuses on four of the most commonly used mass spectrometry techniques in meteoritics: inductively coupled plasma mass spectrometry (ICP-MS), thermal ionization mass spectrometry (TIMS), secondary ion mass spectrometry (SIMS), and noble-gas mass spectrometry (NGMS). For all these methods measurements of the unknown samples are compared to those of known rock or gas standards to determine the absolute abundances and isotopic ratios of elements in the unknown. Accuracy of mass spectrometric methods (e.g., how close a measurement is to the true or accepted value) is reliant on a close fit between the standard and the sample matrix.

State-of-the-art TIMS and ICP-MS can generate very precise isotopic ratios (±5 ppm, where ppm refers to deviations in atom/atom of isotopic ratios in parts per million). These instruments require a minimum sample mass, which typically require element purification by ion exchange chromatography prior to sample introduction into the instrument. This limits isobaric interferences, increases ionization, and thus improves sensitivity of the element being analyzed. In TIMS, a purified sample is loaded onto a thin metal filament which is electrically heated under vacuum until the sample ionizes. A magnetic sector mass analyzer is then used to separate the ions based on their mass/charge ratio, and the resultant ion beams are directed into a detector which converts them into voltages. Inductively coupled plasma-MS couples a radiofrequency inductively coupled argon plasma with a mass spectrometer. Samples are introduced to the plasma via solution, or laser ablation of a solid sample. Inductively coupled plasma-MS is used to measure the isotopic composition of purified solutions, whole rock solutions, or a wide variety of solid materials.

Secondary ion MS is a highly sensitive in-situ technique that allows trace-element and isotopic measurements of specific phases on sub-micrometer to tens of micrometer scales, albeit with generally lower precision (±0.1%, depending on element) than obtained by TIMS or ICP-MS. In SIMS a primary beam (generally $Cs^+$ or $O^-$) is focused onto a sample and sputtered atoms are ionized and weighed with a mass spectrometer. Secondary ion MS instruments are often referred to as ion probes (or microprobes) and a distinction is often made between conventional instruments (e.g., SHRIMP, Cameca ims-1280), which are optimized for high precision but limited to spatial resolutions $>\sim 1$ μm and the Cameca NanoSIMS, optimized for spatial resolution (~0.1 μm).

Noble-gas mass spectrometry differs in the sense that noble gases are extracted from the sample by heating or crushing, so that the matrix effect does not plays any role. Noble gases are also named rare gases as their abundances are generally very low in rock samples, of the order of $10^{-10}$ to $10^{-18}$ mole/g. Their analysis therefore requires a different technique where these elements are extracted from rocks and mineral by heating with furnaces or lasers, or by crushing at room temperature. Gases are then purified under static vacuum (pumps isolated) and analyzed by static mass spectrometry, that is, without pumping during analysis. These conditions require drastic treatment of the ultra-high vacuum lines and analyzers including baking, long term pumping before analysis, and the exclusion of any organic material.

### 2.5. Using small bodies to constrain the composition of Solar System evolution

#### *2.5.1. Approach to defining the Solar System composition*

An important part of understanding the processes and events that define Solar System evolution is to determine the composition of the molecular cloud from which the Sun and planets formed and compare it to the composition of materials we have access to today. Most materials planetary scientists have access to have evolved from their original state. To derive the initial Solar System composition, planetary scientists search for objects that either have not undergone significant chemical modification, such as differentiation or aqueous or thermal processing, or sample very large portions of the Solar System thereby providing an average composition. This limits the selection of samples to a few candidates: the Sun, primitive meteorites, and comets.

The Sun contains ~99.8 % of the mass of the Solar System. Spectroscopic analysis of the Solar photosphere (~100 km deep surface layer of the Sun) is, therefore, a logical starting point to define the original elemental composition of the Solar System. Although fusion reactions are on-going since the Sun formed, their products either remain deep in the star or their surface expressions can be modeled, and those compositions subtracted from the present-day bulk composition of the Sun to derive an original stellar composition. The isotopic composition of the original Solar System is inferred from chondrites because, for the most part, isotopes cannot be measured telescopically.

#### *2.5.2. Goldschmidt's rules and cosmic volatility*

Victor Goldschmidt, colloquially known as the "Father of Geochemistry", determined the elemental composition of the Solar System by compiling vast quantities of chemical data on terrestrial and extraterrestrial rocks (Goldschmidt, 1937). He distilled relationships between elements and isotopes to define a set of principles that describe element behavior under a range of geologic conditions. Elements that tend to form silicates or oxides are called lithophile elements (rock-loving), elements that concentrate in metallic iron or metal alloys are called siderophile elements (iron-loving), elements that react with sulfur are called chalcophile elements (sulfur-loving), and elements that tend to form gaseous species and reside in the atmosphere of planetary bodies are called atmophile (gas-loving). By recognizing these affinities, it became possible to predict where elements would partition during planet formation and evolution. From Goldschmidt's data compilation, he derived a table of "cosmic abundances" (Goldschmidt, 1937). The elemental and isotopic composition of the Solar System has since been derived for all the elements and stable isotopes of the Solar System using Goldschmidt's approach (e.g., Suess and Urey, 1956; Anders and Grevesse, 1989; Lodders, 2003, 2021). Isotopic compositions for elements that are in gas form (e.g., volatile elements) are derived from giant planet atmospheres or from the analysis of the Solar wind either in-situ by spacecraft instruments or implanted in planetary surfaces or directly collected (e.g., the Genesis mission, see subsection 2.2.). The isotopic compositions reported in these compilations are average compositions and do not account for the small (part per million, ppm) isotopic variations that are documented in terrestrial, lunar, Martian, and meteorite samples.

In addition to Goldschmidt's rules, cosmochemistry requires an additional set of chemical principles to predict how elements partition during cooling of the protoplanetary disk. This is known as the cosmic volatility of an element. This involves the condensation of gaseous species to solids (the liquid state is rarely reached due to the very low pressures of space). Based on condensation experiments (Grossman and Larimer, 1974; Davis and Richter, 2007), the solids that formed via equilibrium condensation from a Solar composition gas and their associated temperatures were derived **(Figure 9)**, and most elements condense into solid solution with major element condensates. The 50% condensation temperature of elements are commonly used to describe the temperature where 50% of the element in condensed at a given pressure (e.g., P = $10^{-3}$ to $10^{-4}$ bar). The cosmic volatility of an element is expressed in relative terms. Highly volatile elements have 50 % condensation temperatures below 371 K, volatile elements below 665 K, moderately volatiles between 1335 and 665 K, and refractories above 1335 K (for a gas of solar composition at a total pressure of $10^{-4}$ bar; Lodders, 2003). The concept of condensation sequence works very well to explain the composition of refractory inclusions in chondrites as CAIs are predicted to be the first solids to form during cooling of a nebular gas in the canonical conditions described above. Their chemical and mineralogical compositions are in full agreement with the predicted mineral phases to form from a thermodynamical standpoint (**Figure 9**). As expected, CAIs are the oldest objects that formed in the Solar System (Connelly *et al.*, 2012). This excellent match justifies the concept of the condensation sequence and permits is application over the range of pressure and temperature that prevailed during the formation of planetesimals.

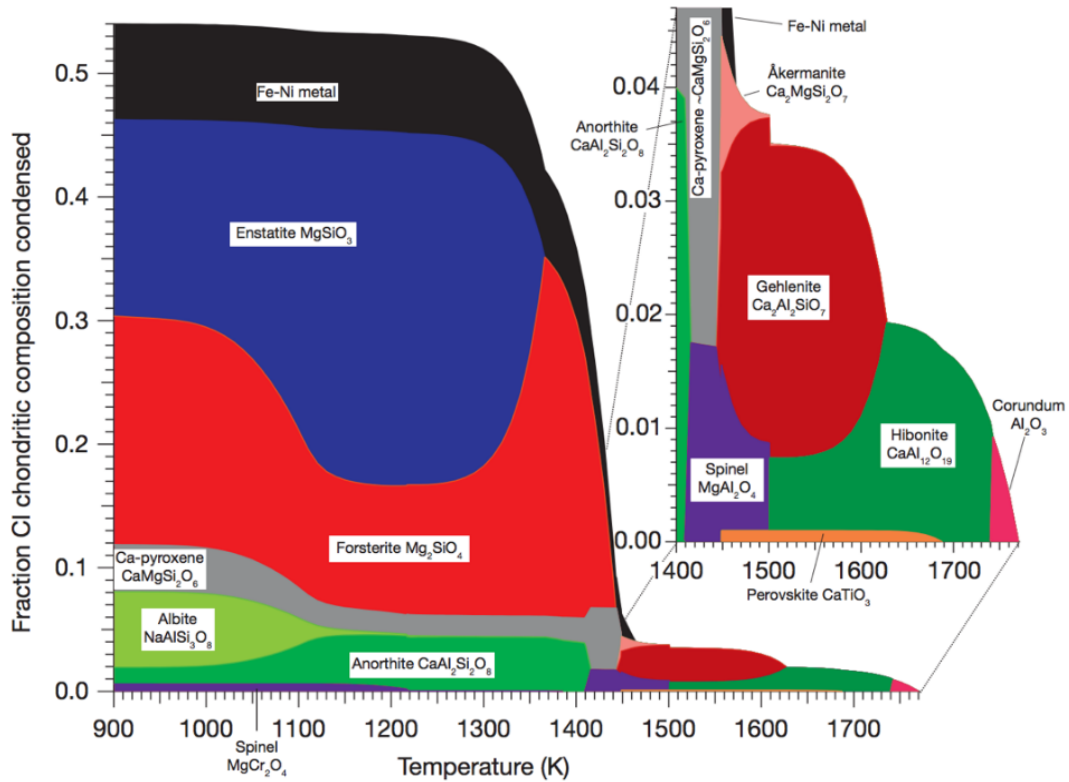

*Figure 9: Condensation of major rock-forming phases from a gas of solar composition at P = $10^{-3}$ bar (modified after Davis and Richter, 2007). The first phases to form upon cooling are Al-Ca-rich that are effectively observed in CAIs found in chondrites.*

### *2.5.3. The Solar System composition*

It has been long recognized that one group of chondrites, Ivuna-type or CI carbonaceous chondrites, have relative elemental abundances that correlate very strongly with those of the Solar photosphere (**Figure 10**). It is noteworthy that the Hayabusa2 mission has returned grains from C-type asteroid Ryugu whose composition matches very well that of CIs (Yokoyama et al., 2022; Tachibana et al., 2022). The exceptions are the most volatile elements that do not condense easily into rocks (e.g., H, He, N, C, O, Ne, Kr, Xe), and Li which has been destroyed in the Sun by nuclear reactions. Consequently, CI chondrites are considered to be the meteorite samples that best define the original Solar System's elemental composition (Wasson and Kallemeyn, 1988). This is despite the fact that these meteorites preserve signatures of extensive aqueous alteration.

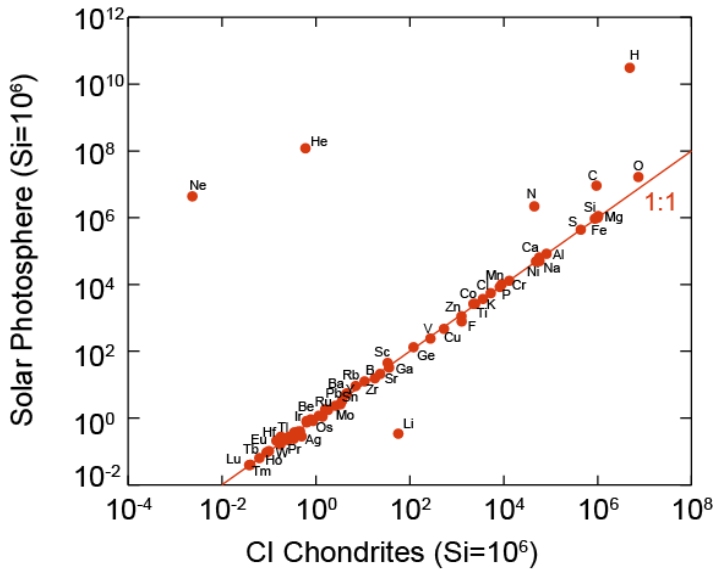

***Figure 10.*** *Abundances of elements as determined spectroscopically in the Solar photosphere are plotted against bulk abundances measured in CI carbonaceous chondrites (both normalized to $10^6$ atoms of silicon). Data from Lodders (2021).*

On a bulk scale, CI chondrites share the same elemental composition as the bulk Solar System (**Figure 10**). All the other groups of primitive meteorites (and hence their parent bodies), however, show elemental compositions that are fractionated to some extent (**Figure 11**).

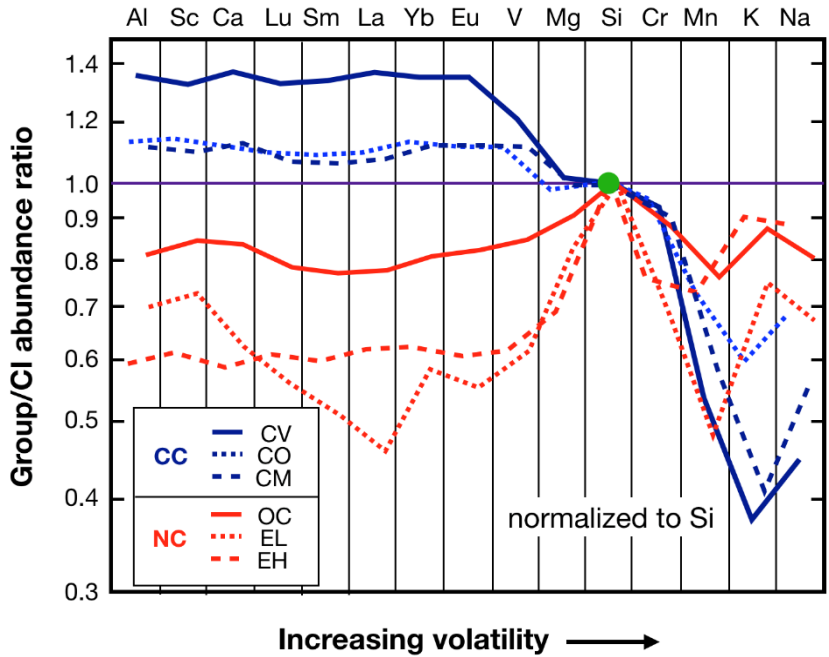

***Figure 11:*** *Si-normalized group/CI abundance ratios for lithophile elements in some of the chondrite groups. The elements are arranged from left to right in order of increasing volatility.*

*Cl chondrites are plotted on a horizontal line at unity. All groups, except EL, have a flat refractory lithophile element pattern, implying that these elements were in the same nebular component. In contrast, group differ on the relative abundances of non-refractory elements, as a result of different condensation environment and/or parent body processing (after Wasson and Kallemeyn, 1988).*

The bulk compositional diversity of small bodies indicated by both observations of asteroids and primitive meteorites indicates that it is highly likely that the rocky planets which accreted from smaller bodies also do not have bulk Solar composition. The estimated bulk composition of Earth has a chondritic refractory element composition, however, there are significant depletions in volatile elements according to their volatility (McDonough and Sun, 1995; Albarède, 2009; Mezger et al., 2021). It remains unclear to what extent the Earth's depleted volatile budget is a consequence of its accretion from differentiated planetesimals and/or if Earth accreted from chondrites but the process of process of differentiation caused a depletion in volatile elements. Although the chondrite analogy holds for refractory elements, it cannot explain the bulk chemistry of the Earth **(Figure 12)**. Further information can be gained from the isotopic signatures of small bodies compared to Earth's, which is the topic of Section 4.

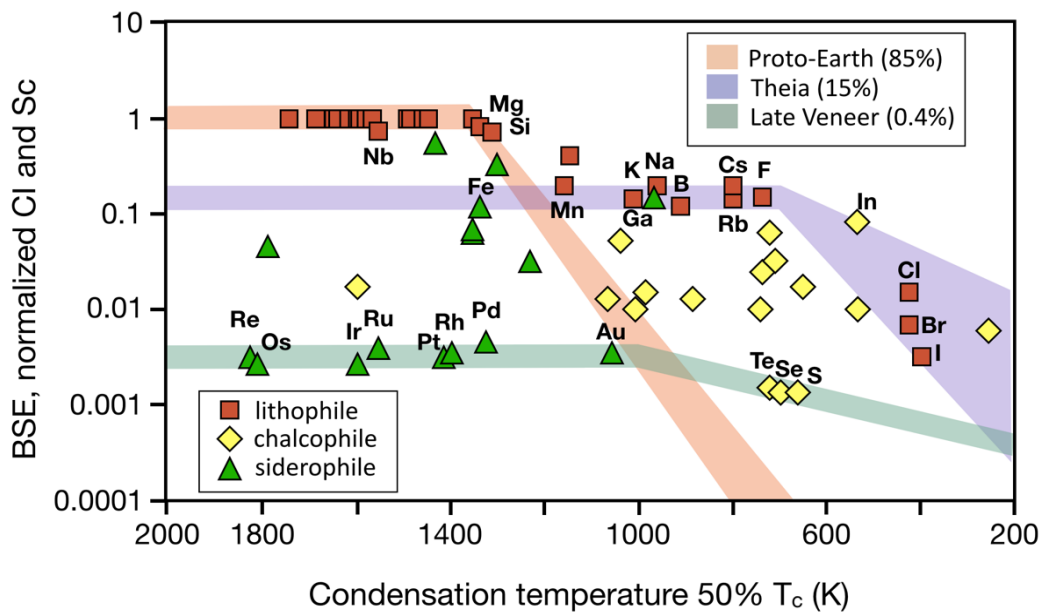

*Figure 12. Abundances of lithophiles (silicate-loving), chalcophile (sulfur-loving) and siderophile (iron-loving) elements of the bulk silicate Earth (BSE: bulk silicate Earth, i.e., bulk Earth excluding core) normalized to CI and Sc (after Mezger et al., 2021). Refractory elements are clearly chondritic in Earth, but non-refractory elements tend to be depleted with increasing volatility (expressed as temperature at which 50% of a given element is condensed in canonical conditions (Solar System composition, P= $10^{-3}$ bar). Note that volatile element depletions are much stronger than for chondrites. The complicated abundance pattern of the BSE could reflect the contribution of several sources, events and processes during Earth's accretion (Albarède, 2009; Mezger et al., 2021). In the model illustrated here, Mezger et al. (2021) propose a three-step process, with a dry proto-Earth to which the Moon-forming event involving collision with an embryo labelled Theia contributes 15 % of Earth's mass, followed by the addition of 0.4 %*

*chondritic material after this event. The last event, labelled the late accretion, accounts for the non-negligible abundance of siderophile elements in the BSE after core formation since the core should have left the mantle barren of siderophile elements, which is not observed (e.g., Wang and Becker, 2013, and refs. therein). This idea that the Moon-forming impactor brought Earth's volatiles is, however, difficult to envision from a dynamical point of view. Other models propose delivery of volatiles throughout Earth's building and/or during the last stages of accretion (see Section 4).*

*2.5.4. Deriving the stellar building blocks of the Solar System*

The estimated elemental composition of the Solar System is presented in **Figure 13** and shows the relative abundance of an element compared to its atomic number (Z). It is characterized by a saw-tooth pattern to the curve which gradually decreases in abundance with increasing atomic number, and peaks at the iron group nuclei and double peaks at A = 80 and 90, 130 and 138, and 194 and 208. The saw-tooth pattern of the Solar elemental abundance curve reflects the structure of atomic nuclei: nuclei with an even number of protons and neutrons are more stable than those with an odd number (Oddo, 1914). This stability translates into higher abundances during the nuclear processes that synthesize elements within stars. When plotting the sequential elemental abundance of a sample, the saw-tooth pattern observed obscures geochemical or cosmochemical trends in the elements which makes it difficult to derive the significance of elemental abundances. The saw-tooth pattern is removed if the composition is divided by or "normalized to" a Solar System composition (e.g., CI).

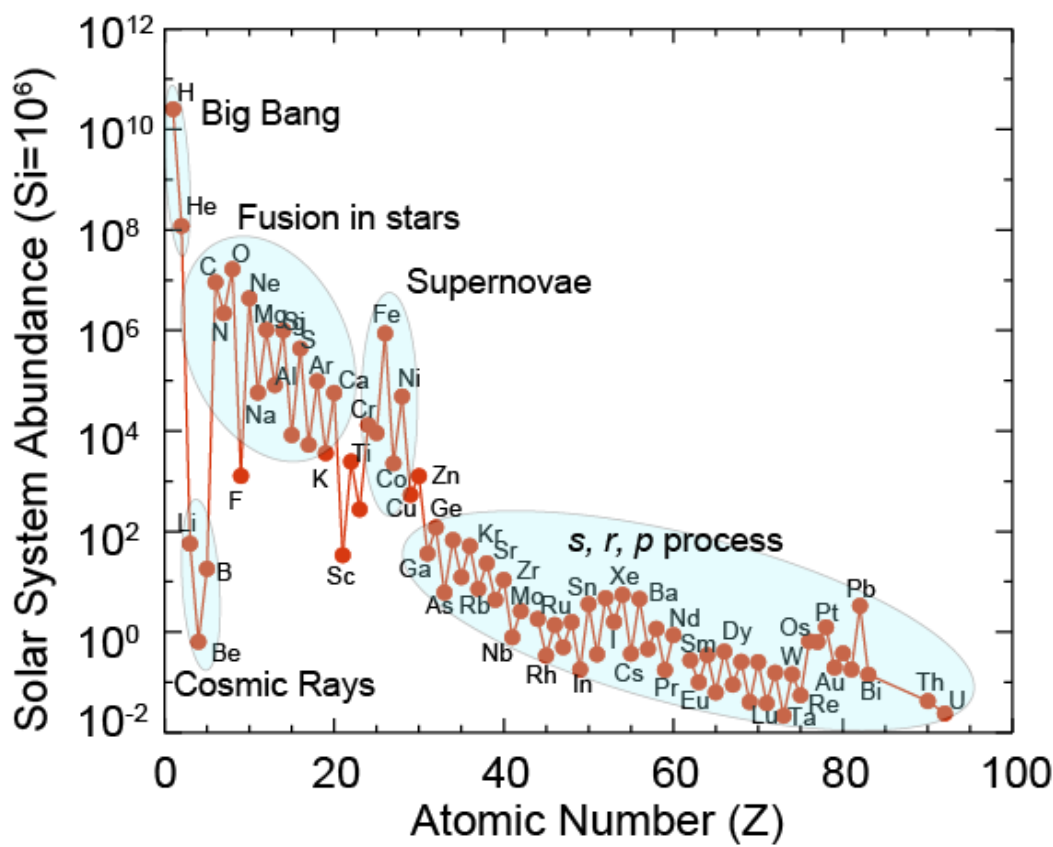

***Figure 13.*** *The Solar elemental abundance pattern. Data from Lodders (2021). Primary nucleosynthesis processes are indicated.*

The significance of most of the Solar elemental abundance curve's structure was realized in the 1950s and summarized in two seminal papers that laid out the basic principles of how the elements are synthesized in the Universe (Burbidge *et al.*, 1957; Cameron, 1957; 1973). Most of the basic ideas suggested in those papers have withstood the test of time, although the scientific community has a progressed to a far more detailed understanding of stellar evolution and nucleosynthesis.

Hydrogen (including deuterium), He, and a small amount of Li were synthesized shortly after the Big Bang as the Universe rapidly cooled, but the rest of the elements were made later, mostly in stars. Stars spend most of their lives powered by the exothermic fusion of H into He (H-burning) and subsequent burning of He in their cores to form $^{12}$C and $^{16}$O (He-burning). The low abundances of Li, Be, and B reflect the fact that these are bypassed during He-burning and in fact these elements are made by spallation, essentially fragmentation of heavier elements in interstellar space when they are bombarded by high-energy cosmic rays. Low-mass stars like the Sun do not reach temperatures high enough for fusion reactions to occur beyond He-burning and end their lives as a cooling white dwarf made of C and O. Thus, most of the elements between C and Fe are made in more massive stars (>~10 $M_\odot$, $M_\odot$ = Solar mass) as their cores burn ever heavier fuels. However, $^{56}$Fe has the highest binding energy of any nucleus and this means that exothermic fusion reactions cannot produce energy once the core has burned to iron. At this point, a massive star's core collapses under its own weight to form a neutron star or black hole and a reverse shock ejects the star's envelope, an explosion known as a Type II supernova. The peak around Fe in the Solar abundance pattern reflects the high stability of nuclei in this range of atomic number; these elements are produced in both Type II and Type Ia supernovae, the latter being thermonuclear explosions of white dwarfs due to mass transfer from a binary companion. The slow (*s*-process) or rapid (*r*-process) progressive addition of neutrons are the dominant processes responsible for creation of most elements heavier than Fe. The former takes place in asymptotic giant branch (AGB) stars, the end stage of evolution of stars < ~8 $M_\odot$, whereas the *r*-process occurs during mergers of neutron stars and possibly in Type II supernovae. Peaks in the heavy element distribution (e.g., at A=50, 82, 126) reflect that these nuclei, with "magic" numbers of neutrons and protons are more stable. A few proton-rich heavy elements are produced by the *p*-process, which is most likely due to photodisintegration of heavier nuclei in supernovae (Burbidge *et al.*, 1957; Arnould and Goriely, 2003).

The processes laid out by Burbidge *et al.* (1957) and Cameron (1957) were based on deconvolution of the bulk Solar abundance pattern and basic principles of nuclear physics. Although it was recognized that the atoms that formed the Solar System originated in diverse astrophysical environments, it was long thought that high temperatures in the inner protoplanetary disk would reprocess and homogenize any distinct presolar nucleosynthetic signatures. It took the discovery of isotopic anomalies in Xe and Ne in primitive meteorites (Reynolds and Turner, 1964; Black and Pepin, 1969) to prove both that the nuclear processes envisioned by the pioneers of nucleosynthesis indeed take place in nature and that preserved presolar grains carrying nucleosynthetic signatures survived Solar System formation (see Section 2.7.1; see chapter by Bergin et al. in this volume). Isotopic variations or anomalies are defined as excesses or depletions of a given isotope in a sample, and they are expressed as positive or negative deviations of their isotope ratios in parts per $10^3$ (permil, ‰-units), $10^4$ (epsilon, ε-units) or $10^6$ (mu, μ-units; or part per million, ppm) compared to terrestrial reference standards (defined as zero, or "normal").

## 2.6. Isotopic compositions of meteorites

Isotopic anomalies in meteorites provide invaluable information on the origin, processing, and transfer of matter in the Solar System, as well as ages of events during its whole history. Several of the most important isotope systems used in cosmochemistry are discussed here. These are divided into (i) isotopes produced or destroyed by nuclear reactions and/or radioactive decay in the nascent Solar System, (ii) stable isotope variations which can depend on isotopic masses and provide information on processes and mixings between Solar System objects and reservoirs, and (iii) signatures of presolar heritage that record the nature of nucleosynthetic sources which contributed to the Solar System (so-called nucleosynthetic isotope anomalies).

### *2.6.1 Radiometric isotopic data*

#### *2.6.1.1. Principle of dating*

Radioactive isotopes, or radionuclides, are those that undergo spontaneous disintegration to form radiogenic isotopes. Stable isotopes are those that are not destroyed by radioactive decay. Radionuclides are produced during nucleosynthesis, from interactions of energetic particles (e.g., cosmic rays) with dust or gas, and by naturally occurring radioactive chains. Conventionally, the radionuclide is termed the "parent" which decays into the "daughter" isotope. Because the rate of radioactive decay is constant and independent of essentially all environmental factors, radionuclides serve as useful clocks to date a wide variety of processes that involve chemical fractionation of parent and daughter elements. Radioactive decay is fundamental to many aspects of Earth and planetary science, but its relevance here is that it is used to determine absolute and relative ages for processes that occurred in the first few million years of Solar System history (including condensation in the protoplanetary disk) and the accretion, alteration, and/or differentiation of meteorite parent bodies. There is a distinction between long-lived radionuclides – those whose lifetimes are longer than the age of the Solar System, and short-lived (or extinct) ones – isotopes which were present when the Solar System formed but which have completely decayed away by today. While long-lived systems can often provide absolute ages, short-lived ones can only provide relative ages, albeit in some cases with very high time resolution. Notably, the existence of short-lived radioactive nuclides requires that their nucleosynthesis occurred shortly before the formation of the protoplanetary disk (see Bergin et al., Chapter 1 of this volume).

The basic concepts of radiochronology are illustrated in **Figure 14** for a long-lived (Rb-Sr) and a short-lived (Al-Mg) system. If a melt solidifies with a certain amount of radioactive $^{87}$Rb (mean-life τ=71 Gyr, where after time t, a radioactive isotope has decreased to $e^{(-t/\tau)}$ of its original abundance) into different minerals with varying Rb/Sr ratios, at some later time, the phase with the highest Rb/Sr would have decayed further into $^{87}$Sr and thus there is an increase in $^{87}$Sr/$^{86}$Sr ratio. The different points lie on a line (an "isochron") whose slope can be used to determine the absolute age of the rock and the y-intercept the initial $^{87}$Sr/$^{86}$Sr ratio of the material. Note that the most precise absolute ages of early Solar System objects come from U-Pb or Pb-Pb dating, which takes advantage of the decay schemes of both $^{235}$U ($\tau$~1 Gyr) and $^{238}$U ($\tau$~6.4 Gyr) into $^{207}$Pb and $^{206}$Pb, respectively. The mathematics is slightly more complicated but fundamentally based on the same concept as **Figure 14A**. For a system containing a short-lived isotope (e.g., $^{26}$Al, $\tau$=1 Myr), phases that concentrate a higher parent/daughter element ratio end up with a higher abundance of the daughter isotope. In this case, however, one plots a stable isotope of the parent element on the x-axis and the isochron slope gives the ratio of radioactive to stable isotope that the system had at the time of crystallization. The intercept provides the initial stable isotopic composition of the daughter element before radioactive decay. Short-lived systems can be directly tied ("anchored") to a long-lived one if the same sample can be precisely analyzed for both. Note that what is dated in both short and long-lived systems is the last time the parent and daughter elements were set in their current proportions in each phase, a time referred to as "isotopic closure." This often refers to the time that a rock crystallized into distinct minerals but could also refer to a chemical fractionation event in the protosolar disk, or a heating event that induced inter-grain diffusion, or loss from a previously formed solid.

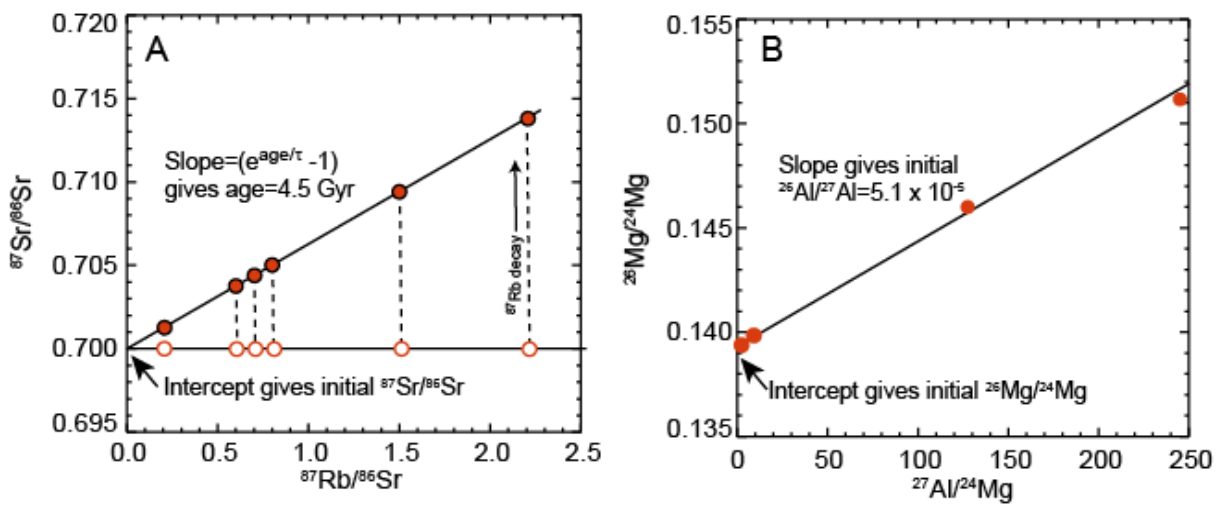

*Figure 14.* Examples of radiometric dating systems. A) In the Rb-Sr system, minerals in a sample with higher Rb/Sr ratios will produce higher amounts of daughter isotope $^{87}$Sr from $^{87}$Rb decay; the slope of the resulting isochron gives the absolute age of the sample. Data are schematic. B) In the Al-Mg system, all initial $^{26}$Al has decayed to $^{26}$Mg. The slope of the isochron gives the initial $^{26}$Al/$^{27}$Al ratio at the time the sample crystallized. Data are from the original CAI measurements of Lee et al. (1977).

Although there are many long-lived and short-lived radiochronological systems used in early Solar System studies, the key results for four of the most important are summarized here: U-Pb, Al-Mg, Hf-W, and Mn-Cr. The first is used to provide high-precision absolute and relative ages of meteorites and their components, the second to provide relative ages of meteoritic objects with sub-Myr precision, the third to provide timing of planetary differentiation and core formation, and the last to provide information on the timing of hydrothermal alteration of meteorite parent bodies.

### 2.6.1.2. Chronology of the early Solar System

All isotope chronometers are consistent with a Solar System age ($t_0$) of ~4.57 Ga, the best estimate being 4,567.30 ±0.16 Ma for CAIs using the U-Pb system (Connelly *et al.*, 2012). The combined use of the U-Pb system with the $^{26}$Al-$^{26}$Mg extinct radioactivity permits high resolution chronology (better of 1 Myr) of events anchored to just after CAI formation. The short-lived $^{182}$Hf-$^{182}$W system ($\tau$=12.8 Myr) is particularly important for dating the time of early planetesimal differentiation. This is because W is highly siderophile while Hf is lithophile, so planetary differentiation leads to fractionation of the W/Hf ratio. The $^{53}$Mn-$^{53}$Cr system ($\tau$=5.3 Myr) has been found to be useful for dating the time of hydrothermal alteration on carbonaceous chondrite parent bodies since carbonate grains formed during the alteration can have extremely high Mn/Cr ratios and thus measurable excesses of daughter $^{53}$Cr. The Mn-Cr system has been also used for dating aqueous alteration of carbonate-free chondrites: H, L, LL, CV, and CO, by measuring hydrothermally formed fayalite and kirschteinite (MacPherson *et al.*, 2017; Doyle *et al.*, 2015)

The main episodes of planetary formation are summarized in **Figure 15,** based mainly on the isotope systems described above. Instead of a succession of events, it appears that planetary accretion and differentiation and the assemblage of primitive bodies took place simultaneously. Some differentiated planetesimals formed very early <1 Myr after CAI, and primitive meteorites were assembled within a few Myr, implying that their formation was contemporary to that of planetesimals. Planetary embryos leading to the accretion of Mars and proto-Earth were also formed within a few Myr after CAI, when the nebular gas was still present in the protoplanetary disk. The early Solar System was a very energetic environment where planetary bodies accreted and were destroyed by collision, and only rare remnants lasted until the present in the form of small bodies providing meteorites to Earth.

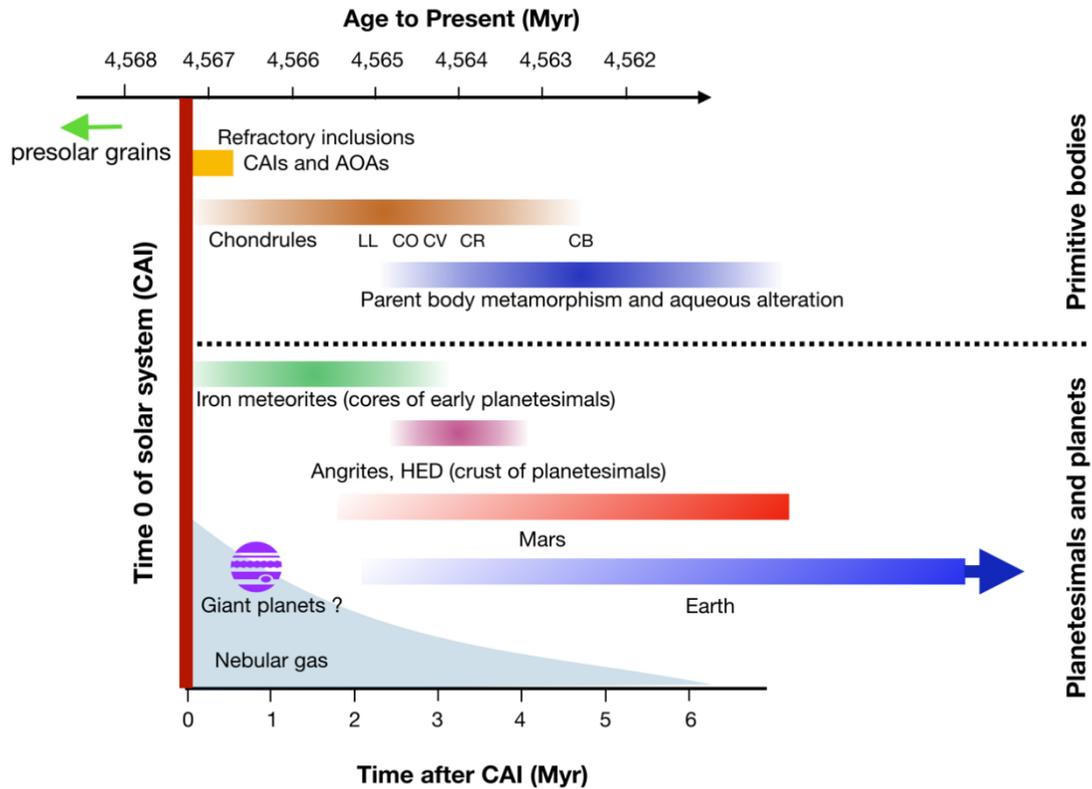

***Figure 15:*** *Chronology of the early Solar System (adapted from Wang and Korotev, 2019). The Solar System formed from the collapse of a molecular cloud core which homogenized stellar dust and gas. Witnesses of past stellar nucleosynthesis can be found in primitive meteorites as refractory presolar grains that survived high enthalpy events. The chronology starts with the condensation of CAIs, the first solids to form during the cooling of the nebular gas (Bouvier and Wadhwa, 2010; Connelly et al., 2012; Connelly et al., 2017). Iron meteorites which are the remnants of disrupted early planetesimals have Hf-W ages (another extinct radioactivity system where $^{182}$Hf decays to $^{182}$W with a mean life of 13 Myr) undistinguishable from the range of CAI ages and extending to a few Myr. Chondrules and by extension chondrites formation intervals extending to ~5 Myr (Krot et al., 2005; Villeneuve et al., 2009; Connelly et al., 2017), implying that chondrule formation was contemporary with the occurrence of early formed planetesimals. Angrites (Zhu et al., 2019) and HED (differentiated meteorites thought to originate from asteroid Vesta; Mittlefehldt, 2015) represent remnants of planetary crusts. Mars, a planetary embryo, was mostly accreted within a few Myr after CAI (Dauphas and Pourmand, 2011). It took much longer time to achieve the accretion of the Earth (≥60 Myr after CAI according to Touboul et al., 2015), although recent models propose early accretion contemporary with that of Mars (Johansen et al., 2021). An early accretion of the proto-Earth is also consistent with the occurrence of solar-like neon in the terrestrial mantle from capture of a nebular-like atmosphere (Yokochi and Marty, 2004). The age of Jupiter is not determined directly, but models of giant planet formation (Guillot and Hueso, 2006) and the purported need to segment the protoplanetary disk early to isolate the NC and CC reservoirs have been interpreted to indicate that Jupiter formed within 1-2 Myr after CAI (Kruijer et al., 2017). The*

*Solar nebula lifetime could have lasted a few Myr before dissipation (Podosek and Cassen, 1994), supplying solar gas to the giant planets and possibly to the terrestrial mantle.*

### *2.6.2. Stable isotopes*

Hydrogen, helium, carbon, nitrogen, oxygen, and neon were the most abundant elements in the molecular cloud from which the Solar System formed. Excepting O, however, these elements are drastically depleted in meteorites and terrestrial planets because of their volatility. Furthermore, H, C, N, and O show substantial, and for two of them (H and N) extreme, variations in their stable isotope compositions. A remarkable feature of H, C, N, and O is that inner planet bodies and reservoirs present isotopic compositions strikingly different from those of the protosolar nebula. Synthesizing these observations, it appears that some major reservoirs of the protosolar nebula, e.g., hydrogen and nitrogen, oxygen, that existed as gaseous molecules ($H_2$, $CO$, $N_2$) were rapidly enriched in their rare and heavy isotopes before, or during, the collapse of the nebula and the formation and early evolution of the protoplanetary disk. Understanding these processes at work will allow insight into the formation of a stellar system and conditions permitting the rise of habitable planets.

Cosmochemists and geochemists commonly use the delta notation ($\delta$) which represents isotopic deviations of an isotopic ratio (R) relative to a terrestrial standard ($R_{std}$) in parts per mil (‰), in linear form expressed as: $\delta\ ^iX = \left[\left(\frac{(R)_{sample}}{(R)_{standard}}\right) - 1\right] \times 1000$. By convention, the rare and heavy isotopes ($^2H$ (D), $^{13}C$, $^{15}N$, $^{17,18}O$) at the numerator are usually normalized to a corresponding abundant isotope ($^1H$, $^{12}C$, $^{14}N$, and $^{16}O$, respectively). Positive $\delta$ values indicate that the sample is enriched in the heavy isotope. In nature, most processes follow a mass-dependent isotope fractionation (MDF) law, where the extent of variation is proportional to the relative mass difference between the two considered isotopes. This means that isotopic fractionation (the change of isotopic proportion due to physical or chemical processing) is larger for hydrogen, which has a relative mass difference close to 100 % between atomic $^1H$ and the heavy hydrogen deuterium (symbol: D, equivalent to $^2H$), than for nitrogen (relative mass difference between $^{14}N$ and $^{15}N$: [15-14]/14 = 7.2 %). There are, however, other types of isotope variations that do not depend on isotopic masses, which are called mass-independent fractionation (MIF). To unambiguously detect MIF effects, one needs elements having more than two stable isotopes, for example as discussed below for the case of oxygen, which has three stable isotopes: $^{16}O$, $^{17}O$ and $^{18}O$, with $^{16}O$ being the most abundant one.

### *2.6.2.1. Oxygen isotopes, the Rosetta stone of the Solar System*

Oxygen is by far the most abundant element in rock-forming minerals of the Solar System (~47%; Lodders, 2003), which makes it a tracer of the origin and evolution of not only volatile-bearing phases in the PSN, but also nearly any solid material. The University of Chicago is where the geochemistry of stable isotopes was developed before WWII under the guidance of Harold Urey. There, Robert Clayton achieved outstanding advances in this field. With Toshiko Mayeda, Clayton recognized that the isotopic composition of oxygen in meteorites does not always follow a mass-dependent isotope fractionation (MDF) (Clayton *et al.*, 1973). On Earth, most oxygen isotope variations are of the MDF type, that is, are proportional to the relative mass difference between two isotopes). In a $\delta^{17}O/^{16}O$ versus $\delta^{18}O/^{16}O$ diagram (**Figure 16**), terrestrial data define a fractionation trend ("terrestrial fractionation line, TFL") with a slope close to ½ (~0.52), because $^{18}O/^{16}O$, which has a relative mass difference $\Delta M/M$ of 2/16, varies twice as much as $^{17}O/^{16}O$ for which the relative mass difference is 1/16. Primitive and differentiated meteorites from a given family originating from a given parent body follow such a trend. In addition to the mass dependent isotope effect, however, different families form parallel trends to the TFL, implying that there is an additional mass-independent isotope fractionation (MIF) effect. Clayton *et al.* (1973) analyzed CAIs as the mineral phases forming CAIs are those predicted to be the first to condense in a cooling solar nebula gas. Oxygen isotopic ratios in CAIs define a correlation with a slope close to 1 (**Figure 16**). This trend implies that these objects were subject to isotope variations of the same extent for both $^{17}O/^{16}O$ and $^{18}O/^{16}O$. Subsequent O isotopic data from bulk meteorites, lunar samples, and Martian samples defined O isotopic variations which required that small bodies accreted regionally with variable mass and independently fractionated oxygen. This discovery permitted genetic relationships amongst meteorite family members to be established leading to the oxygen isotopic compositions of extraterrestrial material becoming one of the most powerful tracers in cosmochemistry.

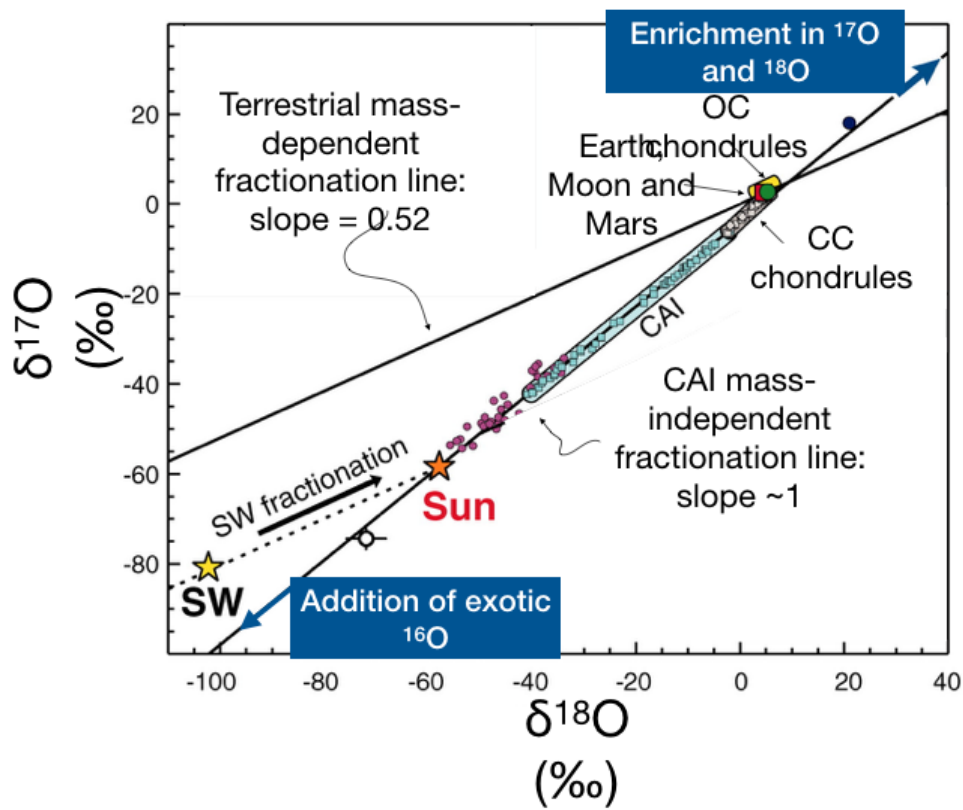

*Figure 16:* Three-oxygen isotope diagram of the Solar System. The coordinates are the $^{18}O/^{16}O$ and $^{17}O/^{16}O$ ratios, in parts per mil relative to terrestrial O (ocean water). Whereas most physical and chemical processes produce mass-dependent fractionations (MDF), here depicted by the slope=0.52 terrestrial line, the Sun, CAIs, chondrules and bulk meteorites define a distinct trend of slope ~1, reflecting mass-independent fractionation (MIF) processes. Modified after McKeegan et al., 2011.

In the O isotope diagram (**Figure 16**), variations could be explained by two different possibilities. They could result from the contribution of an exotic $^{16}$O-rich nucleosynthetic component, possibly from the explosion of a nearby supernova, to a "normal" (Earth-like) component but this possibility was not consistent with the lack of correlation of these oxygen isotopic compositions with those of presolar components (Fahey *et al.*, 1987; Nittler *et al.*, 2008). The alternative view was that the proto-solar nebula *was itself* rich in $^{16}$O, and that a process was able to conjointly enrich $^{17}$O and $^{18}$O relative to $^{16}$O, either by addition of nucleosynthetic products, or by MIF (Thiemens, 1999; Clayton, 2002; Yurimoto and Kuramoto, 2004; Lyons and Young, 2005). The best way to decide between these far-reaching possibilities was to determine where the Sun would plot in the oxygen isotope diagram.

McKeegan *et al.* (2011) succeeded in measuring the three O isotopes of the Solar wind (SW, yellow star in **Figure 16**) in the returned Genesis solar-wind sample. To obtain the Solar value (and therefore that of the proto-solar nebula), the measured SW oxygen isotope composition had to be corrected for isotope fractionation during solar wind acceleration, according to the so-called Coulomb drag model (Bodmer and Bochsler, 1998). The resulting composition (red star) falls perfectly as an end-member of the mass independent fractionation line (McKeegan *et al.*, 2011). This composition is fully consistent with data obtained for CAIs, consistent with the latter being the results of early condensation of the nebular gas. Thus, all Solar System objects and reservoirs except CAI (and some other rare meteoritic mineral phases) were enriched to a comparable extent in $^{17}$O and $^{18}$O relative to $^{16}$O, implying that Solar System oxygen was not contributed by late-stage injection of nucleosynthetic $^{16}$O, but that processes able to enrich solids in the rare, heavy isotopes of oxygen existed. This is further supported by the observation that $^{16}$O-rich presolar dust grains found in meteorites are extremely rare, compared to those enriched in $^{17}$O and/or $^{18}$O (Zinner, 2003; Nittler and Ciesla, 2016).

The observation of oxygen MIF has led to a new line of interpretation invoking photon-gas interactions and substantiated by laboratory experiments (Chakraborty et al., 2008). One of the possibilities is shelf-shielding during photodissociation of CO, as observed in giant molecular clouds (Bally and Langer, 1982). Because $^{16}$O is much more abundant than $^{17}$O and $^{18}$O, ultraviolet photons able to dissociate C$^{16}$O become rapidly depleted by absorption compared to those able to dissociate C$^{17}$O and C$^{18}$O. Consequently, photodissociation of molecules hosting $^{17}$O and $^{18}$O takes place deeper in the cloud than for $^{16}$O. Self-shielding effects could have taken place in the parent molecular cloud illuminated by nearby stars (Yurimoto and Kuramoto, 2004) or at the disk surface of the protosolar nebula with photons from the nascent Sun (Lyons and Young, 2005). Photo-dissociation of gaseous CO could represent a significant source of atomic C and O that could react with H$_2$ to form hydrocarbons (C$_x$H$_y$ molecules) and H$_2$O molecules, respectively. Icy grains formed in this way could then be transported towards the center of the disk where their O isotopes could exchange with those of silicates that eventually got incorporated into primitive meteorites. Hence, CO photodissociation may not only produce the large range of mass independent oxygen isotope fractionation measured across the Solar System, but also provide an effective pathway for the formation of solid phases including ice and organics, which represent the main inventory of volatiles in early forming planetesimals.

The possibility of an irradiation origin for oxygen MIF has been substantiated by key measurements in meteorites. (Sakamoto *et al.*, 2007) used SIMS to analyze an unusual iron oxide/sulfide phase (now known as cosmic symplectite) in the highly primitive ungrouped carbonaceous chondrite Acfer 094 and found similar $\delta^{17}O$ and $\delta^{18}O$ values as high as +180 ‰ **(Figure 17)**. According to these authors, the cosmic symplectite was likely produced by reactions between primordial water and metallic Fe-Ni or iron sulfides and argued that their observation demonstrates the role of water ice as a carrier of the O-isotope signature produced by CO self-shielding. More recently, Nittler *et al.* (2019) reported the discovery of an unusual extremely C-rich clast in an Antarctic CR chondrite, Lapaz Icefield 02432. This highly porous clast contained Na-sulfate minerals with a $^{16}O$-poor isotope composition similar to that of cosmic symplectite. These authors proposed that the clast was a cometary xenolith that accreted water ice which subsequently reacted with other materials to form the sulfate, providing further evidence for a $^{16}O$-poor ice reservoir.

Whatever the origin of O isotope variations, oxygen MIF constitutes an exceptional tool to study genetic relationships between different classes of meteorites and planets. Indeed, meteorites originating from common parent bodies possess a characteristic MIF signature, that allows one to identify genetic relationships and define classes of meteorites (insert in **Figure 17**). The diversity of MIF-O signatures and the grouping of values with meteorite classes strongly suggest that small bodies were imprinted very early with specific oxygen isotope signatures and that different classes of meteorites originate from common parent bodies. Enstatite chondrites, the Moon and Earth have similar MIF-O signatures suggesting they originated from a common pool of matter. This similarity constitutes an important constraint on the origin of the Moon-Earth system.

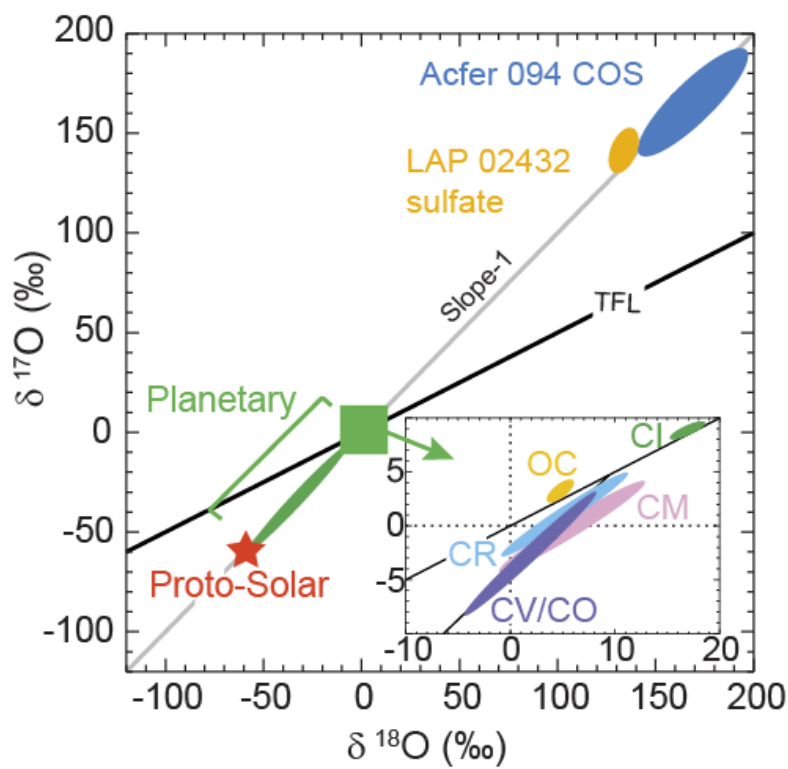

*Figure 17: large-scale variations of oxygen isotopes in the Solar System and relationships between planetary bodies. Proto-Solar is bulk Sun inferred from Genesis solar-wind measurements (McKeegan et al. 2011) and "Planetary" refers to bulk chondrites, Earth, Moon,*

*Mars, asteroids, and CAIs. OC, CI, CM, CR, and CV/CO refer to different groups of chondritic meteorites (Fig. 3). Acfer 094 COS is cosmic symplectite (Sakamoto et al. 2007).*

### 2.6.2.2. *Hydrogen isotope variations*

Hydrogen shows the largest stable isotope variations among the light elements (**Figure 18**). From the comparison of hydrogen isotopes in the atmosphere of Jupiter and Saturn (Lellouch *et al.*, 2001) and from that of the Solar H/He ratio and the $^3$He/$^4$He ratio of the Solar wind (where solar $^3$He has been mainly produced by deuterium burning), it was found that the Earth's oceans as well as meteorites are enriched in D by a factor of about 6 with respect to the protosolar nebula (Geiss and Reeves, 1972). Furthermore, the D/H ratios appear to increase with radial distance from the Sun, with the highest bulk values being found in cometary ice (Bockelée-Morvan *et al.*, 2015; Altwegg *et al.*, 2015). SIMS measurements have revealed much larger D/H variations at the microscale in meteorites (Busemann, 2006) and, especially, cometary interplanetary dust particles (Messenger, 2000) and ultracarbonaceous Antarctic micrometeorites (Duprat *et al.*, 2010). These micrometer-scale variations are largely observed in refractory organic matter (Alexander *et al.*, 2017) but in primitive ordinary chondrites are also seen in hydrated silicates (Deloule and Robert, 1995). Ion-molecule exchange reactions at low temperature in dense molecular clouds, which favor the heavy isotopes with the lowest zero-point energy in the reactants (Aikawa *et al.*, 2018), could explain some of the D enrichments found in molecular cloud ices (Cleeves *et al.*, 2014) and meteoritic organic matter. Thus, hydrogen isotope variations could represent pristine fingerprints of molecular cloud chemistry, with the outer Solar System having better preserved unprocessed (D-rich) material than inner regions. This possibility requires hydrogen now found in planetary bodies to have been isolated from the protosolar nebula gas, probably in the form of icy dust that did not equilibrate with nebular H$_2$ or trapped in refractory organic matter, as suggested by the chondrite data. With respect to the latter, some of the meteoritic classes show large enrichments in deuterium, notably ordinary chondrites (OC in **Figure 18**), which are detected using ion probe techniques at the microscale, interbedded within "normal" material (Deloule and Robert, 1995). CR and some of CM chondrites also present elevated D/H ratios (Alexander *et al.*, 2017), whose origin(s) and timing of delivery are actively debated (McCubbin and Barnes, 2019).

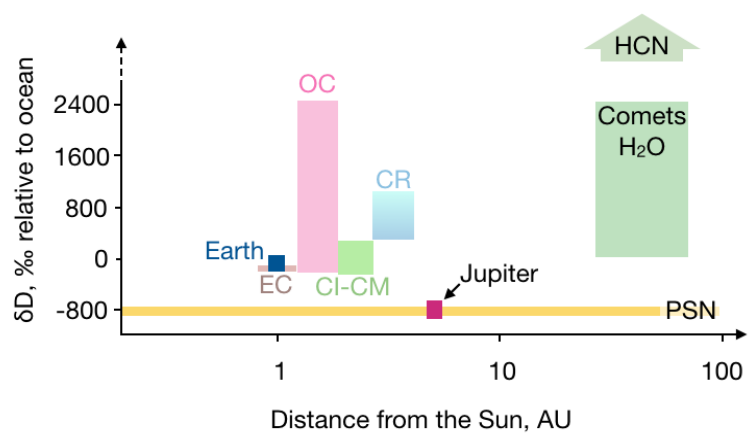

***Figure 18: Hydrogen isotopic variations as a function of present-day radial distance from the Sun (AU).** Meteorite and planetary data are compiled in (https://doi.org/10.24396/ORDAR-66). δD is the deviation from the isotopic composition of the standard which is the D/H of ocean water hydrogen. Jupiter's atmosphere has a δD (Lellouch et al., 2001) similar to that of the protosolar nebula (PSN) (Geiss and Reeves, 1972), indicating that hydrogen in the Jovian atmosphere originated from gravitational capture of the nebular gas. Cometary data are from (Bockelée-Morvan et al., 2015; Altwegg et al., 2015; Lis et al., 2019) and references therein. The different chondrite classes represented here (enstatite chondrites EC, ordinary chondrites OC, CI-CM, CB, CR type carbonaceous chondrites) represent distinct meteorite parent bodies that accreted at different times and locations in the PSN (Desch et al., 2018). Modified after Broadley et al. (2022).*

### 2.6.2.3. Nitrogen isotopic variations

After H, the two isotopes of nitrogen ($^{14}$N and $^{15}$N) present the largest variations among major Solar System objects and reservoirs. The $^{15}$N/$^{14}$N ratio varies by up to a factor of 6 while other volatile elements like carbon present modest variations of a few tens of permil. The nitrogen isotope variability was posited with the return of the Apollo samples in the 1970s, when N isotope analysis of lunar soils revealed variations of several hundreds of permil, which were at that time attributed to secular change in solar wind composition (Kerridge, 1993). A different explanation emerged with the determination of the solar $^{15}$N/$^{14}$N ratio, which had originally been assumed to be similar to that of the Earth. The analysis of Jupiter's atmosphere by spectroscopy (Fouchet *et al.*, 2000) and mass spectrometry onboard the NASA Galileo probe (Owen *et al.*, 2001) revealed that Jovian nitrogen is depleted in $^{15}$N by about 400 permil. The analysis of N in the solar wind-irradiated grains of lunar soils suggested that the Sun is poor in $^{15}$N, with an upper limit for δ$^{15}$N of -250 ‰ (Hashizume *et al.*, 2000). Furthermore, an osbornite (TiN) inclusion in a CAI showed a $^{15}$N depletion of -360 ‰ (Meibom *et al.*, 2007). In contrast, enrichments in $^{15}$N are common in the Solar System, with δ$^{15}$N in some meteorites of the CR-CB clan up to +1,500 ‰ (Grady and Pillinger, 1990) and N-bearing molecules in comets being systematically rich in $^{15}$N by about 1,000 ‰ (Bockelée-Morvan *et al.*, 2015). Nitrogen in chondrites is mainly hosted by organic matter, which could constitute its main carrier from the site of isotope fractionation to the inner Solar System where it was incorporated into primitive meteorites.

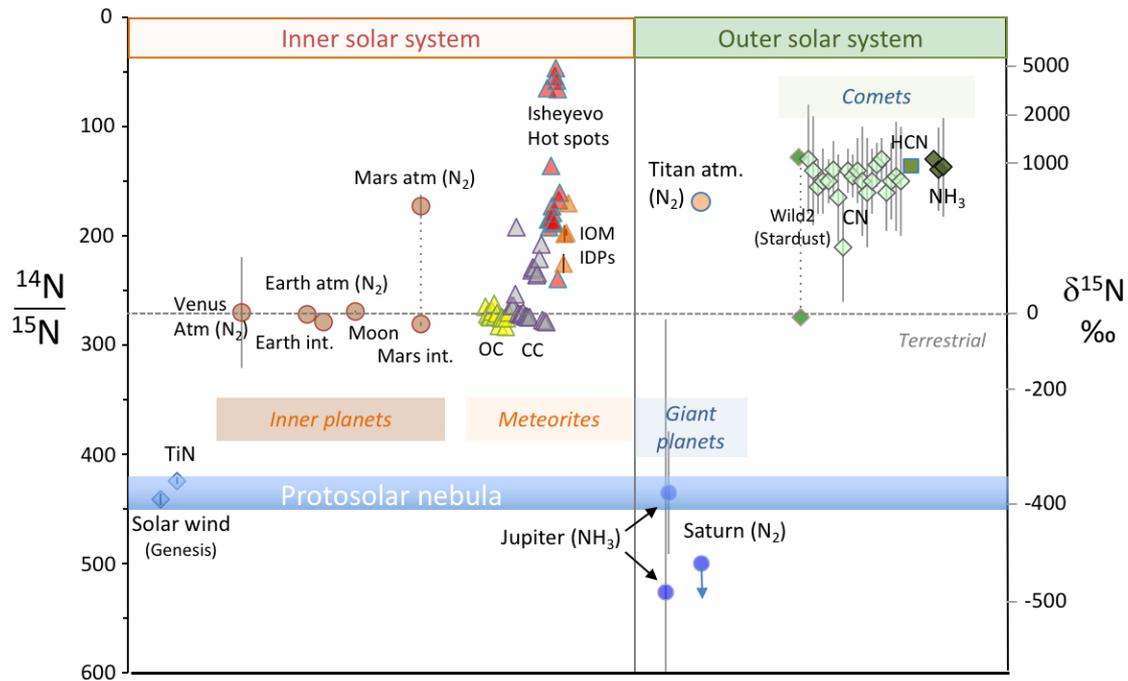

***Figure 19:*** *Nitrogen isotope variation among Solar System objects and reservoirs (adapted from Füri and Marty, 2015, and refs. therein). The left-hand axis gives the values as $^{14}N/^{15}N$, a notation used by astronomers, and the right-hand axis is the $\delta^{15}N$ notation used by geochemists and cosmochemists. Jupiter atmosphere, a TiN inclusion in a CAI and solar wind analyzed by Genesis define the composition of the proto-solar nebula (horizontal blue bar). Other objects and reservoirs are $^{15}N$-rich Note that some meteorites are as rich as comets in $^{15}N$, suggesting a common occurrence of $^{15}N$-rich material. The Martian atmosphere is also $^{15}N$-rich, presumably as a result of isotope fractionation during atmospheric escape.*

The NASA Genesis mission confirmed that the Sun is depleted in $^{15}N$, relative to planetary materials. The Solar wind nitrogen was analyzed at CRPG, Nancy (France) by static mass spectrometry (Marty *et al.*, 2010) and by SIMS (Marty *et al.*, 2011). The results confirmed that Solar wind nitrogen is $^{15}N$-poor, but the precision improved by one order of magnitude ($\delta^{15}N_{SW}$ = -407±7 ‰; 95 % confidence interval). After correction for isotope fractionation related to due to Solar wind generation, (Marty *et al.*, 2011) proposed that the protosolar nebula, now represented by the Sun, had a composition of $\delta^{15}N_{SW}$ = -383±8 ‰. This required that the Sun, which contains more than 99.8 % of present-day Solar System matter, captured the largest share of dust and gas present in the proto-solar nebula. By contrast, all Solar System objects except the atmospheres of giant planets are rich in $^{15}N$ by more than 60% (**Figure 19**). With the new SW N data, N isotope variations among Solar System objects defined an evolution trend with $^{15}N$ excess relative to the Solar composition increasing with the distance from the Sun.

The cause of these remarkable N isotopic variations is debated, and several models have been put forward in the community. Isotopic fractionation during ion-molecule exchange could have taken place in the outer parts of the disk or in the molecular cloud before disk collapse where temperatures of about 10 K would have been required to yield such large isotopic excursions (Rodgers and Charnley, 2008). Another possibility requires illumination of the gas by UV photons, capable of triggering photodissociation of $N_2$ (presumably the main phase of nitrogen in the nebula) and photochemistry to form organic molecules. Self-shielding can occur when stellar photons penetrate a cloud of gas and become progressively absorbed by photoreactions, as observed in giant molecular clouds (Bally and Langer, 1982). Because $^{14}N$ is more abundant compared to the rarer $^{15}N$ ($^{15}N/^{14}N$ = 0.00365) ultraviolet photons able to dissociate $^{14}N^{14}N$ become rapidly depleted by absorption compared to those able to dissociate $^{15}N^{14}N$ and $^{15}N^{15}N$. Consequently, photodissociation of $N_2$ hosting $^{15}N$ takes place deeper in the cloud than for $^{14}N$. Self-shielding effects could have taken place in the parent molecular cloud illuminated by nearby stars and/or at the disk surface of the protosolar nebula with photons from the nascent Sun. Isotopically fractionated N would then be isolated from the gas in a solid form (organics or ice grains), hampering isotope re-equilibration with nebular $N_2$. An alternative possibility has been advanced by Chakraborty *et al.* (2014), while experimentally studying photon-induced chemistry in a gas phase made of $H_2$ and $N_2$, illuminated by photons below 110 nm from a synchrotron light source. They found enrichments up to 1200 % in produced $NH_3$, observed independently of the experiment temperature, that they attributed to preferential exchange of $^{15}N$-bearing molecules in their excited states. Because nitrogen has only two stable isotopes, it is not possible to decipher if MIF, as a result self-shielding, or MDF due to low temperature isotope fractionation are responsible for N isotope variations.

### *2.6.2.4. Carbon stable isotope variations*

A limited number of data suggests that carbon may also present relatively large isotope variations. The most commonly used reference standard for calculating $\delta^{13}C$ values is the $^{13}C/^{12}C$ of terrestrial carbonate (Pee Dee Belemnite). The protosolar nebula value was estimated from measurements of solar-wind-irradiated lunar soil grains (Hashizume *et al.*, 2004) and spectroscopic analyses of photospheric CO (Lyons *et al.*, 2018). Lyons *et al.* (2018) inferred that C in the Solar System is an isotopic mixture of C-rich grains enriched in $^{13}C$ and CO depleted in $^{13}C$. Although $\delta^{13}C$ values have been measured for comets (e.g., 65 ± 51‰ for $CO_2$ in the coma of comet 67P/Churyumov–Gerasimenko; Hässig *et al.*, 2017), the large associated uncertainties do not allow clear comparison to other planetary objects. Carbon isotopic effects may arise from self-shielding of CO molecules in the molecular cloud (Clayton, 2002). The fact that bulk carbon isotope variations are about one order of magnitude lower than those of nitrogen could be due to faster back reactions than for the photodissociation of $N_2$. Some organic particles in chondrites exhibit large C isotope variations (with $^{13}C$ enrichments and depletions of up to hundreds of permil, again likely due to low temperature reactions) but these are much rarer than seen for H and N anomalies (Floss and Stadermann, 2009; Nittler *et al.*, 2018).

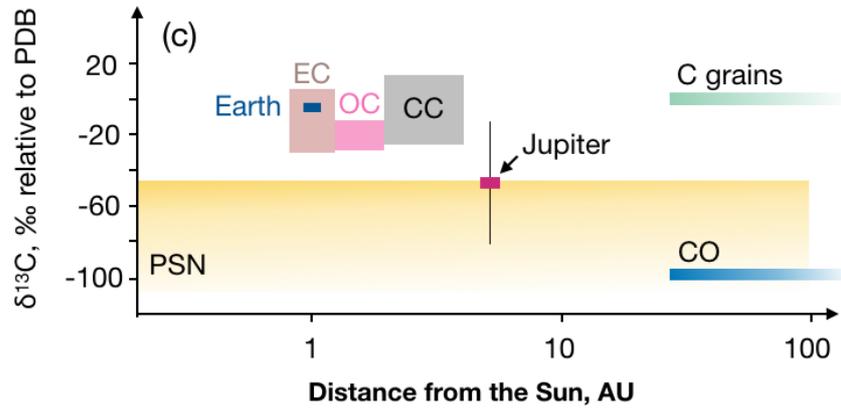

*Figure 20*: Carbon isotope variations in the Solar System (see text for refs.). Modified after Broadley et al. (2022).

### *2.6.2.5. Noble gases: Solar vs. planetary*

There are five stable noble gases: helium (atomic mass: 4.00 u), neon (20.18 u), argon (39.95 u), krypton 83.80 u) and xenon (131.29 u), all of which are excellent tracers of the origin(s) and evolution of small bodies. They are chemically inert in natural environments and their isotopes and abundances can only vary through physical processes such as phase changes, isotopic fractionation, and nuclear reactions. The stable noble gases cover the whole spectrum of masses and their behaviour in natural systems, such as solubility in minerals, magmas, is mainly governed by their contrasting atomic sizes. Some of their isotopes are primordial (inherited from the stellar environment), other are produced by nuclear reactions, such as spallation, neutron activation, radioactive decay, and fission. Given these characteristics, they have the potential to act as geochronometers as well as excellent tracers of the origin and history of their host phases (Ozima and Podosek, 2002; Ott, 2014). In this section, we show that noble gases trapped in small bodies are non-Solar, that is, their abundance and their isotopic compositions cannot be related directly to the protosolar nebula gas.

In the protosolar nebula, light noble gases were among the most abundant elements **(Figure 13)** (Lodders, 2003). The analysis of gases sublimated from the coma 67P/Churyumov Gerasimenko has shown that they are also abundant in cometary ice, as these elements were cryo-trapped in forming icy dust depending on the environmental (cold) temperature (Balsiger *et al.*, 2015; Marty *et al.*, 2017; Rubin *et al.*, 2018). By contrast, noble gases are "rare" on Earth and in meteorites, and for this reason they are also labelled the rare gases (Ozima and Podosek, 2002). Their scarcity in silicates and metal is the result of their chemical inertness that prevented them to be efficiently trapped in solids at temperatures typical of the inner part of the disk. Noble gases are measured by static mass spectrometry which permits to quantify isotopes in amounts as low as a few million atoms. One of the main geochronological tools is the potassium-argon dating system, where $^{40}$K, a rare isotope of potassium (0.01167 %) decays to $^{40}$Ar ($T_{1/2}$ = 1.25 Gyr). Because $^{40}$Ar was barely produced by stellar nucleosynthesis, almost all $^{40}$Ar found in planetary bodies including meteorites has been produced by the decay of $^{40}$K and the K-Ar dating method and its elaborated so-called Ar-Ar method are among the most sensitive geochronological tools to establish the timeframe of Solar System formation and evolution (Lee, 2015).

The so-called extinct radioactivities were first discovered thanks to the study of noble gases, in particular xenon, in meteorites. Nucleosynthetic models predicted that some of the isotopes synthesized in stars, having half-lives shorter than the age of the Solar System, could have been alive when the Solar System started to condense. Iodine-129 is a radioisotope that decays to $^{129}$Xe with a half-life of a few Myr ($t_{1/2}$ = 15.7 Myr) and should have been produced in massive stars together with the stable isotope of iodine, $^{127}$I. A monoisotopic excess of $^{129}$Xe (with respect to the "normal" Xe isotope composition) was found in the Richardton meteorite and correctly attributed to the in-situ decay of $^{129}$I (Reynolds, 1960). This was the first experimental evidence that extinct radioactivities were alive when the Solar System formed. The $^{129}$I-$^{129}$Xe dating method became one of the most important tools to date the first tens of million years of planetary evolution including the terrestrial atmosphere (Wetherill, 1980). Noble gases were subsequently used as tracers for the occurrence presolar material.

Other examples of isotopically unusual noble gases were found soon thereafter in primitive meteorites and interpreted as signs of surviving refractory presolar grains bearing stellar nucleosynthesis signatures (Reynolds and Turner, 1964; Black and Pepin, 1969) In order to identify the carriers of exotic noble gases, meteorite samples were subjected to sequential chemical attacks and after each step, noble gases were analyzed (Lewis *et al.*, 1987; Amari *et al.*, 1990; Lewis *et al.*, 1994; Huss *et al.*, 2003). This approach eventually permitted the identification of acid-resistant presolar grains including nanodiamonds, graphite, silicon carbide etc. that were subsequently isolated and analyzed for stable isotopes with ion probes (see Section 2.7.1) (Zinner, 1998).

The remarkable noble gas isotope variations seen in presolar material do not reflect their composition in bulk meteorites, which is instead dominated by a component common to most meteoritic classes. As for the stable isotopes of H, C, N, and O, noble gases trapped in meteorites and inner planets differ from the Solar composition. Not only are their isotopic ratios non-Solar, but also their relative abundances are fractionated relative to the solar composition, as approximated by the noble gas composition of the Solar wind (see Ott, 2014, and refs. therein).

In primitive meteorites as well as in inner planet atmospheres (Mars and Earth), noble gases are depleted by several orders of magnitude with light noble gases being the most depleted ones **(Figure 21).** They are ubiquitous in primitive meteorites, being hosted by a common phase labelled Phase Q and linked to refractory organic matter. Phase Q noble gases form a rather homogenous reservoir of noble gases having comparable elemental abundance patterns and isotopic ratios which dominate the inventory of Ar, Kr and Xe in chondrites (Busemann *et al.*, 2000). In addition to Phase Q, several exotic components hosted by presolar phases (nanodiamond, silicon carbide etc.) are also present in chondrites as discussed above. The origin of phase Q has been debated for a long time. Ionization experiments are able to reproduce some of the characteristics of Q noble gases (elemental and isotopic fractionation), suggesting that ionization was the main process (apart for cryo-trapping) for incorporating efficiently noble gases in solids (Bernatowicz and Hagee, 1987; Matsuda and Maekawa, 1992). Ionization experiments involving organics also reproduced some of the features of meteoritic organic matter in addition to fractionating the isotopes of noble gases (Kuga *et al.*, 2015; 2017; Marrocchi *et al.*, 2005). Therefore, ionization of a reduced gas rich in $H_2$, CO, $N_2$ and noble gases has the potential of not only synthesizing organic matter reminiscent of that trapped in primitive meteorites, but also of trapping efficiently noble gases with the required elemental and isotopic fractionation. However, so far these experiments did not reproduce the thermal resistance of meteoritic organics. Q noble gases are often extracted at temperatures > 1300 K whereas in experimental runs they often degas below 800 K, and additional processing appears necessary. All these data converge with an important role of gas-photon interactions as the main driver of stable isotope and noble gas fractionation.

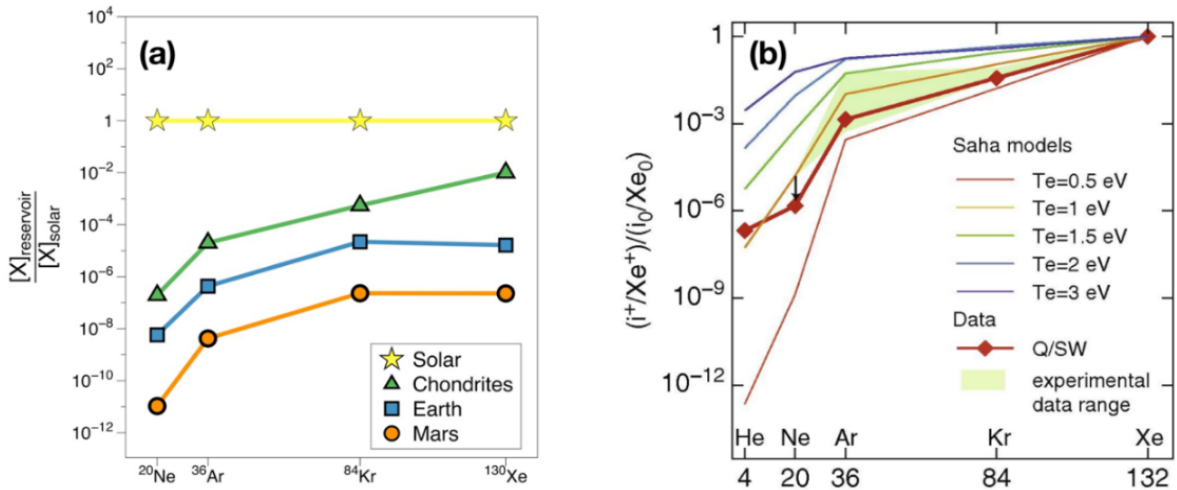

*Figure 21: **Noble gas elemental fractionation of planetary bodies with respect to solar composition**. **(a)** The Solar composition is approximated by that of Solar wind (Meshik et al., 2014; Vogel et al., 201); yellow stars). Chondrites (Pepin, 1991), Earth's atmosphere (Ozima and Podosek, 2002) and Mars' atmosphere (Swindle, 2002) are depleted relative to solar, with neon (and helium, not represented) being the most depleted elements. The depletion pattern of chondrites and inner planets, sometimes labelled misleadingly "planetary", was attributed to the poor retention of chemically inert noble gases into solids at temperatures typical of the inner Solar System. The somewhat less pronounced depletion of the heavy noble gases (e.g., Xe) compared to the light ones (e.g., Ne) was attributed to the better "sticking" efficiency of the former due to van der Walls bonding. Note that the atmospheres of Earth and Mars (and Venus, but Venusian data are imprecise) are depleted in xenon relative to krypton and chondritic, which defined the "mission Xe" problem. It was recently shown that xenon was lost from the terrestrial atmosphere by ionization-related escape to space over geological periods of time (Avice et al., 2018) and is therefore not a primordial feature. **(b)** The noble gas fractionation pattern of chondrites, normalized to xenon and solar (thick red line) is compared to fractional ionization in a Saha-type plasma with electron energies of a few eV (Kuga et al., 2017).*

The specific isotopic compositions of noble gases in Solar System objects and reservoirs permit the investigation of processes of planetary formation and evolution. An example is given in **Figure 22** which represents a comparison of the xenon isotopic compositions between Solar and planetary components. Here, meteoritic and Earth's atmospheric Xe present excesses in radiogenic $^{129}$Xe (an extinct radioactivity permitting fine chronology of the first hundreds of Myr) as well as mass-dependent isotope fractionation (MDF), plus nucleosynthetic anomalies. In particular, the terrestrial atmosphere has high MDF (which has been regarded as preferential escape of atmospheric Xe over geological periods of time (Pujol *et al.*, 2011) as well as depletions in $^{134}$Xe, $^{136}$Xe, a likely nucleosynthetic anomaly. Such isotope variations allow one to investigate genetic relationships between primitive and evolved materials. For instance, the contribution of cometary matter to the Earth's atmosphere was identified for the first time thanks to the specific isotopic composition of xenon measured in comet 67P/Churyumov Gerasimenko (Marty *et al.*, 2017). Likewise, noble gases trapped in SNC meteorites were the smoking gun for a Martian origin of this class of meteorites (Becker and Pepin, 1984; Treiman *et al.*, 2000). Some of the noble gas isotopic structures were inherited from previous generations of stars, and noble gases were among the first isotopic systems used to identify presolar material preserved in meteorites. For example, the specific neon isotopic composition of the terrestrial atmosphere is probably a remnant of the contribution of exotic nanodiamonds trapped in carbonaceous chondrites (Marty, 2012; Mukhopadhyay and Parai, 2019).

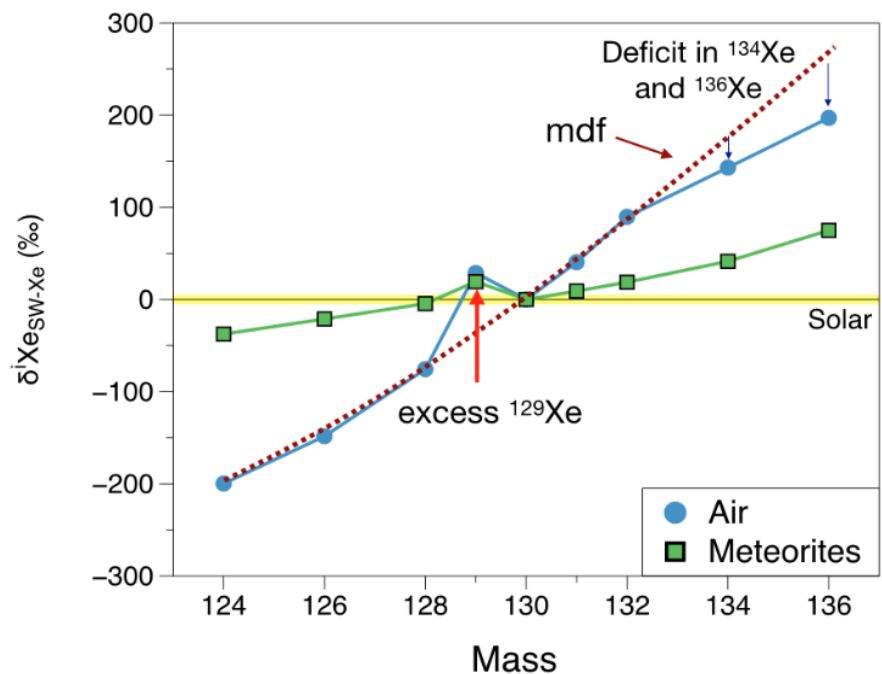

*Figure 22: Xenon isotope diagram for Solar System reservoirs (data from Ott, 2014). The nine Xe isotopes are normalized to one of them, here $^{130}$Xe, and to a specific reference composition, here solar wind (SW) Xe, assumed to represent the composition of the solar nebula from which planetary bodies formed. The isotopic deviations from the reference composition (solar) are given in the classical δ notation: δ$^{i}$Xe = [($^{i}$Xe/$^{130}$Xe)$_{sample}$/[($^{i}$Xe/$^{130}$Xe)$_{SW}$ -1] x 1000 (part per mil; ‰). In this format, a solar-like isotope composition yields a horizontal pattern similar to SW (δ$^{i}$Xe=0‰). Earth's atmospheric xenon differs from SW by three important features: (i) it contains a monotopic excess of $^{129}$Xe from the decay of $^{129}$I (extinct radioactivity $T_{1/2}$ = 15.7 Myr), (ii) it is isotopically fractionated by 35 ‰ per atomic mass unit relative to SW, and (iii)*

*it is depleted in heavy Xe isotopes: $^{134}$Xe and $^{136}$Xe, relative to a line that passes through the other isotopes, and which characterizes mass-dependent isotopic fractionation. The MDF-Xe has been attributed to atmospheric escape (Pujol et al., 2011) and the deficit in $^{134}$Xe and $^{136}$Xe is a nucleosynthetic feature inherited from comets (Marty et al., 2017).*

As for stable isotopes described above, noble gases are isotopically non-Solar in primitive meteorites and, as for N and partly H, they are associated with organics. Their elemental and isotopic fractionation might have resulted from trapping in organics during interactions between stellar photons, from the protosun or from nearby stars, and the nebular gas rich in $H_2$ and CO (Kuga *et al.*, 2015; 2017). Such an origin implies large scale transport in the protoplanetary disk from the regions of organic production to those of primitive meteorite building up.

### 2.6.3. Nucleosynthetic isotopic anomalies and the NC-CC isotopic dichotomy

#### 2.6.3.1. Nucleosynthetic isotopic anomalies

The so-called nucleosynthetic isotope anomalies are a special class of mass-independent isotope variations found in meteorites and on the planetary scale. Isotopes associated with a growing number of elements (e.g., He, Ne, Kr, Xe, Ca, Ti, Cr, Fe, Ni, Sr, Zr, Mo, Ru, Pd, Ba, Nd, Sm, Er, Yb, Hf, W, and Pt) have been found to show nucleosynthetic isotope anomalies at the bulk meteorite scale (for reviews, see Ott, 2014; Qin and Carlson, 2016; Valdes et al., 2021; Bermingham and Kruijer, 2022). These subtle isotopic anomalies arise from the imperfect homogenization of presolar grains throughout the protoplanetary disk, and they are expressed as positive or negative deviations of their isotope ratios in parts per $10^4$ (epsilon, ε-units) or $10^6$ (mu, μ-units; or part per million, ppm) from terrestrial reference standards.

The nucleosynthetic isotope composition of a sample is used to constrain the bulk (or average) composition of the parent body, and thus the nebular region from which it accreted (e.g., Walker *et al.*, 2015). As discussed above Subsection (*2.6.2.5.*), noble gas isotopic variations led to the identification of many specific presolar grain types. As nucleosynthetic isotope anomalies are similarly caused by the heterogeneous distribution of presolar grains, these isotope effects were anticipated to be on the same scale as those recorded by noble gases. The bulk effects for noble gases, however, are much larger due to the rarity of these elements in most Solar System solids. By contrast, many of the refractory elements that display nucleosynthetic isotope anomalies are much more abundant such that the nucleosynthetic signatures of individual presolar grains are highly diluted **(Figure 23)**. Nucleosynthetic isotope anomalies are also found in individual meteorite components including CAIs and refractory hibonite inclusions, and in successive acid leachates of different meteorites. These anomalies are, however, typically highly variable and more anomalous than bulk meteorite samples (e.g., Ti isotopes Davis *et al.*, 2018, Mo isotopes, Brennecka *et al.*, 2020, Ca isotopes, Valdes *et al.*, 2021), which is expected because the bulk samples are mixtures over large numbers of individual components.

Not all elements display nucleosynthetic isotope variations on the bulk sample scale. For example, nucleosynthetic Os isotope variations are not detected in most bulk meteorites. Ureilites are an exception, however, it has been suggested that Os was mobilized as a result of parent body thermal-based processes rather than this being a bulk characteristic of the parent body (Goderis et al., 2015). Osmium isotopic variations are also documented in acid residues of some chondrites and the extent of an Os isotopic anomaly in an acid residue can correlates with the degree of aqueous alteration of the host chondrite (Yokoyama *et al.*, 2011; 2007). These data indicate that Os isotope anomalies can also be caused by selective destruction/modification of presolar grains carrying Os during progressive aqueous alteration on parent bodies. The generally non-anomalous Os isotopic composition of most bulk samples measured to date (except ureilites), however, demands that the isotopically anomalous Os released from presolar phases during processing, was re-incorporated into phase(s) which was not lost from the part of the parent body where most meteorites were sampled (Yokoyama et al., 2011).

The existence of subtle nucleosynthetic isotope heterogeneity indicates that the protoplanetary disk became reasonably, but not perfectly, homogenized by the time meteorite parent bodies accreted. The cause of the heterogeneous distribution of presolar grains in the disk remains debated. It may be a consequence of imperfect mixing of these grains throughout the protoplanetary disk (e.g., Clayton, 1982), their selective destruction via thermal chemical processing in the protoplanetary disk (e.g., Trinquier *et al.*, 2009), and/or size sorting processes in the risk (for review see Dauphas and Schauble, 2016). Regardless of the cause of the heterogeneous distribution of presolar grains in the disk, the unique nucleosynthetic isotope composition each parent body possesses can be harnessed to trace communication between regions of the Solar System during planetary accretion.

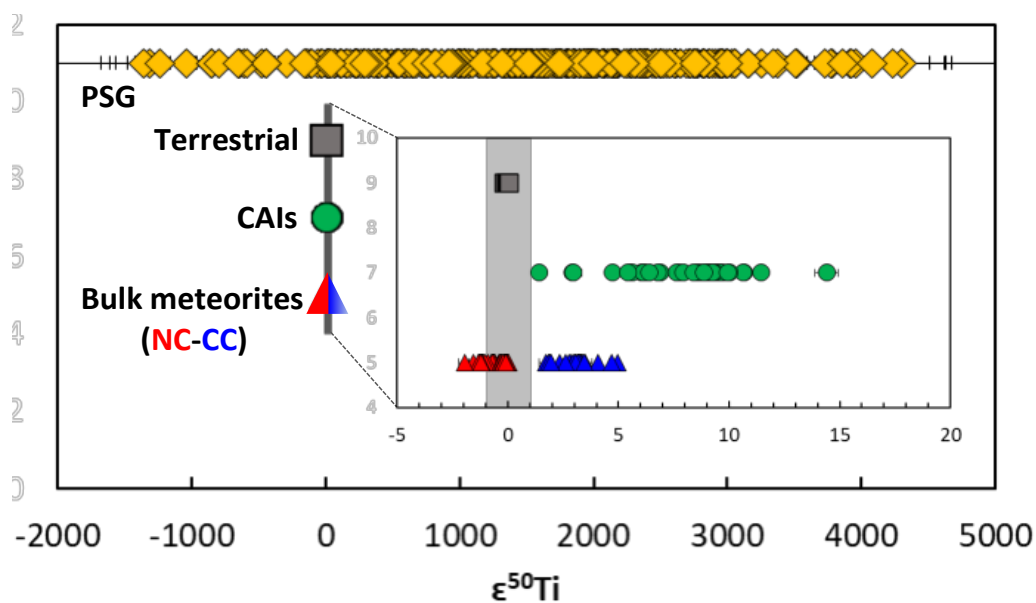

***Figure 23***: *$\varepsilon^{50}Ti$ data for individual presolar grains (PSG; yellow), individual CAIs, and bulk terrestrial samples, and NC-CC meteorite samples. The inset figure plots a zoomed in display of the whole rock and CAI data. The grey field corresponds to the range of terrestrial sample compositions and analytical precision reached in the respective studies. (Torrano et al., 2019 ; Alexander and Nittler, 1999 ; Amari et al., 2001 ; Davis et al., 2018; Gerber et al., 2017; Gyngard et al., 2018; Hoppe et al., 1994; Huss and Smith, 2007; Ireland et al., 1991; Render et al., 2019; Trinquier et al., 2009; Zhang et al., 2012; Zinner et al., 2007). Figure adapted from Bermingham et al. (2020).*

*2.6.3.2. The NC-CC isotope dichotomy*

The noncarbonaceous – carbonaceous chondrite (NC-CC) isotope dichotomy has become one of the most utilized and researched applications of nucleosynthetic isotope anomalies in Cosmochemistry. Early indication of the NC-CC isotope dichotomy was reported by Trinquier et al. (2007; 2009). Subsequently, Warren (2011) observed that nucleosynthetic isotope composition of meteorites consistently fell into one of two compositional groups **(Figure 24)**. The first group comprises primarily carbonaceous chondrites (CO, CK, CV, CM, CB, CR, CI, Tagish Lake), an ungrouped meteorite (NWA 011), and Eagle Station group pallasites. The second group comprises meteorites and other bodies that are not carbonaceous chondrites (e.g., Earth, Moon, enstatite chondrites, ordinary chondrites, angrites, HED, main group pallasites, mesosiderites, and ureilites). Warren (2011) labelled the first group as "carbonaceous chondrite (CC)" and the second group as "noncarbonaceous (NC)". The isotopic bimodality has since been extended to siderophile elements (e.g., Mo, Ru, W) in iron meteorites (e.g., Budde *et al.*, 2019; Kruijer *et al.*, 2020; Poole *et al.*, 2017; Worsham *et al.*, 2017; Bermingham *et al.*, 2018a), and several lithophile elements (e.g., Ca, Ti, Cr, Ni, Sr, Zr, Mo, Ru, Ba, Nd, Sm, Hf, W), for reviews, see (Scott *et al.*, 2018; Burkhardt *et al.*, 2019; Bermingham et al., 2018b). Using "carbonaceous chondrite" to describe iron meteorites can be confusing because iron meteorites are not carbonaceous chondrites. The NC-CC terminology, however, describes two isotopic or "genetic" reservoirs in the protoplanetary disk that are sampled by meteorites. It is now considered the highest taxonomic division in meteorite/planetary classification. To date almost all meteorites fall into either the NC or CC group. Thus far, there is only one exception to this rule, the ungrouped iron meteorite Nedagolla, for which the Ni-Mo-Ru isotopic composition of this sample are intermediate between NC and CC (Spitzer *et al.*, 2021). Its $^{182}$Hf-$^{182}$W based chronology indicates that the mixed NC-CC composition was established >7 Myr after CAI formation, likely from collisional mixing of preexisting NC-CC bodies.

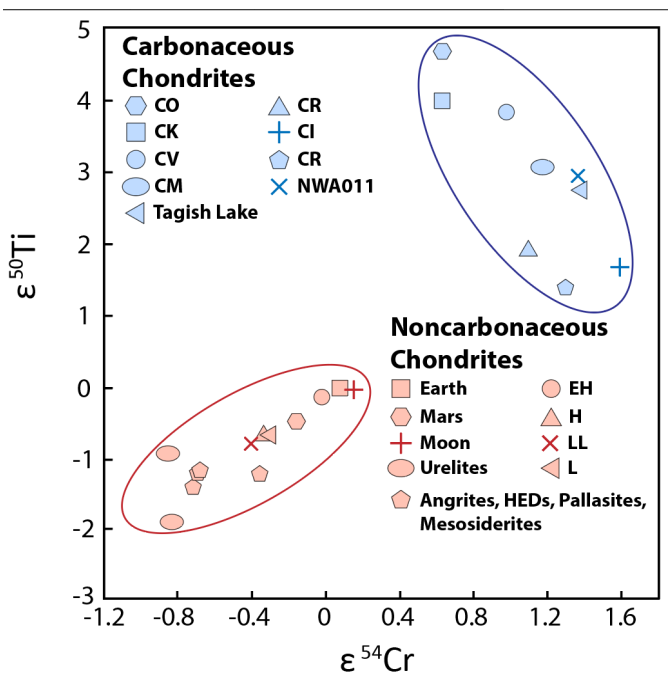

***Figure 24:*** *$\varepsilon^{54}Cr$ vs. $\varepsilon^{50}Ti$ in bulk noncarbonaceous (NC, red) and carbonaceous (CC blue) meteorites (figure modified after Warren, 2011). Noncarbonaceous meteorites include HED, ordinary chondrites, angrites, enstatite chondrites, mesosiderites, pallasites, lunar meteorites, Martian meteorites, and ureilites. CCs include CI, CO3, CM2, CV3, CR2, CH3, Tagish Lake (C2-ung), CK, CBa.*

Using $^{182}$Hf-$^{182}$W derived core-segregation ages from iron meteorites, Kruijer *et al.* (2017) proposed that the NC and CC groups possess distinct accretion ages, NC (<0.4 Myr after CAI formation) and CC (0.9 +0.4/−0.2 Myr after CAI formation). Undifferentiated chondrites have not undergone metal-silicate fractionation and thus their accretion ages must be inferred by dating meteorite components that formed before accretion into parent body (e.g., chondrules, Krot et al., 2018; Alexander et al., 2008). Robust chronology comes from convergence of ages determined using several chronometers (for a review of this approach see Bermingham and Kruijer, 2022). Accretion ages for differentiated achondrites (e.g., angrites and eucrites) likely underwent multiple stages of Hf-W fractionation. Accordingly, their accretion age must be inferred using Hf-W isotope evolution of mantle sources of these samples (Kleine *et al.*, 2012; Kruijer *et al.*, 2012; Touboul *et al.*, 2015), thus these ages can be less certain than core-segregation ages defined by the $^{182}$Hf-$^{182}$W system. Accretion ages for differentiated achondrites constrain core formation to within the first 1-2 Myr of Solar System formation on the angrite and eucrite parent bodies (Baker *et al.*, 2005; Kleine *et al.*, 2012; Touboul *et al.*, 2015). The recent discovery of an andesitic meteorite, presumably from the crust of a protoplanet and having a $^{26}$Al age of 2.2 Ma after CAI, attests for melting and differentiation of the parent body within ~1 Ma after CAI (Barrat *et al.*, 2021).

A more recent compilation of age data for multiple bodies, including new data for meteorites that had not previously been studied, showed that the accretion timescales of NC and CC meteorite groups overlap in their uncertainties, meaning the accretion ages of some NC and CC irons can be indistinguishable (Scott *et al.*, 2018; Hellmann *et al.*, 2019; Hilton *et al.*, 2019; Kruijer *et al.*, 2020) (e.g., **Figure 25**). Kruijer et al. (2020) interpreted a compilation of chronological data to indicate that some iron meteorite parent bodies accreted before chondrite parent bodies (**Figure 25**, see also **Figure 15** for a general chronological framework). This dataset indicates that some chondrite parent bodies accreted later than the iron meteorite parent bodies, starting at ~ 2 Myr in the NC reservoir and continuing until at least ~ 4 Myr after CAI formation in the CC reservoir. Combining these data with nucleosynthetic $^{50}$Ti, $^{54}$Cr, $^{58}$Ni, and Mo isotope data that discriminate between NC and CC, CC irons and chondrites appear to be enriched in nuclides produced in neutron-rich stellar environments compared to the NC irons and chondrites. It was concluded that NC and CC reservoirs represent two spatially separated reservoirs of the disk that co-existed for several million years.

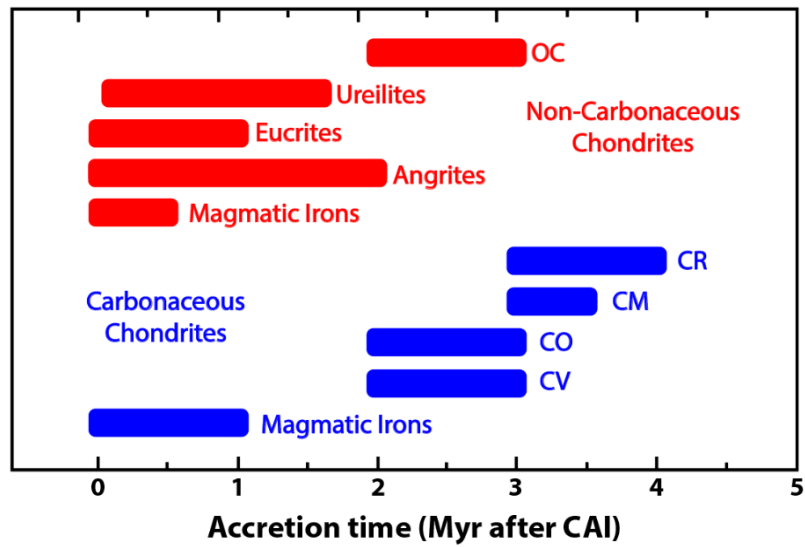

*Figure 25. Accretion timescales in Myr after CAI formation of CC (red) and NC (blue) meteorite parent bodies, where the horizontal bars reflect the uncertainty of the accretion age estimates. Accretion timescales for chondrites are based on the chronology of alteration products and chondrule-derived Al-Mg, Hf-W, Pb-Pb and/or U-Pb ages, which are integrated using thermal modeling. Accretion ages for iron meteorites are model ages derived for differentiated and thermal modeling for internal heating by $^{26}Al$ of small parent bodies. Figure based on Kruijer et al., (2020).*

## 3: FROM STARS TO PLANETS

### 3.1. A summary of the astrophysical context based on studies of small bodies

The astrophysical context of the Solar System can be constrained using the elemental and isotopic compositions of meteorites. Volatiles, including H, C, N, O, and noble gases trapped in chondrites and comets, have elemental and isotopic compositions that are markedly different from those of the protosolar nebula, precluding a direct genetic relationship between the building blocks that make up the terrestrial planets and the gaseous Solar nebula. In addition to having a non-Solar volatile composition, these materials show large variations in volatile and isotopic compositions **(Figures 17-22)**. These isotopic fingerprints were likely inherited from processes having taken place in the interstellar medium or in the molecular cloud. Notably, however, that interactions between stellar photon and the nebular gas rich in $H_2$, CO, and noble gases have the potential to trigger elemental and isotopic fractionation able to account for some of the observed chemical signatures. It has been proposed that such processes took place when the molecular cloud was illuminated by nearby stars (Yurimoto and Kuramoto, 2004), knowing that the Solar System was probably born in a cluster of several hundreds to thousands of stars (Adams, 2010). Another potential site of irradiation is the surface of the disk illuminated by the protosun (Lyons and Young, 2005). These possibilities are illustrated in **Figure 26**. Icy and organic dusty grains would then be transported along the forming disk by molecular cloud infall, turbulence and disk spreading.

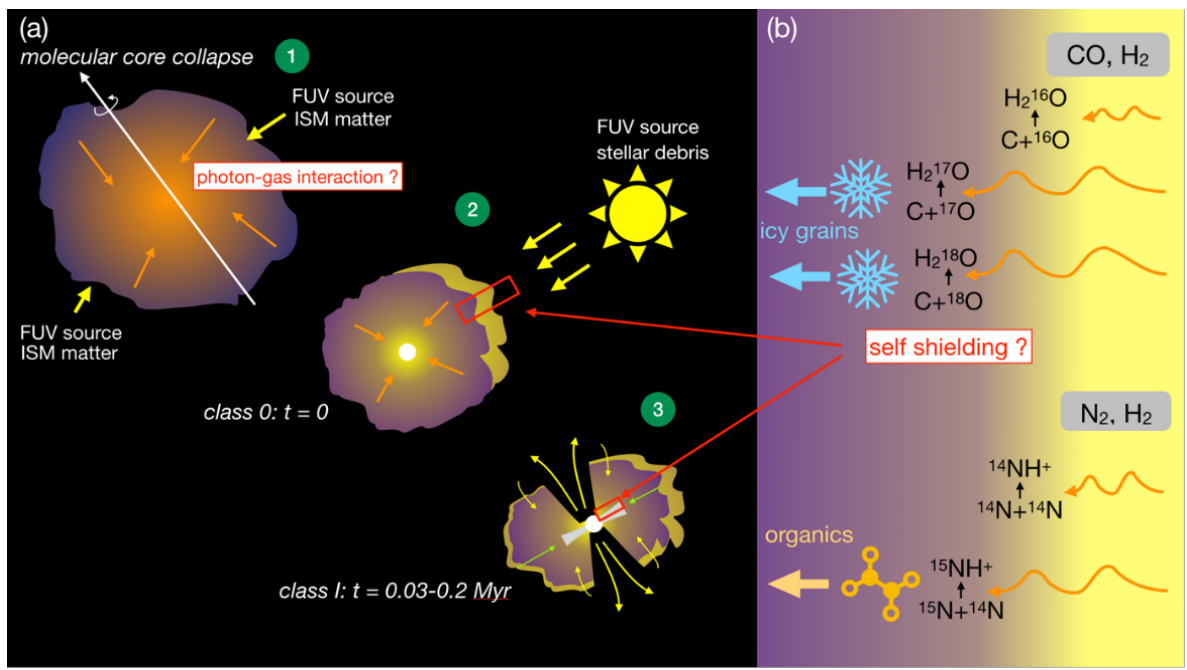

*Figure 26: Possible origins of isotopic signatures of H, C, N, O, and noble gases observed among Solar System bodies (adapted from Lee et al., 2008). (a.1) Heritage from the interstellar medium (ISM). (a.2) Reactions between stellar photons from nearby stars and the nebular gas. Photodissociation of $H_2$, CO and $N_2$ results in the formation (i) of water ice carrying MIF oxygen, and (ii) of organics hosting excesses of $^{15}N$ and possibly of D. Noble gases can also be fractionated and trapped in organics at this stage. Synthesized ice and organic dusty grains are then transported to the planet forming regions where they exchange their isotopic signatures with forming solids. (a.3) Photochemistry could have also taken place at the disk surface illuminated by protosolar photons. (b): Possible reactions between stellar photons and nebular gas (made of $H_2$, CO, $N_2$) leading to the formation of icy grains and organic dust both hosting isotopic fractionation.*

The molecular cloud also hosted presolar grains and their preservation in meteorites indicates that not all of them were destroyed and homogenized with the gas during Solar System formation. When and how the isotopic bimodality defined by nucleosynthetic isotope anomalies (NC and CC isotopic dichotomy) **(Figure 24)** was established is debated. It could indicate a heterogeneous distribution of presolar grains in the molecular cloud itself such that the composition of infalling material changed with time. Evidence for early formed refractory grains with large Ti isotope anomalies but no $^{26}Al$ (e.g., Kööp *et al.*, 2016) may provide evidence for this as it implies late addition of $^{26}Al$ to the protosolar nebula, possibly contained in presolar grains or direct input from a nearby supernova. Alternatively, it could reflect differential processing of different types of presolar grains (e.g., by evaporation) in different parts of the disk. The early formation of Jupiter might have kept a barrier against transport keeping the segmentation alive for several Myr. If heterogeneous infall occurred, it was present when the first planetesimals formed (< 1 Myr after CAI formation), as the NC-CC bimodality is recorded in early-formed iron meteorites. The infall is required to have lasted several Myr, as indicated by the formation of NC and CC bodies **(Figure 25)**. How this dichotomy could have been maintained for such periods of time is fueling multiple models of Solar System evolution. Below some of these models are developed in the context of dynamical models of planetary formation.

### 3.2. Formation of planets

#### *3.2.1: General overview*

In this section, models of planetary formation in the light of data obtained from the study of small bodies are reviewed. The main processes of planetary growth as understood today are summarized in **Figure 27** and the following subsections. A variety of processes are at play in the growth of micrometer-sized dust grains to mm- to cm-sized 'pebbles' that drift rapidly within the disk, up to 100 km-scale 'planetesimals', which are generally considered the macroscopic building blocks of the planets (for a review, see Johansen *et al.*, 2014). Planetesimals may grow by mutual collisions and by accreting drifting pebbles into planetary embryos or cores. The gas giants are constrained to have formed very quickly, within the few million-year lifetime of the gaseous disk. In contrast, cosmochemical measurements indicate that Earth probably took ~100 Myr to complete its formation (e.g., Kleine et al., 2009), although Mars' growth was complete much faster (Dauphas and Pourmand, 2011).

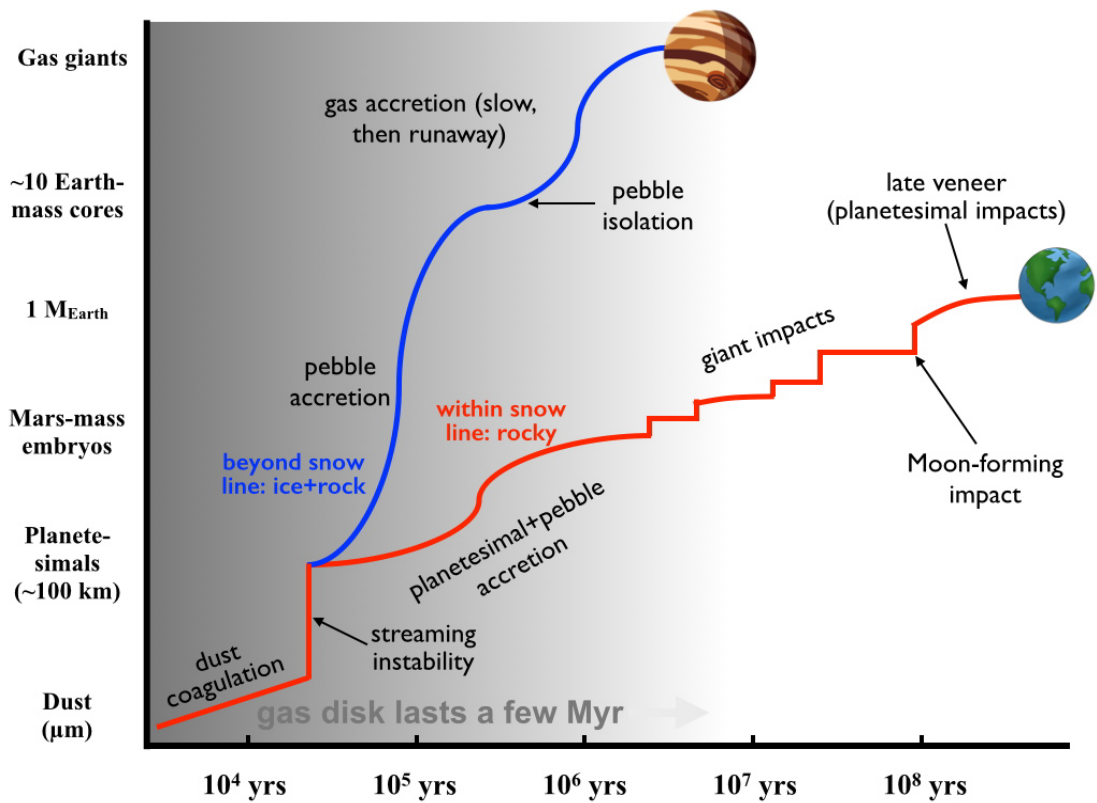

*Figure 27:* *General framework of planetary formation (adapted from Raymond & Morbidelli 2022). This image illustrates plausible growth pathways of Earth and Jupiter, highlighting the main processes responsible. Current thinking is that the primary branching point between the two growth modes comes from the fact that Jupiter's core accreted beyond*

*the water snow line, where the efficiency of pebble accretion was drastically higher (Lambrechts et al., 2014; Morbidelli et al., 2015).*

### *3.2.2. Processes of accretion*

Although only a minor (~1%) constituent of the protoplanetary disk mass, the dust is of vital importance to us as it represents the solid building blocks of the planets. The first phases of growth involve coagulation of dust by low-velocity collisions (e.g., Dominik and Tielens, 1997). As dust grains grow and collide there is a rich variety of outcomes, which has been studied in laboratory experiments and using numerical simulations (Blum and Wurm, 2008; Güttler *et al.*, 2010; Zsom *et al.*, 2010). Collisional growth is efficient at small sizes but is stifled as dust grows to mm- to cm-sizes, simply because collisions no longer lead to growth (e.g. Birnstiel *et al.*, 2016). In addition, it is at this size scale that interactions with the gaseous disk increase in importance.

The force balance in the disk is such that, at a given orbital radius, gas orbits the Sun at a speed that is slower than the local Keplerian (purely gravitational) velocity. This is because the force of gravity is slightly counteracted by thermal pressure, as gas closer to the star is hotter and, therefore, pushes outward. Small dust grains are simply entrained with the gas, but as they grow the pressure force weakens. At a critical size, dust grains start to decouple from the gas' motion. Because the gas is sub-Keplerian, large dust grains feel a headwind. This headwind acts to extract orbital energy and cause the dust to spiral inward (Weidenschilling, 1977; Birnstiel *et al.*, 2012). Drifting dust grains are commonly referred to in the literature as 'pebbles', although it is noted here that the sizes of what is considered a pebble depends on its aerodynamic properties and thus on the local gas density in the disk (Johansen and Lambrechts, 2017; Ormel, 2017).

Pebble drift was once considered a catastrophe because all macroscopic solids should simply spiral into the Sun on short timescales (Weidenschilling, 1977). Yet newer models show that drifting pebbles are likely the lifeblood of planet formation. The first step is the formation of planetesimals, which can form directly from pebbles by gas-particle instabilities such as the streaming instability (Johansen *et al.*, 2014). When solids are locally concentrated relative to the gas, they collectively create a back-reaction on the gas, accelerating its orbital speed in that location. This removes the headwind and stabilizes the orbits of pebbles in a ring. Other pebbles that drift inward are then trapped in the ring and can enhance this feedback on the gas, increasing the local pebble-to-gas ratio. Once this ratio exceeds a critical value, the streaming instability naturally acts to clump pebbles directly into 100 km-scale planetesimals that follow a characteristic mass distribution (Carrera *et al.*, 2015; Simon *et al.*, 2016; Yang *et al.*, 2017; Abod *et al.*, 2019). Simulations show that planetesimals are likely to form first just outside strong condensation fronts such as the water snow line (Armitage *et al.*, 2016; Schoonenberg and Ormel, 2017; Drążkowska and Alibert, 2017). While this remains a vigorous area of study, several recent studies have proposed that planetesimals may commonly form in relatively narrow rings rather than broad disks (Drążkowska et al., 2016; Charnoz et al., 2019 Morbidelli et al., 2022, Izidoro et al., 2022). The Solar System appears to require at least three such rings to explain the orbital structure of the planets and small bodies. Izidoro et al. (2022) suggested that such rings could have been linked with pressure bumps associated with the condensation fronts of silicates, water, and carbon monoxide (corresponding to the inner-outer rings, respectively).

Once large planetesimals have formed, they continue to grow by both colliding with other planetesimals and by accreting pebbles that continue to drift through the disk. Planetesimal accretion has been studied for decades (e.g., Safronov and Zvjagina, 1969; Greenberg *et al.*, 1978; Wetherill and Stewart, 1989; Ida and Makino, 1992), and is generally thought to be the dominant growth mode of the terrestrial planets (Chambers, 2016; Izidoro *et al.*, 2021). Starting from a broad disk of planetesimals assumed to have formed early, simulations show that roughly Mars-mass planetary embryos naturally form in the terrestrial region on a million-year timescale, with closer-in embryos forming significantly earlier (Kokubo and Ida, 1998; Kokubo, 2000; Leinhardt and Richardson, 2005; Walsh and Levison, 2019). The final phases of growth then involve giant impacts between planetary embryos as well as with remnant planetesimals (e.g., Chambers, 2004; Morbidelli *et al.*, 2012; Raymond *et al.*, 2014; see discussion below). This timeline is consistent with the cosmochemically-derived rapid formation of Mars (Dauphas and Pourmand, 2011), implying that it can be interpreted as a stranded planetary embryo.

### *3.2.3. Giant planet formation and migration (during gas disk phase), and instability*

Planetesimal accretion is too inefficient to explain the rapid growth of the gas giants, whose several Earth-mass cores must have grown within the few million-year lifetime of the gaseous disk (Thommes *et al.*, 2003; Levison *et al.*, 2010). Pebble accretion is far more efficient beyond the snow line. As long as there is a sufficient reservoir in mm- to cm-sized bodies, pebble accretion can grow giant planet cores to several Earth masses in a fraction of the gaseous disk lifetime (Ormel and Klahr, 2010; Lambrechts and Johansen, 2012; Levison *et al.*, 2015). Pebble accretion is self-limiting: at a critical mass, a growing core creates a pressure bump exterior to its orbit that traps pebbles and halts further growth (Morbidelli and Nesvorny, 2012; Lambrechts *et al.*, 2014; Bitsch *et al.*, 2018). The 'pebble isolation' mass depends on the properties of the gaseous disk, but for characteristic values it is roughly 20 Earth masses in the Jupiter-Saturn region.

The cores of the gas giants must have undergone orbital migration for the simple reason that they must have grown massive while the gaseous disk was still dense. Decades of analytical and numerical modeling have shown that planet-disk interactions invariably lead to angular momentum exchange between a planet's orbit and the gaseous disk, causing the planet's orbit to shrink or sometimes grow (see reviews by Kley and Nelson, 2012; Baruteau *et al.*, 2014). The direction and speed of migration of the gas giants' cores depends on the detailed structure of the disk. In simple viscous-disk models' migration is directed inward in the outer part of the disk but outward closer-in, creating zones of convergent migration at a few au (Lyra *et al.*, 2010; Bitsch *et al.*, 2015a; Bitsch *et al.*, 2014). Jupiter and Saturn's cores may have converged at such a location, allowing them to accrete gas from the disk at roughly constant orbital radii. Once Jupiter's gaseous envelope reached a critical mass it underwent runaway gas accretion, rapidly growing to hundreds of Earth masses in just ~$10^5$ years (Pollack *et al.*, 1996; Ida and Lin, 2004; Lissauer *et al.*, 2009). It gravitationally carved an azimuthal gap in the gaseous disk, thus transitioning to a different mode of migration directly linked to the disk's viscous evolution (Lin and Papaloizou, 1986; Ward, 1997)) Saturn likely underwent the same rapid accretion at a later time, when the gaseous disk was less dense, and so ended up at a lower mass. This view is supported by the different isotopic signatures in H and N between Saturn and Jupiter (e.g., Mousis *et al.*, 2018, and refs. therein).

The evolution of the gas giants' orbital radii is uncertain. One school of thought proposes that, given the propensity for inward migration, they must have originated much farther from the Sun – at ~15-25 au – to end up at their current radii (Bitsch *et al.*, 2015b). However, it is also possible that they underwent limited migration. Hydrodynamical simulations find that, on its own, each planet would migrate inward (e.g., D'Angelo *et al.*, 2005). Yet both planets together can share a common gap within the disk and enter a different migration. Depending on the disk properties the gas giants can migrate slowly inward, remain on roughly stationary orbits, or even migrate outward mode (Masset and Snellgrove, 2001; Morbidelli and Crida, 2007; Zhang and Zhou, 2010; Pierens *et al.*, 2014). The possibility of outward migration is the cornerstone of the Grand Tack model (Walsh *et al.*, 2011) in which Jupiter grew first and migrated inward on its own, and then migrated back outward along with Saturn.

At the end of the gas disk phase, the gas giants were likely in a resonant configuration regardless of their exact migration history. When two planets migrate toward each other, they invariably become trapped in orbital resonances, in which the planets' orbital periods form the ratio of small integers – for instance, in 3:2 resonance the outer planet completes two orbits in the time that the inner planet completes three. The most probable resonances for Jupiter and Saturn were the 3:2 or 2:1 (Pierens and Nelson, 2008). The ice giants were likely also trapped in a chain of orbital resonances extending out to 10 to 15 au (Morbidelli and Crida, 2007; Izidoro *et al.*, 2015). This is a far cry from their current orbital locations, with Jupiter and Saturn just inside the 5:2 orbital resonance. The giant planets likely reached their present-day orbits as a result of a dynamical instability (Tsiganis *et al.*, 2005). This dynamical instability, which is colloquially known as the 'Nice model,' was originally proposed as a delayed event to explain the terminal lunar cataclysm (Tera *et al.*, 1974), a perceived spike in the bombardment rate on the Moon roughly 500-700 million years after the planets formed (Gomes *et al.*, 2005). New analyses have called into question the existence of a terminal lunar cataclysm (e.g., Boehnke and Harrison, 2016), and several studies propose that the giant planets' instability must have taken place much earlier, anytime within ~100 million years of the start of planet formation (Zellner, 2017; Morbidelli *et al.*, 2018; Mojzsis *et al.*, 2019). If the instability took place early enough, it may even have played an important role in sculpting the inner Solar System (Clement *et al.*, 2019; Clement *et al.*, 2021a); see discussion in Section 4.2). This may very well have been the case: Liu et al. (2022) proposed that the instability was triggered by the inside-out dispersal of the Sun's gaseous disk, which would nail down the timing to within 5-10 Myr of CAIs (e.g., Hunt et al.,2022).

The dynamical instability involved close gravitational encounters between the ice giants, gas giants, and a massive outer disk of planetesimals. The instability spread out the giant planets' orbits, massively depleted the outer planetesimal disk, and likely resulted in the ejection of one or two additional ice giants into interstellar space (Nesvorný, 2011; Batygin and Brown, 2016). The instability can explain the orbital structure (and in some cases, the very existence) of a number of small body populations in the Solar System including Jupiter's Trojan asteroids (Morbidelli *et al.*, 2005; Nesvorný *et al.*, 2013) the giant planets' irregular satellites (Nesvorný *et al.*, 2007), the asteroid (Roig and Nesvorný, 2015; Deienno *et al.*, 2018) and Kuiper belts (Levison *et al.*, 2008; Nesvorný, 2015).

The evidence for the giant planet instability is remarkably strong (although admittedly circumstantial; see review by Nesvorný, 2018), yet the exact evolution of the giant planets' orbits is hard to pin down. Dynamical instabilities are inherently stochastic such that tiny changes in the planets' positions during individual gravitational encounters substantially affect the outcome. The present-day giant planets' orbits are consistent with a broad range of initial conditions and evolutionary paths during the instability (Nesvorný and Morbidelli, 2012; Clement *et al.*, 2021a; Clement *et al.*, 2021b). Connecting the exact evolutionary pathway of the giant planets with the orbital structures of small body populations remains another active area of study (e.g., Clement *et al.*, 2020; Nesvorný, 2021).

The gas giants' growth likely affected the compositions and orbital distribution of small bodies in several ways. First, once the giant planets' cores reached the pebble isolation mass they blocked the flux of pebbles into the inner Solar System and providing a barrier between the reservoirs of pebbles (e.g. Budde *et al.*, 2016; Kruijer *et al.*, 2017). Second, the gas giants' rapid gas accretion represented a drastic change in the gravitational environment for small bodies. For objects already present in the asteroid belt, Jupiter's runaway gas accretion greatly increased the collision velocities and may have induced collisional erosion (Turrini *et al.*, 2012). The gas giants' growth also destabilized the orbits of any planetesimals in a broad region between roughly 4-10 AU and scattered them in all directions. Under the action of aerodynamic gas drag, a fraction of scattered objects were implanted into the asteroid belt on stable orbits, with a distribution that matches that of C-type asteroids (Raymond and Izidoro, 2017a). Third, the gas giants' migration may have shepherded planetesimals across the Solar System, inducing large-scale radial mixing (Walsh *et al.*, 2011; Raymond and Izidoro, 2017b; Pirani *et al.*, 2019). Finally, while the instability cleared out the majority of the outer Solar System's planetesimals, it also sculpted the present-day comet reservoirs, both by scattering them onto very wide orbits in the Oort cloud and by trapping them in select locations within the Kuiper belt and Scattered disk (see chapter by Kaib and Volk in this volume). In the process, a fraction of planetesimals was implanted in the inner Solar System – notably in the outer asteroid belt and Jupiter's Trojan swarm (Levison *et al.*, 2009; Nesvorný *et al.*, 2013; Ribeiro de Sousa et al.,2022).

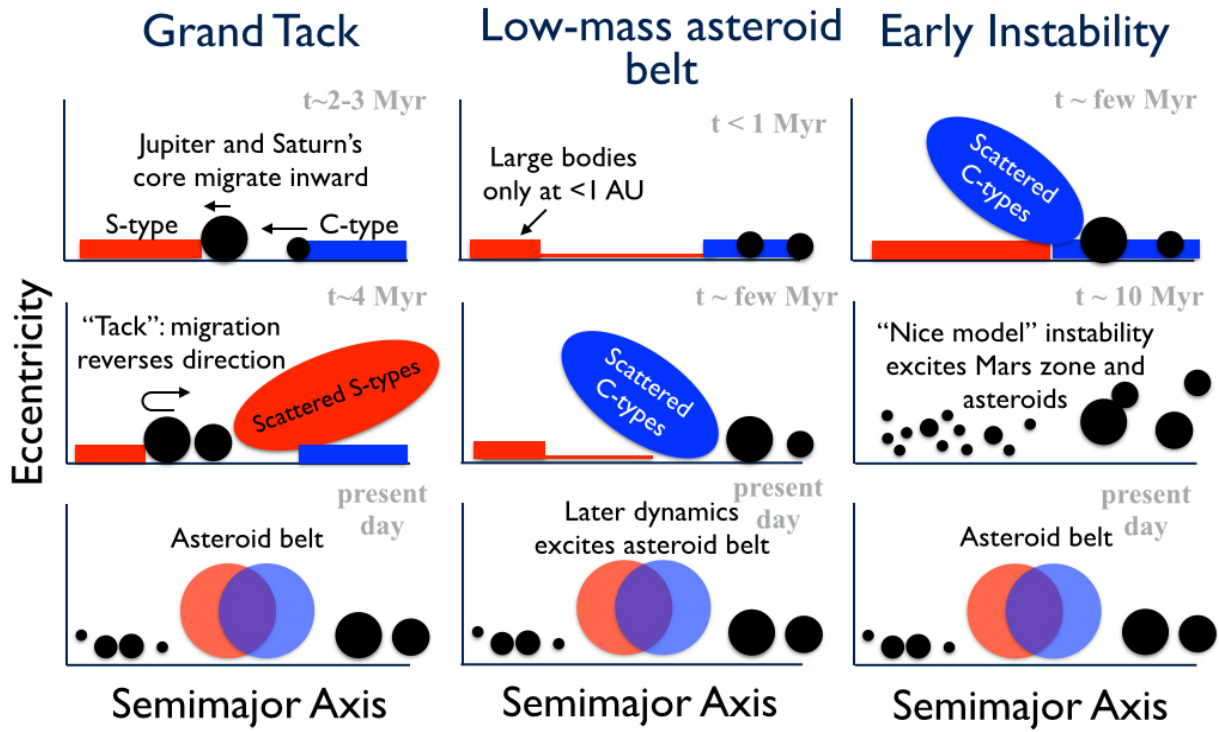

*Figure 28: Cartoon of different global models that can match the orbital architecture of the Solar System (adapted from Raymond et al., 2020). They are discussed in section 3.2.4. In context, comets are thought to have originated from ice-rich planetesimals that accreted beyond the giant planets' orbits (Izidoro et al.,2022; see Simon et al., Chapter 2, this volume). Comets were scattered by the giant planets into their present-day reservoirs (the Kuiper belt and Oort cloud) during the giant planets' growth and (perhaps) migration (see discussion in Sec. 3.2.6), and most strongly during the giant planet instability ((Nesvorný, 2018; Kaib & Volk, Chapter 4, this volume). Red and blue objects may correspond to the parent bodies of NC and CC meteorites.*

### 3.2.4 How many pathways could match the sample-based constraints?

There is currently a multitude of plausible terrestrial planet formation models. These models fall in two broad categories depending on whether the planets' growth is dominated by planetesimal- or pebble accretion. The planetesimal accretion models include the Grand Tack, Early Instability, and Low-mass Asteroid belt models (**Figure 28**; see Raymond et al., 2020, for a detailed discussion). Each of these models was created in part to solve the so-called 'small Mars problem', so named because simulations of the classical model tend to form Mars analogs that are roughly an order of magnitude more massive that the real Mars (Wetherhill, 1991; Chambers, 2001; Raymond *et al.*, 2009; Fischer and Ciesla, 2014). Successful formation models thus include mechanisms to deplete the planetesimal reservoir between roughly Earth's and Jupiter's present-day orbits.

The Low-mass Asteroid belt model proposes that planetesimals never formed efficiently in the present-day belt. As such, the depletion in Mars' feeding zone was primordial. Models that include pebble drift and a prescription for planetesimal formation within evolving gaseous disks do indeed find that rings of planetesimals may form (Drazkowska et al., 2016; Charnoz et al., 2019; Izidoro et al., 2022). If the terrestrial planets formed predominantly from a ring of planetesimals between ~0.7-1 au, their masses and orbits would naturally be reproduced (Hansen 2009, Kaib & Cowan 2015, Raymond & Izidoro 2017b). The outer asteroid belt would have been populated in large part with planetesimals scattered inward during the giant planets' growth, and some of those planetesimals would have delivered water to the terrestrial planets (Raymond & Izidoro 2017a).

The Grand Tack model (Walsh et al., 2011, 2012, O'Brien et al.,2014, Brasser et al.,2016) invokes Jupiter's migration to deplete Mars' feeding zone. In this model, Jupiter is assumed to have formed at a few au and migrated inward, then turned around (or "tacked") after Saturn migrated inward and became trapped in resonance with Jupiter, triggering the outward migration mechanism discussed in Section 2.4.2 (Masset and Snellgrove 2001). If this "tack" happened when Jupiter was at 1.5-2 au then the inner disk of planetesimals and planetary embryos would have been truncated into a narrow ring similar to the one invoked in the Low-mass asteroid belt model (Walsh et al., 2011, Brasser et al., 2016). The Grand Tack model can match the orbital structure of the terrestrial planets and asteroid belt and appears consistent with cosmochemical constraints (see Raymond & Morbidelli 2014).

The Early Instability model invokes an early giant instability (Clement et al., 2018; 2019; 2021, Nesvorný et al., 2021). The instability would have excited the orbits of planetesimals exterior to roughly 1 au, depleting the asteroid belt and stunting Mars' growth. While the detailed evolutionary pathway of the giant planet instability is hard to pin down (see Section 2.4.2), simulations that best match the orbits of the giant planets are the same ones that best match the orbits of the terrestrial planets (Clement et al., 2021b). The Early Instability model can also match the terrestrial planets' and asteroids' orbits.

Two pebble-driven models appear capable of matching the terrestrial planets' orbits. Johansen et al. (2021) showed that the terrestrial planets were consistent with the combined effects of pebble accretion and inward migration if planetesimals only formed at a specific location a little exterior to Mars' orbit (at ~1.6 au). Under such an assumption there is a correlation between the distance from the starting location and the planet mass, which can be matched assuming that this process ended up forming Earth in two parts: the proto-Earth and the Moon-forming body (often called Theia), each of which was less massive than Venus.

Another pebble-driven model invokes convergent orbital migration toward ~1 au (Broz et al., 2021). In that model the most massive planets (Earth and Venus) would have grown closest to the location of convergent migration whereas the smaller planets (Mercury and Mars) would have been at the outskirts.

The distribution of present-day comet orbits is largely independent of exactly how the terrestrial planets formed. This is because the Oort cloud, Kuiper belt and Scattered Disk were all mainly populated during the giant planet instability (Brasser & Morbidelli 2013; Nesvorný 2018; see Kaib & Volk, Chapter 4, this volume). Current thinking is that the giant planet instability took place, regardless of how the terrestrial planets formed – although, as discussed above, if the instability took place early then it must have affected the rocky planets' growth (Clement et al., 2018; Liu et al., 2022).

### *3.2.5. Disk evolution informed by the NC-CC dichotomy*

How the NC-CC isotopic dichotomy was established is debated. The consistency of meteorites falling into the NC or CC group suggests that the NC-CC isotope dichotomy is a pervasive feature of the protoplanetary disk. Two primary unknowns are (1) what caused the formation of two distinct isotope reservoirs which appear to have been isolated from each other in the protoplanetary disk? (2) What caused the compositional difference between the NC and CC reservoirs?

Regarding (1), Warren (2011) proposed that because the NC and CC groups did not overlap in isotopic composition, the dichotomy reflected the sampling of a temporal or spatial compositional dichotomy in the disk. It was speculated that the CC group represents material sourced from the outer Solar System and the NC group from the inner Solar System. This is because at that time, members of the CC group were understood to be typically more volatile rich than NC bodies (e.g., Wood, 2005; Gounelle *et al.*, 2006b). The author proposed that the two regions may have been physical isolated from each other by the formation of Jupiter; thus, NC meteorites would be sourced from regions inward of Jupiter and CC meteorites from regions outward of Jupiter.

Combining Mo and W nucleosynthetic isotope data with $^{182}$Hf-$^{182}$W model-derived accretion ages, Kruijer et al. (2017) concluded that NC bodies (<0.4 Myr after CAI formation) formed before CC bodies (0.9 +0.4/−0.2 Myr after CAI formation) and inferred an early age for Jupiter (Kruijer *et al.*, 2017). The strict age difference between NC and CC parent bodies is no longer supported following the addition of new data (e.g., **Figure 24**; Scott *et al.*, 2018; Hilton *et al.*, 2019), as some NC parent bodies are as old as those of CCs (**Figure 25**). The cause of segregation has also been revisited. A pressure maximum or a dust trap in the disk near the location at which Jupiter later formed, rather than the early formation of proto-Jupiter, has been proposed as the event that separated the NC and CC reservoirs (Brasser and Mojzsis, 2020). Izidoro *et al.* (2021) found that such a barrier was consistent with the structure of the inner Solar System, with the terrestrial planets forming from planetesimals rather than pebbles. In contrast, Johansen *et al.* (2021) invoked planetesimal formation at very specific, disjoint orbital radii to explain the dichotomy without invoking a barrier between the inner and outer Solar System. Lichtenberg *et al.* (2021) explained the origin of NC and CC planetesimal reservoirs as coming from independent bursts of planet formation separated in both time and orbital radius.

Regarding (2), the compositional difference between the NC and CC reservoirs was originally proposed to result from the addition of *r*-process-rich material to the CC reservoir which did not infiltrate the coexisting, yet spatially separated, NC reservoir (Kruijer *et al.*, 2017). For example, in Mo isotope compositional space (e.g., $\mu^{94}$Mo *vs.* $\mu^{95}$Mo), the NC and CC groups lie on two close-to parallel lines (e.g., Budde *et al.*, 2016; Worsham *et al.*, 2017; Poole *et al.*, 2017; Kruijer *et al.*, 2017; Budde *et al.*, 2019; Yokoyama *et al.*, 2019). Budde *et al.* (2016), Worsham *et al.* (2017), and Poole *et al.* (2017) proposed that the offset between the lines could be explained by the CC region containing more *r*-process and/or *p*-process material than the NC region. Stephan and Davis (2021) contrasted Mo isotopic data for presolar SiC grains and meteorites and determined that there is a fixed ratio between *p*- and *r*-process contributions in all sample data. This was interpreted to indicate that the NC-CC dichotomy, defined by Mo isotopes, can be explained by variations in the isotopic makeup of the *s*-process contribution to the samples. Desch *et al.* (2018) used an extensive isotopic and age data compilation of NC and CC meteorites to develop a numerical disk evolution model that constrains why carbonaceous chondrites have an overabundance of CAIs relative to NC meteorites. They determined that inside Jupiter's orbit, CAIs could be depleted by aerodynamic drag which could result in an offset in bulk sample isotopic composition, potentially identifying why NC and CC meteorites had different isotopic compositions. Alexander (2019a; 2019b) reported a comprehensive study of the elemental and isotopic characteristics of CC and NC meteorites and identified that different mixtures of the same four meteorite components could reproduce most bulk compositions of CC meteorites, but NC meteorites require different mixtures. Burkhardt *et al.* (2019) used an isotope database that included bulk meteorites, meteorite acid leachates, and presolar grain data and concluded that by mixing "CAI-like" material into an NC-like composition, the chemical and isotopic composition of the CC reservoir could be produced. This conclusion is similar to early reports (e.g., Gerber *et al.*, 2017) that concluded the Ti isotopic compositions of chondrules and CAIs can be attributed to the addition of isotopically heterogeneous CAI-like material to enstatite and ordinary chondrite-like chondrule precursors.

Deriving a unifying theory to explain the NC-CC dichotomy remains elusive. Attempts include those of Burkhardt *et al.* (2019), Nanne *et al.* (2019), and Spitzer *et al.* (2020) who proposed variations of so-called *infall* models. Authors proposed that the change in isotope composition between NC and CC reflects a change in the composition of infalling material from the parental molecular cloud into the disk, coupled with variable mixing and subsequent isolation (by proto-Jupiter) of reservoirs within the disk (**Figure 29**). Although infall models are currently the most frequently used models to address questions regarding the origin of the NC-CC isotope dichotomy, their feasibility depends on central assumptions that are yet to be verified (see Bermingham *et al.*, 2020 for further discussion).

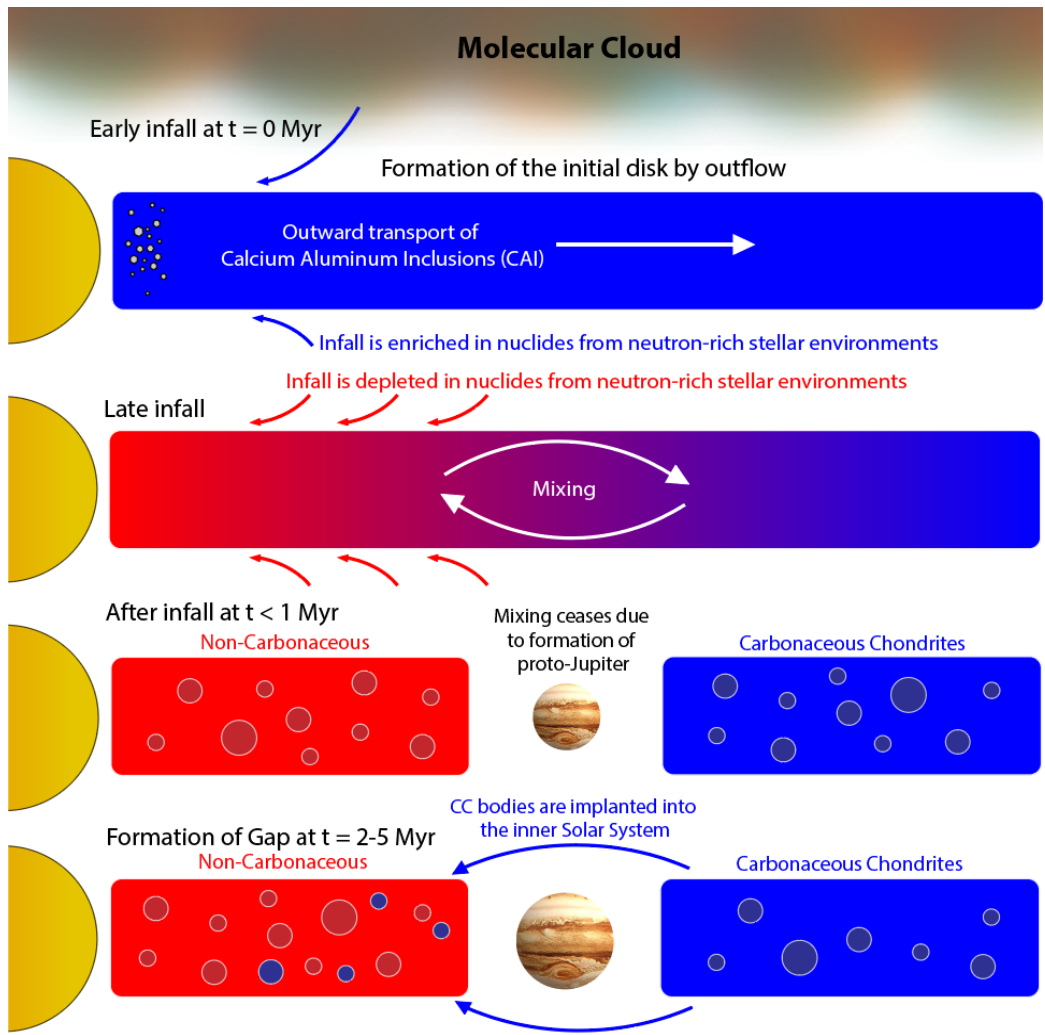

*Figure 29: Infall model of the protoplanetary disk. Stages I and II (~$t_0$) the disk grew which facilitated rapid transport and radial mixing of nebular dust and infalling material from the parental molecular cloud. The growth of proto-Jupiter to the "pebble isolation mass" (see **Figure 31**; Stage III, t<1 Myr) may have segregated inner and outer protoplanetary disk material, preserving any pre-existing isotope difference between the domains which sourced NC and CC materials. Stage IV, (t~2–5 Myr) saw Jupiter's migration which facilitated the inward scattering of planetesimals and their implantation into the main asteroid belt region (Raymond and Izidoro, 2017b). Figure modified after Nanne et al. (2019), Kruijer et al. (2020), and Bermingham and Kruijer (2022).*

Despite the lack of certainty about the cause of the NC-CC isotopic dichotomy, the NC-CC character of parent bodies is used as a tool to cosmolocate material in the protoplanetary disk to recreate the physico-chemical evolution of the protoplanetary disk (e.g., Kruijer *et al.*, (2017; 2020; Desch *et al.*, 2018; Nanne *et al.*, 2019; Burkhardt *et al.*, 2019). This approach contends that NC and CC isotope signatures in meteorites reflect the unique regional characteristics of the inner and outer disk, respectively. Consequently, it is purported that the NC-CC isotope compositions and the timing of accretion can be used to trace communication between the inner and outer regions of the disk, and potentially the source of water and other volatiles for Earth and the terrestrial planets. Questions remain, however, about the significance and utility of the NC and CC dichotomy to cosmolocate material in the protoplanetary disk. To use the NC-CC dichotomy as an indicator of volatile element distribution throughout the Solar System, a reliable link between the NC-CC isotope signature of a meteorite, the volatile content of its parent body, and the volatile content of the protoplanetary disk from which it accreted is required (Bermingham et al., 2020). Although there may be an increase in volatile element abundances when comparing some stony NC and CC meteorites and their inferred accretion locations within the disk, this is not necessarily the general rule. Bermingham et al. (2020) contrasted the nucleosynthetic isotope compositions of whole rock meteorite samples with major volatile element compositions (i.e., C, N, and O), and found that CC parent bodies are not always "volatile-rich" relative to those that formed in the NC. Based on these data, it becomes challenging to assume that all CC meteorites accreted outward and NC meteorites accreted inward of Jupiter. Further research is required to determine how faithfully the volatile element composition of the disk is preserved in bulk meteorite element abundances and isotopic compositions, and thus if bulk meteorite volatile compositions can be used to assess if the NC-CC dichotomy extends to volatile elements.

### *3.2.6. Radial mixing of small bodies driven by planet formation and dynamics*

Radial mixing of solids likely took place at several different phases of planet formation. If the NC-CC isotopic dichotomy reflects inner *vs.* outer Solar System materials, this dichotomy would not exist if that mixing had been complete, nor would the large differences in compositions between the planets, different types of asteroids, comets, and other small body populations. Large-scale radial (and perhaps temporal) gradients in composition were imprinted onto the constituents of the planets during planet formation.

The emerging paradigm of pebble drift invokes large-scale radial mixing from the outer parts of the Solar System inward. The source of inward-drifting pebbles, or 'pebble production front', is thought to expand outward in time because the timescale for dust to coagulate to pebble sizes correlates strongly with distance (Lambrechts and Johansen, 2012; Lambrechts *et al.*, 2019; Ida *et al.*, 2016). The radial source region of pebbles at a given location is thus continually expanding outward in time. Given their small sizes, inward-drifting pebbles are generally thought to rapidly lose their ices and to 'forget' their formation location (e.g., Morbidelli *et al.*, 2015). The bulk compositions of pebbles accreted by a growing planet are thus likely to be dictated by the local temperature. Pebbles' isotopic compositions, however, are unlikely to be altered by their inward drift in the disk. This means that isotopic analyses may be the most promising tool to constrain the role of pebble accretion of different Solar System bodies. Yet piecing together the full picture of Earth's growth and whether it was built mostly from planetesimals, or pebbles will require isotopic studies of a number of different elements including simultaneous consideration of constraints from lithophiles, siderophiles, and volatiles.

Additional radial mixing took place at planetesimal sizes, driven by dynamical interactions. Once Mars-mass planetary embryos formed, they excited the orbits of nearby planetesimals (e.g., Kokubo and Ida, 1998). Most planetesimals are accreted by an embryo close to their starting location, but a fraction is scattered multiple times by many embryos and can traverse the inner Solar System. Simulations in the framework of the 'classical model' of terrestrial planet formation find that the growing Earth accretes a significant budget of planetesimals from the outer asteroid belt (Morbidelli *et al.*, 2000; Raymond *et al.*, 2007; Raymond *et al.*, 2009). Scattering can also transport material outward, and planetesimals from the terrestrial planet-forming region can in some cases be scattered outward and trapped on stable orbits in the asteroid belt (Bottke *et al.*, 2006; Raymond and Izidoro, 2017a).

The giant planets were likely the dominant driver of radial mixing of planetesimals. As the giant planets' cores migrated (see Section 3.2.3), they shepherded planetesimals along with them, either inward (Raymond and Izidoro, 2017b; Pirani *et al.*, 2019) or possibly outward (Raymond *et al.*, 2016). The giant planets' rapid gas accretion destabilized the orbits of planetesimals in a wide belt, leading to gravitational scattering in all directions. The orbits of some scattered planetesimals were re-circularized under the action of aerodynamic gas drag, trapping them on stable orbits in the outer asteroid belt (Raymond and Izidoro, 2017b; Ronnet *et al.*, 2018). A much less substantial fraction of planetesimals could have been stabilized on orbits exterior to the growing gas giants' (Raymond and Izidoro, 2017b). The orbital migration of Jupiter and Saturn (as well as the ice giants') certainly caused widespread orbital mixing of planetesimals. In the context of the Grand Tack model, Jupiter's inward migration would have scattered out (or pushed inward) all homegrown asteroidal planetesimals (assumed to be associated with present-day S-type asteroids; see Walsh *et al.* (2011). The belt would have been repopulated during Jupiter and Saturn's subsequent outward migration from a combination of previously scattered S-types and planetesimals originating past the giant planets' orbits (assumed to be associated with present-day C-types). Finally, a side-effect of the giant planet instability was to implant a population of asteroids into the outer parts of the main belt, as well as in Jupiter's co-orbital region (the D-types and Jupiter's Trojans; (Morbidelli *et al.*, 2005; Levison *et al.*, 2009; Nesvorný *et al.*, 2013; Vokrouhlický *et al.*, 2016)).

## 4. EARTH'S ACCRETION

### 4.1. Stages of Earth's accretion

Contrary to giant planet formation, it took several tens of million years for Earth to reach its actual size **(Figure 30)**. Earth's accretion of Moon-to-Mars-sized planetary embryos can be broadly divided into two stages. The "main-stage accretion" occurred before core formation ceased. The Moon-forming event came at the end of this stage when a Mars-sized body (Theia) collided with proto-Earth (Benz *et al.*, 1986; Canup and Asphaug, 2001). Material from Theia merged with the proto-Earth's core and mantle and the Moon formed from the resultant debris cloud (**Figure 30**). The timing of the final giant impact on the Earth can be constrained using Hf/W isotopes (Kleine, 2005; Fischer and Nimmo, 2018). Most estimates find a timing of 40-100 million years after CAIs (e.g., Touboul *et al.*, 2007; Thiemens *et al.*, 2019). This marks the end of the giant impact phase in the inner Solar System. Late accretion (sometimes referred to as "late veneer") followed, during which ~0.5 to 2 wt. % of Earth's mass was accreted during this stage (Walker, 2009; Bottke *et al.*, 2010; Marty, 2012; Marchi *et al.*, 2018). Late accretion of planetesimals enriched Earth's crust and mantle with highly siderophile elements that, had they been accreted earlier, would have been sequestered in the core (Walker, 2009). Meteorite measurements indicate that Mars accreted ~9 times less than Earth during late accretion, and that the Moon accreted 200-1200 times less than Earth (Day *et al.*, 2007; Walker, 2009).

Reconciling the differences in highly siderophile element abundances directly constrains the dynamics of terrestrial planet formation and the properties of its leftovers (Bottke *et al.*, 2010; Schlichting *et al.*, 2012; Raymond *et al.*, 2013; Morbidelli *et al.*, 2018).

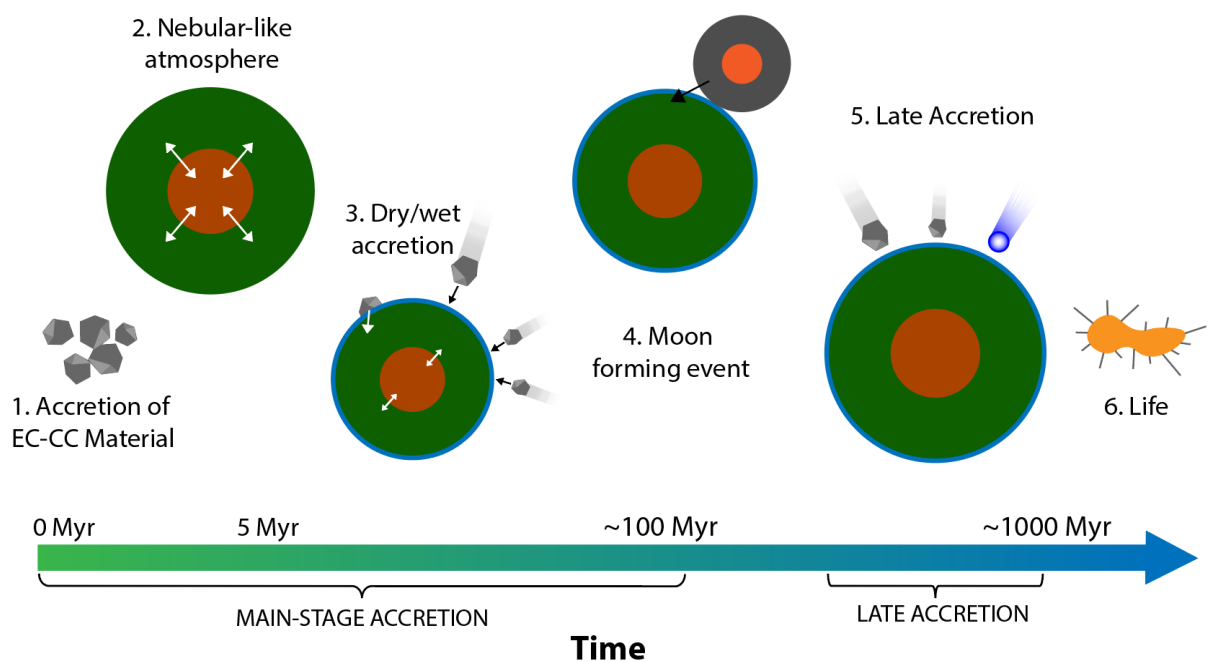

*Figure 30:* A schematic depicting Earth's accretion history. (1) and (2) Early accreted material formed a proto-Earth with a nebular-like atmosphere. (3) Accretion of planetesimals occurred during on-going core formation. (4) The Moon-forming impact was the final major addition of mass to proto-Earth (10 to 20 wt. % of Earth's current mass), after which final core segregation occurred. (5) A small fraction (0.5 to 2 wt. % of Earth's current mass) was added to the mantle after final core segregation during Late Accretion. (6) At some point following this, life began to emerge.

### 4.2. Major element (N, C, N) and noble gas constraints on the origin of terrestrial volatiles

#### 4.2.1. Major elements (N, C, N)

The causes of D and $^{15}$N isotope enrichments among Solar System objects and reservoirs (**Figures 18 and 19**) are not fully understood, but their variations can be used as tracers of provenance and of exchange between objects and reservoirs. The N and H isotope covariations with heliocentric distance constrain the origin of terrestrial water and the composition of associated volatile elements (Marty, 2012; Alexander et al., 2012). The Earth, the Moon, and the interior of Mars have D/H and $^{15}$N/$^{14}$N compositions within the range of values defined by primitive meteorites, and different from either those of the proto-solar nebula or of comets (**Figure 31**). This similarity indicates a mainly inner Solar System, rather than cometary or nebular, origin for the terrestrial atmosphere and oceans. It does not exclude minor contributions from either nebular gas (e.g., D-poor hydrogen in the mantle (Hallis et al., 2015;

Olson and Sharp, 2019); Solar-like neon in the terrestrial mantle) or comets (estimated from the Xe and Kr isotopic compositions of the terrestrial atmosphere, see subsection 4.4.).

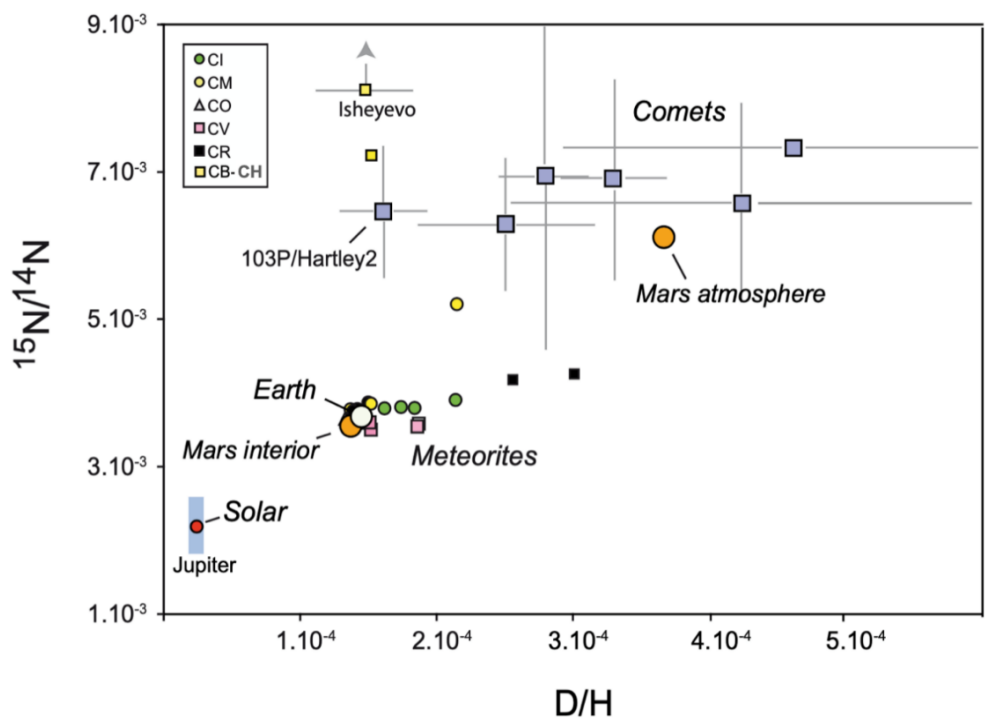

*Figure 31: Nitrogen and hydrogen isotope covariations among Solar System objects and reservoirs (adapted from Marty, 2012). Data define three reservoirs: (i) proto-solar nebula (Solar and Jupiter), (ii) inner planets and chondrites, and (iii) comets. This distribution was taken as evidence for a chondritic origin of volatile elements on Earth.*

### *4.2.2. Noble gases*

#### *4.2.2.1. Earth's mantle inventory*

The composition of terrestrial noble gases permits further identification of cosmochemical ancestors of terrestrial volatiles. In particular, the neon isotope composition of mantle volatiles trapped in oceanic basalts and mineral assemblages of mantle derivation points to the presence of a Solar Ne component at depth. Neon has three isotopes, $^{20}Ne$, $^{21}Ne$ and $^{22}Ne$, and, in a two-component mixing, $^{20}Ne/^{22}Ne$ versus $^{21}Ne/^{22}Ne$ variations define straight lines linking the component end-members **(Figure 32)**. One of them is atmospheric Ne ("Air" symbol in **Figure 32**; $^{20}Ne/^{22}Ne = 9.80$). Data for mantle-derived samples define different straight lines having variable slopes and pointing to components rich in $^{21}Ne$ and $^{20}Ne$. "Nucleogenic" $^{21}Ne$ has been produced in the mantle by neutron and alpha reactions on O and Mg isotopes over eons and is not a primordial component. Neon-22 is also produced by nuclear reaction, but to a lesser extent than $^{21}Ne$. Thus, deviations towards increasing $^{21}Ne/^{22}Ne$ ratios are a function of mantle residence time and not a primordial feature. In contrast, there is no known nuclear process producing quantitatively $^{20}Ne$ in Earth, and the data define a mantle end-member composition with $^{20}$Ne/$^{22}$Ne up to ~13.0 which is associated with mantle plumes (Yokochi and Marty, 2004; Williams and Mukhopadhyay, 2019) signifying a primordial composition. This value is notably higher than the range observed in chondrites (8.2 to 12.7; Ott, 2014), suggesting trapping in the mantle of a Solar gas component (Solar $^{20}$Ne/$^{22}$Ne = 13.36 ± 0.18; Heber *et al.*, 2012). A way to incorporate Solar neon in the mantle would be dissolution of nebular gas gravitationally trapped by the proto-Earth and dissolved in molten silicates during magma ocean episodes (Sasaki, 1990; Yokochi and Marty, 2004; Marty and Yokochi, 2006; Williams and Mukhopadhyay, 2019). This possibility implies that the proto-Earth might have been massive enough to capture a small nebular, Jupiter-like atmosphere, that was subsequently blown away by the active proto-Sun. Notably, xenon in the Martian mantle appears to have a Solar isotopic composition, and could have been trapped from a nebular-like atmosphere before gas dissipation, a possibility consistent with both the lifetime of the gas (a few Myr; Haisch *et al.*, 2001; Mamajek *et al.*, 2009) and the timeframe for Mars accretion (Dauphas and Pourmand, 2011). However, Péron and Mukhopadhyay (2022) recently argued that krypton in the Martian mantle is chondritic and not Solar, casting doubt on a Solar origin for Xe trapped in the Martian mantle.

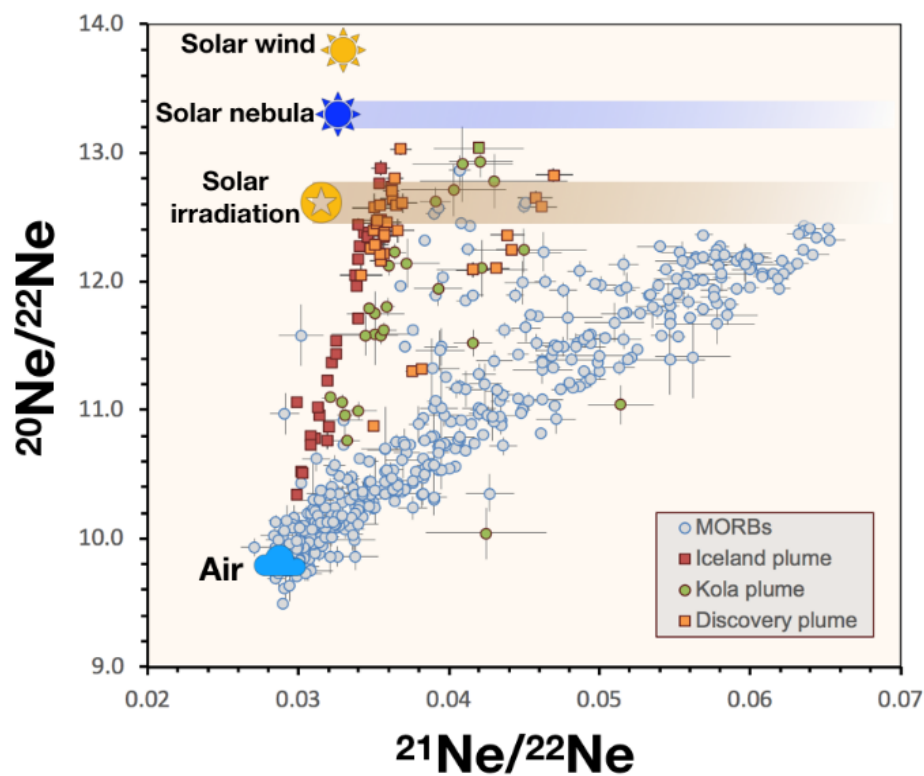

*Figure 32*: *Neon isotope data of mid-ocean ridge basalts (MORBs) and mantle plume samples. Data source: MORB data from compilation by M. Moreira (personal communication), Icelandic subglacial glass data from Mukhopadhyay (2012), Kola mantle plume carbonatite data from Yokochi and Marty (2004) and submarine glasses from the Discovery plume area data from Williams and Mukhopadhyay (2019). The solar wind value was measured on the Moon and more recently in Solar wind ions collected by the Genesis mission (Heber et al., 2012, and refs. therein). The Solar nebula value is inferred from the composition of the Sun, corrected*

*for isotope fractionation during Solar wind generation (Heber et al., 2012). The solar irradiation value is the Ne-B component, representing the end-member value found in lunar soils and solar gas-rich meteorites (Black and Pepin, 1969; Wieler et al., 1989). Ne-B is isotopically fractionated relative to Solar wind by surface processing on the Moon and on asteroids. In this format, a two-component mixing yields straight lines. In both cases the arrays defined by sample data result from mixing between the composition of atmospheric neon contaminating the samples and a mantle end-member composition. The slopes of correlations relate to the degassing state of the respective mantle reservoirs.*

Contrary to mantle plumes, the mantle end-member seen in mid ocean ridge basalts (MORBs), presumably sampling the shallower convective mantle, appears to have a lower $^{20}$Ne/$^{22}$Ne end-member ratio of 12.50-12.75 (Moreira, 2013; Péron *et al.*, 2016) than mantle plume neon. The MORB Ne signature appears comparable to that of the so-called Ne-B component (12.5±0.2) (Black and Pepin, 1969). This component, found in planetary regoliths including lunar soils, is thought to be original Solar wind Ne that was implanted in regolithic grains and isotopically fractionated during regolithic processing (Moreira, 2013; Trieloff and Kunz, 2005). Moreira (2013) and Moreira and Charnoz (2016) argued that the carrier of mantle Ne was dust irradiated by the nascent Sun as turbulence in the disk transported grains in regions that were unshielded from Solar irradiation. A dual origin for mantle neon cannot be discarded: dissolution from a primordial atmosphere in what is now the deep mantle, and contribution of irradiated dust at a later stage of accretion. Resolving the origin of mantle Ne will have far-reaching implications for understanding the evolution of the proto-Earth.

Contrary to Ne (and He), the Kr and Xe isotopic signature of the mantle is dominated by a chondritic component (Holland *et al.*, 2009; Broadley *et al.*, 2020), in line with H and N isotopes. This dichotomy could be due to the lower concentrations of Kr and Xe in the protosolar nebula compared to He and Ne, which, coupled with their lower solubility in basaltic melt, would result in Kr and Xe being less efficiently degassed than He and Ne (Olson and Sharp, 2019).

Recently, Piani *et al.* (2020) reported the analysis of hydrogen and its D/H ratio in a suite of enstatite chondrites (EC) and argued that EC could constitute an important source of mantle water. This possibility is consistent with the isotopic composition of several key elements like oxygen, titanium and others, which call for a EC-like source for the Earth. However, it also raises a mass balance problem: CC (CI and CM) (5~10% H$_2$O by mass) are one order of magnitude richer in equivalent water than EC (~0.5 % H$_2$O by mass), but only a factor of ~3 richer in trapped noble gases. If, therefore, mantle water was from an EC-like source and assuming closed system condition, the mantle should contain one order of magnitude more EC-like noble gases than CC-like noble gases. This possibility cannot be excluded and could be tested with heavy noble gas composition (since light noble gases, e.g., Ne, are dominated by a superimposed Solar component). The mantle, however, is drastically depleted in noble gases, by orders of magnitude compared to both EC and CC contents and has evidently lost its noble gas cargo presumably during accretion and partial melting. One possibility to retain EC-like hydrogen while degassing noble gases could be a low oxygen fugacity, consistent with the reduced character of EC-like material. The situation might have evolved with the subsequent delivery of oxidized material to the proto-Earth (Javoy *et al.*, 2010; Rubie *et al.*, 2015) which might have also supplied CC-like noble gases.

*4.2.2.2. Earth's surface inventory*

The noble gas and stable isotope signatures of the atmosphere and the oceans are markedly different from those of the mantle. This difference is best illustrated by the variations of the $^{20}Ne/^{22}Ne$ ratio as a function of the elemental $^{36}Ar/^{22}Ne$ ratio in different classes of chondrites compared to the atmosphere (Marty, 2012; Williams and Mukhopadhyay, 2019; Marty, 2020; 2022). In this format **(Figure 33)**, a two-component mixing results in a straight line, joining the component endmembers. Enstatite chondrite data cannot account for mixing between the atmosphere and the mantle (panel on the left), Noble gases in ECs often present enrichments in $^{36}Ar$ compared to other chondritic classes, a pattern labelled "subsolar" (Crabb and Anders, 1981). Argon-36 enrichments result in a large spread of data in the figure, which means they are collectively unable to account for the atmospheric composition atm in the figures). Some of the ECs are also rich in solar neon (Ne-B), and/or contain the so-called Ne-A (Black and Pepin, 1969), also labelled "planetary" component (Mazor et al., 1970), a presolar Ne component found in nanodiamonds and SiC grains. "Q-gas" is the Phase Q component ubiquitously found in chondrites. Error bars are generally large due to correction for cosmic ray effects and relevant error propagation.

In contrast, carbonaceous chondrite data define a mixing trend starting at the Solar-like mantle Ne component that encompasses nicely the atmospheric composition (**Figure 33**, right). Data can be well explained by mixing between a solar-like component and Ne-A (or planetary Ne) with low $^{20}Ne/^{22}Ne$ ratio (Mazor et al., 1970). The composition of the latter is dominated by presolar $^{22}Ne$ trapped in nanodiamonds. Remarkably, atmospheric and mantle noble gases are intermediate between Solar and Planetary, which strongly suggests a CC-like origin for terrestrial volatiles (Marty, 2012; 2022).

Another important implication of this diagram is that atmospheric escape to space (outlined by the dotted curve, right) cannot account for the composition of atmospheric noble gases, as otherwise often advocated (Pepin, 1991).

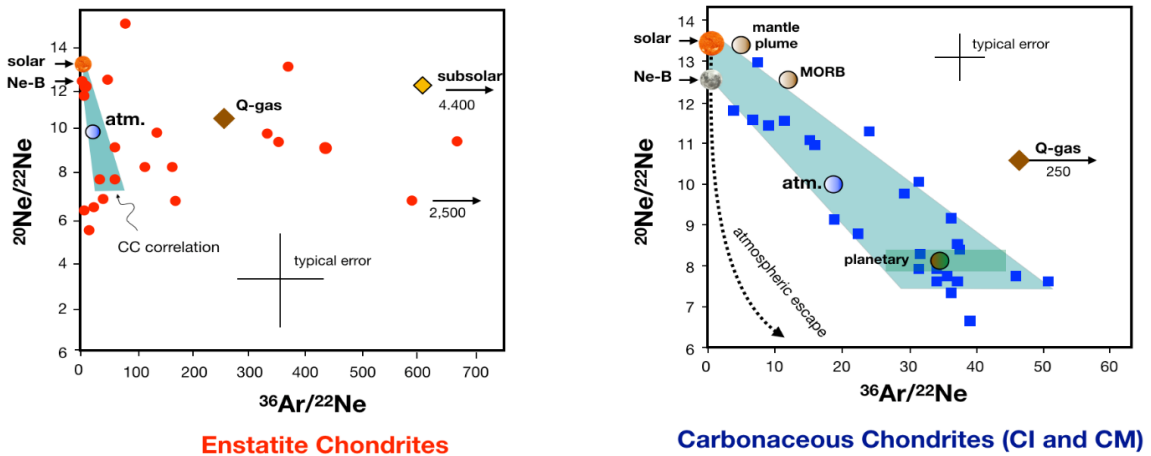

***Figure 33***: *$^{20}Ne/^{22}Ne$ ratio as a function of the elemental $^{36}Ar/^{22}Ne$ ratio in different classes of chondrites compared to terrestrial values. Left: Compilation of enstatite chondrite (EC) data. The above diagrams are based on a compilation of noble gas data in enstatite (NC-type) and*

*carbonaceous (CC-type, here CM and CI) chondrites (relevant literature listed at: https://zenodo.org/record/3984898#.XzZU7i3M2iA). Data were corrected for contribution of cosmic-ray produced Ne and Ar isotopes using the $^{21}Ne/^{22}Ne$ ratios. Adapted from Marty (2022).*

In summary, light noble gases (Ne and Ar) point to a carbonaceous chondrite-like origin for the Earth's surface inventory. In addition to a limited Solar-like Ne (and He, possibly H) component, mantle noble gases (Kr and Xe) are also dominated by a chondritic component, but not necessarily of the CC type. Stable isotope ratios of H and N are also consistent with the occurrence of an EC-like component: H might have been already present in NC-like material (Piani et al., 2020) as could have been C and N hosted by refractory organics. Hence terrestrial volatiles might have been sourced by different types of primitive material both within and outside the snowline.

Deciphering the fractional contributions of cosmochemical sources is difficult, and it depends on which element or isotope system is considered (see also Subsection 4.4.2.). The bulk Earth (mantle plus surface) H, C, N, noble gas, and halogen inventory is roughly equivalent to the contribution of ~2±1 % CC-type material to a dry proto-Earth (Marty, 2012), although estimates are variable depending on the considered volatiles (Hirschmann and Dasgupta, 2009; Hirschmann, 2018). Mass balance involving H and N isotopes suggest a NC/CC mix of about 75-95%/25-5% (Piani et al., 2020; Marty, 2020). Establishing the terrestrial inventory of volatiles also requires accounting for the core and its potential to store H (Wu *et al.*, 2018), C (Fischer *et al.*, 2020), N (Roskosz *et al.*, 2013; Dalou *et al.*, 2017) and possibly noble gases (Bouhifd *et al.*, 2013). Accurately anticipating the repartition of volatile elements between the major terrestrial reservoirs requires knowledge of the mode of accretion, the evolution of planetary atmospheres and the mode of core formation (e.g., Grewal *et al.*, 2019). This is presently an active area of research involving geochemistry, high pressure mineralogy and planetary dynamics. Given these uncertainties, isotopic tracers are often regarded as being more compelling than elemental abundance ratios to address the origin of terrestrial planetary building blocks.

## 4.3. Origin of Earth's volatiles (oceans, atmosphere) inferred from numerical modeling

Several numerical models have been developed to explain the origin of Earth's water (see Meech and Raymond, 2020). Some models invoke a local source of water at Earth's orbital distance, either via adsorption of hydrogen onto silicate grains (e.g., Sharp, 2017) or by oxidation of a primordial hydrogen-rich atmosphere (Genda and Ikoma, 2008). The possibility of Earth having trapped nebular-like water has been substantiated by the claim that the terrestrial mantle contains D-poor hydrogen (Hallis *et al.*, 2015). In addition, the recent discovery of abundant hydrogen in enstatite chondrite meteorites with Earth-like D/H ratios (Piani *et al.*, 2020) begs the question of whether Earth's water was locally sourced. Many local models, however, struggle to match chemical and isotopic constraints (see discussion in Meech and Raymond, 2020), and enstatite chondrites cannot account for Earth's full volatile budget **(Figure 33)**. Indeed, stable isotopes of volatile elements point to the water delivery by chondritic-like material (Alexander *et al.*, 2012; Marty, 2012) (see **Figure 10**). A such, here attention is focused on models in which water is 'delivered' from more distant parts of the planet-forming disk.

Water delivery is efficient in the context of the classical model, but the model has fallen out of favor. The classical model assumes that the terrestrial planets grew from a broad disk of planetesimals and embryos with minimal intrusion from the giant planets (discussed in (Morbidelli *et al.*, 2012; Clement *et al.*, 2020). Radial mixing of planetesimals is driven by scattering by planetary embryos, assumed to have formed throughout the inner Solar System (see Section 3.2). Simulations find that water is efficiently delivered by this mechanism and the growing Earth typically receives an order of magnitude more water than the present-day budget (Morbidelli *et al.*, 2000; Raymond *et al.*, 2009; Raymond and Izidoro, 2017a; Izidoro *et al.*, 2013). The framework of the classical model, however, has a fatal flaw in that it systematically fails to match the terrestrial planets' coupled mass and orbital distribution (see discussion in Clement *et al.*, 2020, and the discussion of the 'small Mars' problem in Section 3.2.4).

In the context of newer terrestrial planet formation models that can match the orbital structure of the terrestrial planets and asteroids (see Raymond *et al.*, 2020)), water delivery is driven by the giant planets. When Jupiter and Saturn underwent rapid gas accretion, they scattered nearby planetesimals in all directions; many were implanted into the outer asteroid belt as C-types (see Section 3.2). Some planetesimals were scattered past the asteroid belt toward the growing terrestrial planets, delivering water (Raymond and Izidoro, 2017a). In the context of the Grand Tack model, Jupiter's outward migration scattered planetesimals inward – again, some were trapped in the main asteroid belt as C-types and others were scattered past the belt to the terrestrial planet-forming region (Walsh *et al.*, 2011; O'Brien *et al.*, 2014). Each of these scenarios can deliver a few times Earth's current water budget, and each predicts that the water delivered should have the same chemical makeup as that of C-type asteroids. Indeed, the D/H ratios and the $^{15}N/^{14}N$ ratios of Earth's water are a good match to that of carbonaceous chondrite meteorites (e.g., Marty and Yokochi, 2006; Alexander *et al.*, 2012; Marty, 2012).

Water could in principle have been delivered to the growing Earth in the context of pebble accretion. As the gaseous protoplanetary disk cooled, the snow line swept inward. In most current disk models, the snow line is located interior to 1 au in the late phases of the disk's lifetime, and in some models for the majority of the disk's life (e.g., Sasselov and Lecar, 2000; Lecar *et al.*, 2006; Martin and Livio, 2012; Bitsch *et al.*, 2015b; Savvidou *et al.*, 2020). If the terrestrial planets grew mostly by accreting inward-drifting pebbles, then that accretion would include any ices contained in the pebbles. Models find that this mechanism can indeed operate in a hypothetical evolving disk (Sato *et al.*, 2016; Ida *et al.*, 2019), and it has been invoked in recent pebble-driven terrestrial planet formation models (Johansen *et al.*, 2021; Brož *et al.*, 2021). This is a delicate balancing act, however, given that most carbonaceous chondrite meteorites contain 10-100 times more water by mass than the Earth (e.g., Marty, 2012; Alexander *et al.*, 2012; 2018). In addition, the total flux in pebbles must remain low enough to avoid having the terrestrial planets transition to a super-Earth growth mode (Lambrechts *et al.*, 2019). To match the terrestrial planets' masses and orbits and simultaneously deliver the appropriate amount of water requires a disk with a fine-tuned evolution of the snow line and pebble flux.

### 4.4. Late losses and deliveries: chondritic and cometary contributions to Earth

*4.4.1. Evidence for volatile loss during Earth's accretion*

The main accretion of Earth ended with the Moon-forming impact (Giant Impact) 40-100 Myr after CAI (e.g., Lock *et al.*, 2020 and refs. therein), after which our planet evolved towards a habitable world. The Giant Impact was an extremely energetic event, which likely resulted in the partial, or complete, melting of the proto-Earth. To what extent existing volatiles survived this event is a matter of debate (Porcelli *et al.*, 2001; Genda and Abe, 2005). A dramatic loss of terrestrial volatiles during the Moon-forming event is supported by the observation that our planet contains at most a few percent of $^{129}$Xe produced from the decay of short-lived $^{129}$I ($t_{1/2}$ = 15.6 Myr) (Pepin, 1991; Allègre *et al.*, 1995; Avice and Marty, 2014), indicating that > 90% of terrestrial $^{129}$Xe was lost before or during the Moon-forming event. The noble gas content of bulk Earth, mostly concentrated in the atmosphere, corresponds to about ~2±1 % CC (Marty, 2012), attesting independently for efficient volatile loss. On the other hand, the delivery of chondritic material to Earth after the moon forming event (the late accretion) may amount for less than 1% Earth's mass, so that some of the noble gases (and presumably other volatiles) survived the Moon-forming event (Marty and Yokochi, 2006).

### *4.4.2. Tracing volatile accretion using nucleosynthetic isotope anomalies*

Deciphering the cosmochemical character and heliocentric source location of Earth's building blocks is complicated by mantle convection which unquestionably mixes these components in the mantle over time. Studies have shown, however, that the nucleosynthetic isotope composition of Earth building blocks can be constrained by examining the isotopic composition of elements with differing affinities for metal in mantle-derived materials (Dauphas, 2017). The lithophile (rock-loving) elemental composition of the mantle (e.g., O, Ca, Ti, and Nd) reflects the average composition of materials accreted by the Earth throughout its history (Walker *et al.*, 2015; Dauphas, 2017). The siderophile (iron-loving) elemental composition of the mantle, however, is biased towards material accreted during late stages of growth (e.g., final core segregation onwards). This is because during core formation the siderophile elements partitioned predominately into the core during core segregation. Their accumulation in the mantle thus partly reflects addition after final core formation. Accordingly, siderophile element isotopic compositions of the mantle can be used to trace the identify of building blocks towards the end of Earth's accretion, i.e., late-stage accretion (e.g., Dauphas et al., 2004; Walker *et al.*, 2015; Dauphas, 2017; Fischer-Gödde and Kleine, 2017; Bermingham *et al.*, 2018a; Budde *et al.*, 2019; Hopp *et al.*, 2020; Fischer-Gödde *et al.*, 2020). This approach contrasts the nucleosynthetic isotope compositions of meteorites with mantle-derived geochemical data that track different stages of Earth late-stage accretion history (e.g., Dauphas, 2017).

Siderophile elements Mo and Ru are well-suited to tracing late-stage accretion due to four primary reasons. (1) Most of the Mo and Ru mantle budgets preserve the chemical signatures of different stages of accretion. The Mo bulk silicate Earth (BSE) budget was likely established during the final 10 to 20 wt.% of Earth's accretion (i.e., during core formation), while the Ru BSE budget was likely established after core formation (i.e., during late accretion) (**Figure 34**) (cf. Dauphas et al., 2004; Dauphas, 2017). (2) Nucleosynthetic isotope variations in Mo and Ru discriminate between meteorites and the NC-CC character of a parent body (e.g., Budde et al., 2016; Worsham et al., 2017; Poole et al., 2017; Bermingham et al., 2018a). (3) The Mo and Ru nucleosynthetic isotope compositions of many meteorites correlate ("Mo-Ru Cosmic Correlation"), indicating that the presolar carriers responsible for Mo and Ru isotopic anomalies were the same (Dauphas et al., 2004; Bermingham et al., 2018a; Worsham et al., 2019; Hilton et al., 2019; Hopp et al., 2020; Tornabene et al., 2020). (4) Recent development of high precision isotopic methods for Ru and Mo (e.g., Fischer-Gödde et al., 2015; Nagai and Yokoyama, 2016; Bermingham et al., 2016; Worsham et al., 2016) permit resolution of small (~10 to 30 ppm) variations in Mo and Ru isotopes in terrestrial materials. With these state-of-the-art analytical tools to hand, the search for remnants of Earth's building blocks has begun.

Few terrestrial samples have been analyzed for Mo and Ru isotopic compositions (Fischer-Gödde et al., 2015, 2020; Bermingham et al., 2016, 2018a; Bermingham and Walker, 2017; Budde et al., 2019). Dauphas et al. (2004) recognized that in a Mo *vs.* Ru isotopic space, the BSE and meteorites define a common correlation. Subsequent studies have extended the meteorite and terrestrial datasets, and they largely confirm this conclusion (e.g., Dauphas, 2017; Bermingham and Walker, 2017; Bermingham et al., 2018a). Most terrestrial Ru isotopic data suggest that the BSE is most similar to IAB irons or enstatite chondrites (Fischer-Gödde et al., 2015; Bermingham and Walker, 2017; Bermingham et al., 2018a). These data also indicate that late accretion did not provide a significant source of volatiles to Earth, in contrast to the discussion above based on other elements. Fischer-Gödde and Kleine (2017) analyzed several chondrites and concluded that all chondrites, including carbonaceous chondrites, have Ru isotopic compositions distinct from that of the Earth's mantle. Their data appeared to refute an outer Solar System origin for late accretion and indicated that late accretion was not the primary source of volatiles and water on the Earth.

Recently, Fischer-Gödde et al. (2020) reported the first finding of Ru isotope anomalies in the mantle, in the form of a 22 ppm excess in $^{100}$Ru/$^{101}$Ru of Eoarchean ultramafic rocks from Southwest Greenland (3.8-3.0 Ga). These authors concluded that these materials sampled a part of the mantle containing a substantial fraction of Ru that was accreted before late accretion and that the mantle beneath Southwest Greenland did not fully equilibrate with late accreted material. This interpretation requires that an *s*-process-enriched reservoir contributed to Earth's growth (**Figure 34**), which is contrary to thus far all reported bulk sample meteoritic Ru isotopic data. The excess in $^{100}$Ru/$^{101}$Ru Archaean materials compared to modern mantle was reconciled by the authors if late accretion contained substantial amounts of carbonaceous chondrite-like materials with their characteristic $^{100}$Ru/$^{101}$Ru deficits. This finding thus relaxed previous constraints on the composition of late accretion and brings them in line with previous studies that propose a volatile-rich material late accretion composition (Marty, 2012). More Mo and Ru isotopic data are required, however, to ascertain if the southwest Greenland sample suite is representative of the pre-late accretion mantle, if nucleosynthetic fingerprints observed in the isotopic compositions of other elements in the mantle, and what the identity is of the thus far undetected *s*-process enriched meteorite group (Bermingham, 2020).

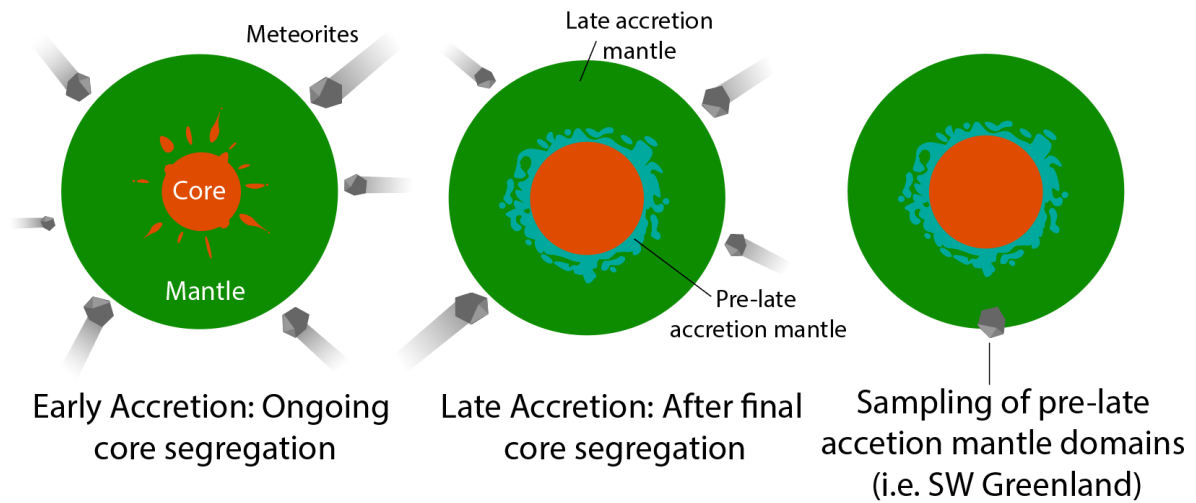

*Figure 34. Possible preservation of pre-late accretion material in the mantle indicated by nucleosynthetic isotope compositions of terrestrial materials. A) Primary accretion saw most siderophile elements segregate to the core. B) After final core formation, ~0.5 to 2% of the total percentage weight of Earth's mass accreted from meteorites during late accretion. C), Fischer-Gödde et al. (2020) report anomalous isotopic compositions in ancient rocks from Southwest Greenland which they conclude is due to the presence of pre-late-accretion mantle material in the rocks. Figure based on (Bermingham, 2020).*

### *4.4.3. Late cometary contribution to Earth?*

Comets, which are rich in water ice (10-50 %) and organic materials, have long been proposed as potential suppliers of water and biologically relevant molecules to early Earth (Bar-Nun and Owen, 1998). The D/H value of cometary water is generally higher than the ocean value (Bockelée-Morvan *et al.*, 2015; Altwegg *et al.*, 2015; Lis *et al.*, 2019), and remote measurements of nitrogen isotopes in cometary coma have shown that at least some molecules in comets are $^{15}$N-rich as well (e.g., Füri and Marty, 2015; Bockelée-Morvan *et al.*, 2015). From mass balance, H and N isotopes of the atmosphere and the oceans, together with argon measurements in 67P/Churyumov-Gerasimenko (67P/C-G) (Balsiger *et al.*, 2015) constrain the cometary water contribution to Earth to one percent or less (Marty *et al.*, 2016). One important finding from the Rosetta mission, however, was that the Xe isotope composition of comet 67P/C-G exhibits a drastic depletion in the two heaviest isotopes ($^{134,136}$Xe), a feature that had been recognized in the terrestrial atmosphere (Pepin, 1991) but had remained unexplained. Based on a binary mixing model between cometary and chondritic components, the cometary contribution to atmospheric Kr and Xe was estimated to be 20 ± 5 % (Marty *et al.*, 2017; Rubin *et al.*, 2018). The relative abundances of noble gases fit well such a mixing trend between cometary and chondritic materials **(Figure 35)**.

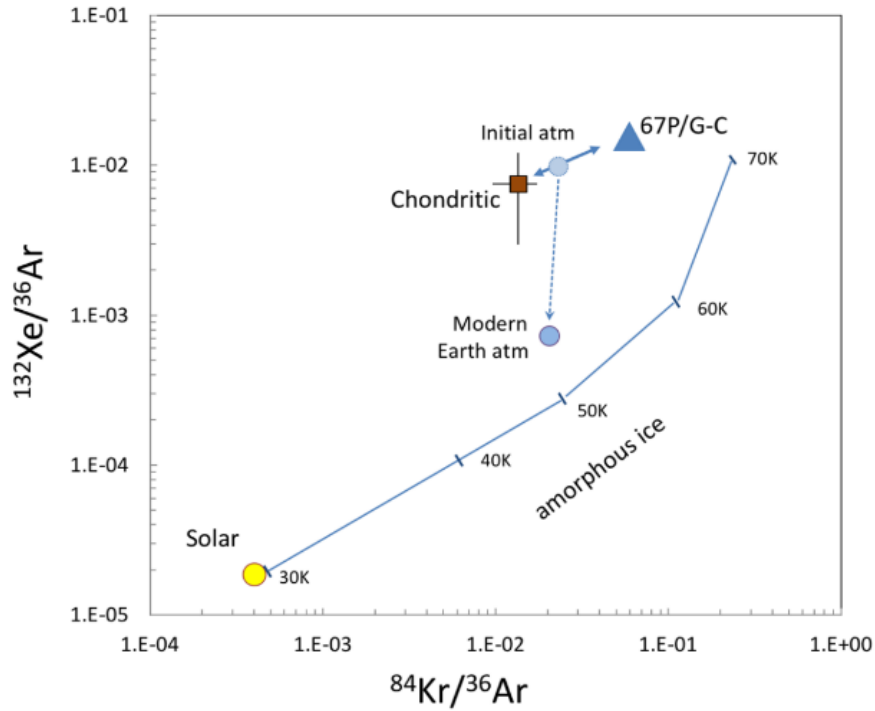

*Figure 35:* *Noble gas elemental ratios of Solar System reservoirs. The solar composition is from Lodders, (2003). The amorphous ice data as a function of temperature are from extrapolations of Dauphas (2003) done using data from Bar-Nun and Owen (1998). 67P/C-G data are from Rubin et al. (2018). Average chondrite data are from compilation in (https://doi.org/10.24396/ORDAR-66). The modern Earth's atmosphere (data from Ozima and Podosek, 2002) data point is off any mixing relation between Solar, chondritic and cometary. With respect to a mixing between chondritic and cometary (double arrow line), the initial atmosphere of the Earth, corrected for loss of xenon to space (Bekaert et al., 2020) fits well with chondritic-cometary mixing. This possibility is substantiated by Xe isotopes measured in 67P/C-G indicating the contribution of ~20 % cometary Xe to ~ 80 % chondritic Xe to form the initial atmosphere of our planet (Marty et al., 2017).*

Comets are noble gas-rich, as such their contribution to the terrestrial inventory can only be distinguished with these elements. For other volatiles, the relative contribution of comets to the water and nitrogen to the surface inventory is low, less than 1% of the oceans and the atmosphere, whatever the composition of the chondritic end-member is considered (e.g., EC or CC; **Figure 36)**.

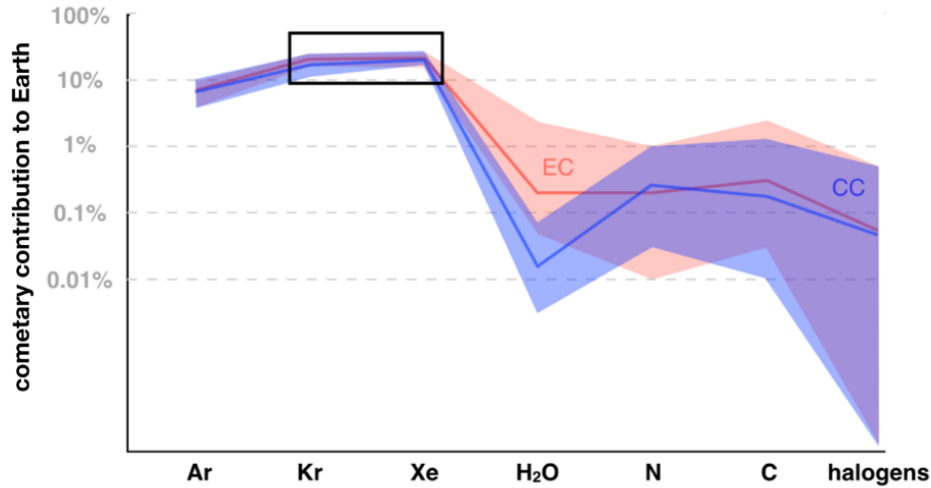

*Figure 36:* Contribution of comets to the inventory of terrestrial volatiles (adapted from Bekaert et al., 2020). The figure is based on a two-component mixing between cometary and chondritic (either EC or CC) compared to bulk Earth inventory. Comets only contributed significantly to noble gases. The colored areas correspond to 95% CI.

### 5. FUTURE WORK

Determining the terrestrial mantle nucleosynthetic isotopic composition and identifying fingerprints of our building blocks will provide the most robust sample-based constraints on our planets cosmochemical origins. With the extensive nucleosynthetic isotope database from meteorites and high precision analytical techniques sufficiently developed, the community is now able to undertake these investigations. The exploration of the Solar System is tending towards targeting more and more sample return missions which will greatly aid in corroborating remote observations with sample-derived constraints. Samples already returned from a primitive carbonaceous chondrite type asteroid, Ryugu, by the Haybusa2 mission from JAXA are presently being analyzed by several hundreds of researchers worldwide and the OSIRIS-REx mission from NASA will also return in 2023 samples from another CC-like asteroid, Bennu. These missions document the most pristine materials yet returned to Earth. These studies will undoubtedly shed light on the formation of Solar System building blocks, as well as on radial mixing that took place early in Solar System history. Returning samples from outer Solar System bodies, such as comets and D-type asteroids would address burning questions such as the origin (Solar or not ?) of outer Solar System bodies, degree of presolar grain homogenization and the NC-CC character of the outer Solar System, dynamics of accretion, and potential contributions to inner Solar System.

Planet formation numerical models are in a state of rapid diversification. New global models are being assembled at a rapid rate by connecting various physical mechanisms – for instance, connecting disk evolution models with streaming instability constraints (e.g., Drazkowska & Alibert 2017), combining pebble accretion with embryo migration (Broz et al., 2021), and connecting the giant planets' instability with a full dynamical picture of the orbital evolution and growth of the terrestrial planets and asteroids (Clement *et al.*, 2018). Evaluating, comparing, and refuting planet formation models requires a combination a cosmochemistry, disk studies (including modeling and observations), and dynamical simulations. Differentiating between planetesimal- and pebble-driven models requires an understanding of the relative contribution of CC-like material to the growing terrestrial planets, which itself must be based on isotopic studies of different elements in a range of samples. Determining where and when planetesimals formed is of paramount importance and will itself require confronting detailed disk models (combined with dust coagulation, drift and streaming instability models) with cosmochemical and astronomical constraints. The Grand Tack model is based on a migration mechanism whose robustness is unclear. New complex hydrodynamical modeling can determine whether it is plausible. Finally, the Early Instability model can best be constrained using a cosmochemical marker of the timing of the instability itself. Indirect markers have already been used (e.g., impact-reset ages like those used in Mojzsis et al., 2019) and perhaps others will be able to pin down the timing of the instability more clearly in the future.

## ACKNOWLEDGEMENTS

This study was supported by the European Research Council (ERC) under the European Union's Horizon 2020 research and innovation program (PHOTONIS Advanced Grant # 695618 to BM). K.R.B. received support from NASA (80NSSC20K0997), NSF (EAR-2051577), and the Department of Earth and Planetary Science Rutgers University. H. Tornabene (Rutgers University) is thanked for assistance in creating figures. S.N.R. thanks the CNRS's PNP and MITI/80PRIME programs for support. L.R.N thanks the Carnegie Institution for support. We are grateful to Mike Zolensky and the reviewers for helpful suggestions and corrections.